\documentclass[aps,preprint,showpacs,superscriptaddress]{revtex4} 

\usepackage{graphicx}
\usepackage{amssymb}
\usepackage[dvips]{epsfig}
\usepackage{subfigure}
\usepackage{amsmath}
\usepackage{float}
\DeclareMathOperator{\sech}{sech}
\newcommand{\scl}{0.30}

\begin{document}

\title{Solitary Wave Formation under the Interplay between \\ 
Spatial Inhomogeneity and Nonlocality}       
\author{Konstantinos Dragonas}
\author{Yannis Kominis} 
\affiliation{%
 School of Applied Mathematical and Physical Science,\\
			 National Technical University of Athens,\\
			 Zographou GR-15773, Greece}%

\date{\today}

\begin{abstract}
The presence of spatial inhomogeneity in a nonlinear medium restricts the formation of Solitary Waves (SW) on a discrete set of positions whereas a nonlocal nonlinearity tends to smooth the medium response by averaging over neighboring points. The interplay of these antagonistic effects is studied in terms of SW formation and propagation. Formation dynamics is analyzed under a phase space approach and analytical conditions for the existence of either discrete families of Bright SW or continuous families of Kink SW are obtained in terms of Melnikov's method. Propagation dynamics are studied numerically and cases of stable and oscillatory propagation as well as dynamical transformation between different types of SW are shown. The existence of different types and families of SW in the same configuration, under appropriate relations between their spatial width and power with the inhomogeneity and the nonlocality parameters, suggests an advanced functionality of such structures that is quite promising for applications.
\end{abstract}
\maketitle

\section{\label{sec:level1}	INTRODUCTION}
Solitary Waves (SWs) are self-localized wave entities resulting from the interplay of mutually counterbalancing linear and nonlinear effects. In the case where the model describing the physical system is integrable, such wave entities are well known as spatial solitons \cite{trillo2013}. In nonlinear optics, such is the case of the NonLinear Schr{\"o}ndiger Equation (NLSE), which models electromagnetic wave propagation in a dielectric material with nonlinear response due to intensity dependent change of the refractive index. Among various types of nonlinearities, only Kerr (cubic) nonlinearity results in an integrable system supporting solitons. In all other cases of local nonlinearities \cite{rottwitt2014} robust SW exist which are not solitons in the strict mathematical sense and have been extensively studied for decades \cite{yang2010}. A Kerr type of local nonlinearity is also present in Bose-Einstein Condensates (BEC) matter wave realizations, when taking into account collisions between atoms in the mean field approximation \cite{kivshar2001nonlinear}. \

Recent technological advances have focused on both theoretical and experimental research towards wave self-localization in spatially inhomogeneous nonlinear systems \cite{2019frontiers1,christodoulides2003,kevrekidis2007}. In this context lies the increased interest for NLS-type partial differential equations with periodic potentials, also known as Gross-Pitaevskii equations, which arise in the modeling of various physical systems in different scientific areas like optics \cite{trillo2013}, Bose-Einstein condensation \cite{abdullaev2006} and plasma physics \cite{kuznetsov1986}, due to their common underlying models of wave propagation. In nonlinear optics, the periodic potential signifies the periodic transverse variation of the refractive index which can be either optically induced in photorefractive \cite{efremidis2002,fleischer2003,neshev2003,neshev2004, chen2005,rosberg2005,song2006}  and Kerr \cite{kominis2004,kominis2005,tsopelas2006,tsopelas2007} materials or prefabricated as in waveguide arrays and photonic crystals \cite{trillo2013}. Regarding Bose-Einstein condensation, the periodic potential corresponds to an optical lattice which is induced by the interference of coherent laser beams in order to confine the condensate \cite{pitaevskii2006,brazhnyi2004,morsch2006,alfimov2002,kevrekidis2005,pelinovsky2004,konotop2002,kartashov2011snl}. Moreover, the introduction of a spatial  modulation of the BEC's atomic scattering length is modeled by a periodic modulation of the respective NLSE-type cubic term \cite{kartashov2011snl}. 
Overall, the presence of spatial inhomogeneity in a nonlinear medium increases the complexity of Solitary Wave formation and dynamics and introduces new interesting phenomena as a result of a unique interplay between the medium inhomogeneity and nonlinearity. Among them are: the formation of a rich set of all kinds of solitary waves with a quite robust behavior under propagation \cite{kominis2013,kominis2008,kominis2007,kartashov2005cpl,zhong2012}, their stabilization in multidimensional settings by the use of spatially periodic potentials induced by photonic lattices or arrays \cite{2019frontiers1} and the strong dependence of the position and stability of such spatially localized structures on the characteristics of the medium \cite{kominis2008}. These features are directly related to the breaking of the translational invariance of wave propagation in the inhomogeneous medium; in contrast to the homogeneous case, the inhomogeneity introduces a set of discrete preferential positions where SW can be formed and propagate in a stable fashion.

However, the nonlinear response of several optical media is characterized by nonlocality. For example, in Bose-Einstein condensate physics, the importance of this phenomenon is addressed in the approach to include dipolar effects and signifies the long-range character of the dipolar interaction \cite{lahaye2009} as well as in the study of laser-induced attractive interactions which result in the stabilization of condensates \cite{o2000bose}. In nonlinear optics, an underlying mechanism adding a nonlocal contribution to the nonlinear response has been addressed in photorefractive crystals in the presence of diffusion effects \cite{krolikowski1996,christodoulides1995,carvalho1995,christodoulides1996}, in optical systems with thermal nonlinearity \cite{shi2012} and liquid crystals \cite{conti2003,conti2004,papazisimos2013}, accounting for a given point's refractive index dependence on the distribution of light's intensity in it's vicinity. 
Several studies have revealed that the nonlocality of nonlinear response affects solitary waves' stability and mobility \cite{krolikowski2000} with an asymmetric nonlocal response having an even stronger impact on the latter \cite{kartashov2004, xu2006gap,zhang2014}, leads to prevention of catastrophic collapse \cite{bang2002collapse} and suppression of modulation instability (MI) \cite{wyller2002}. These features result from the fact that nonlocality tends to smooth the nonlinear response of the medium at a specific position, due to its dependence on the average field and medium characteristics in a neighborhood of each position.  

In the present work, we consider an NLS-type of equation with asymmetric nonlocal nonlinear response along with the presence of spatial inhomogeneity. The action of these two features is antagonistic in the sense that the inhomogeneity tends to ``discretize'' the response of the medium by restricting the SW formation only at a specific set of positions, whereas the nonlocality tends to ``smooth'' its response due to averaging over nearby positions. In order to study this interplay we perform a phase space analysis of the corresponding dynamical system that provides a clear geometric intuition for the SW formation dynamics. Moreover, we utilize an analysis based on Melnikov's perturbation theory and we obtain analytical conditions for the formation of different types of either discrete or continuous families of SW in terms of bifurcations in the parameter space of the system. The phase space analysis enables the systematic categorization of a large variety of qualitatively different SW as well as their profile calculation. The latter are also studied numerically in terms of their propagation dynamics.

\section{\label{sec:level2}	INHOMOGENEOUS NONLOCAL NLSE AND MELNIKOV'S METHOD}
A general model for the description of solitary waves' propagation in a nonlocal nonlinear Kerr type medium with transverse modulation of the linear refractive index is described by the following modified NLSE \cite{wyller2002}
\begin{equation} \label{eq:1} 
i\frac{\partial u}{\partial z}+\frac{\partial^2 u}{\partial x^2}+\eta(x) u+2\Delta n(I)u=0 
\end{equation}
where $u$ is the wave field envelope, $z$ is the normalized propagation distance, $x$ is the scaled transverse coordinate and $\eta(x)=-A\sin(kx)$ describes a harmonic modulation of the linear refractive index along the transverse spatial dimension, with the parameters $A$ and $k$ standing for the amplitude and the spatial frequency (wavenumber) of the refractive index's modulation respectively. $\Delta n(I)$ expresses the change of the refractive index induced by light's intensity $I(x,z)=|u(x,z)|^2$ having the nonlocal form
\begin{equation} \label{eq:2} 
\Delta n(I)=\int\limits_{-\infty}^{\infty} R(x'-x)I(x',z)dx'
\end{equation}
The function $R(x)$ is the response function of the medium which is assumed to be real and $L^1$ integrable. Besides that, depending on the physical system that is being modeled it may also be symmetric or asymmetric. Based on the latter distinction and under the hypothesis that the response function is small compared to the field's intensity, eq.(\ref{eq:1}) can be approximated by models with a second or a first order derivative of field intensity respectively, omitting this way the integral term \cite{wyller2002}. 
By considering the first order nonlocal contribution of the nonlinear response of the medium \cite{xu2006gap,zhang2014}, the equation is written as 
\begin{equation} \label{eq:3} 
iu_z+u_{xx}+2|u|^2u+\eta(x) u+2\gamma\frac{\partial|u|^2}{\partial x} u=0
\end{equation}
where the parameter $\gamma=\int\limits_{-\infty}^{\infty} xR(x)dx$ is related to the medium properties and describes the magnitude of the first order nonlocal component of nonlinear response.

In general, SW profiles can be considered as spatially localized transitions between two asymptotic states. Depending on the values of these asymptotic states at infinity, a SW is characterized as Bright, Kink, Dark or anti-Dark \cite{trillo2013}. Thus, in terms of our dynamical system analysis, the existence of SWs is equivalent to the existence of solutions constituting localized transitions between hyperbolic solutions corresponding to their asymptotic states \cite{kominis2013}. As a consequence, the first step towards discovering the existence and the types of SWs supported by the system, is the study of the existence of hyperbolic solutions resulting from any possible transverse intersections or smooth connections between the stable and unstable invariant manifolds either of the same or different basic hyperbolic trajectories. 

The formation dynamics of SW profile in such configurations, is completely described by the reduced dynamical system resulting from the consideration of stationary wave solutions supported by eq.(\ref{eq:3}), having the form
\begin{equation} \label{eq:4} 
\\ u(x,z)=\psi(x) e^{i\beta z},
\end{equation}
with $\psi(x)$ a real function describing the transverse wave profile and $\beta$ being the real propagation constant. The respective stationary equation is the following:
\begin{equation} \label{eq:5}
-\beta\psi+\psi_{xx}+2\psi^3+4\gamma\psi^2 \psi_x = A sin(kx) \psi,
\end{equation}
and corresponds to a non-autonomous dynamical system due to the explicit dependence on the spatial variable $x$ playing the role of ``time''. By setting $\phi=kx$ and $\dot{\psi}=v$, where $\dot{\psi}\equiv\frac{d\psi}{dx}$, eq.(\ref{eq:5}) can be rewritten as an autonomous system, 
\begin{equation} \label{eq:6}
\begin{gathered}
\dot{\psi} = v\\
\dot{v} = \beta\psi-2\psi^3+\epsilon(-4\gamma\psi^2v+Asin(\phi)\psi)\\
\dot{\phi} = k
\end{gathered}
\end{equation}
where the nonlinear terms of nonlocality and inhomogeneity can be considered as perturbations, by being multiplied with a dimensionless perturbation parameter $\epsilon$, that can be set equal to unity in the final results. 

For $\epsilon=0$, the dynamical system (4) coincides with the nonlinear Duffing oscillator, which is a one degree of freedom integrable system with Hamiltonian \cite{wiggins2003}
\begin{equation} \label{eq:7}
H=\frac{1}{2}\dot{\psi}̇^2-\frac{\beta}{2}\psi^2+\frac{\psi^4}{2}
\end{equation}

Considering the case $\beta>0$, the saddle of the origin $(\psi,v)=(0,0)$ of the phase space, posseses a pair of smoothly joined stable and unstable invariant manifolds 
forming a homoclinic solution given as
\begin{equation} \label{eq:8}
\begin{split}
q_0^{\pm}(x)\equiv(\psi_0^{\pm}(x),v_0^{\pm}(x))=(\pm\sqrt{\beta}\sech{\sqrt{\beta}x},\mp\sqrt{\beta}\sech{\sqrt{\beta}x}\tanh{\sqrt{\beta}x})
\end{split}
\end{equation}
which corresponds to the stationary Bright soliton solution of the unperturbed NLSE, where the propagation constant $\beta$ also determines the amplitude and the spatial width of the soliton. However, this smooth join of the stable and unstable manifolds of the origin of the unperturbed system, is a highly degenerate structure that is expected to break under perturbation. More specifically, the presence of the nonlinear terms due to nonlocality and spatially-dependent terms due to inhomogeneity leads to the lack of integrability and as a consequence the analytic determination of SWs profiles is no longer possible. 

For a sufficiently small $\epsilon$, the saddle of the origin transforms to an unstable hyperbolic orbit in the extended three-dimensional phase space $\gamma_\epsilon(x)$ with $(\psi,\dot{\psi})=(0,0)$ for all $x$, whose invariant manifolds may either intersect transversely or not, depending on the values of the parameters of the system. Such an existence of transverse intersections, gives rise to the existence of a discrete set of Bright SWs, whose profiles can be located by the determination of the homoclinic solutions of the reduced system of ODEs (\ref{eq:6}). Likewise, in the case where these two invariant manifolds do not intersect, there is the possibility of Kink SWs formation. In terms of our phase space approach, the asymmetric profiles of Kink solutions, can be formed either by the smooth joining or transverse intersections of an invariant manifold of the origin's hyperbolic trajectory, with the corresponding opposite stability's invariant manifold of a different hyperbolic trajectory. The cases of smooth joining and transverse intersections result in either continuous or discrete families of Kink SW. In addition, in the case where transverse intersections take place, the existence of homoclinic points on the Poincare Section $\Sigma^{\phi_0}$ implies the existence of chaotic dynamics according to Moser or Smale-Birkhoff theorem \cite{wiggins2003}. It is worth emphasizing that the condition for the existence of Bright SW is directly related to the condition for chaotic dynamics. In fact, the complexity of the homoclinic structure that gives rise to chaoticity also gives rise to a rich set of Bright SW with spatially complex profiles corresponding to different homoclinic points of an uncountable set. The interplay between inhomogeneity and nonlocality suggests that the presence of the latter tends to reduce homoclinic chaos.\

In order to give a quantitative form in the above qualitative analysis, we are going to investigate the existence of homoclinic points and transverse intersections of the stable and unstable manifolds of the hyperbolic orbit at the origin with the utilization of Melnikov's perturbation theory. The general form of Melnikov function that corresponds to our problem is given as follows \cite{wiggins2003}
\begin{equation} \label{eq:9}
M(x_0,\phi_0)=\int_{-\infty}^{\infty} DH(q_0(x))\cdot g(q_0(x),kx+kx_0+\phi_0) dx
\end{equation}
where  $DH=(\frac{\partial H}{\partial \psi},\frac{\partial H}{\partial v})$, $g=(g_1,g_2)$ is the perturbative part and $x_0$ is the evolution interval that takes the point $(q_0(-x_0),\phi_0)$ of the parameterized unperturbed homoclinic trajectory to reach $(q_0(0),\phi)$. Under proper substitutions, Melnikov's function for system of eq.(\ref{eq:6}) takes the following form
\begin{equation} \label{eq:10}
M^{\pm}(x_0,\phi_0)=\int_{-\infty}^{\infty} v_0^{\pm}(x)(-4\gamma \psi_0^{\pm}(x)^2 v_0^{\pm}(x)+
A sin(kx+kx_0+\phi_0)\psi_0^{\pm}(x)) dx
\end{equation}
and is equal with
\begin{equation} \label{eq:11}
\begin{split}
M(x_0,\phi_0)=-\frac{16\gamma\beta^3}{15\sqrt{\beta}}-\frac{\pi A k^2 \cos{(k x_0+\phi_0)}}{2\sinh{\frac{\pi k}{2\sqrt{\beta}}}}
\end{split}
\end{equation}
By keeping $\phi_0$ constant, eq.(\ref{eq:11}) describes the separation of the stable and unstable manifolds of the hyperbolic orbit on the Poincare Section $\Sigma^{\phi_0}=\big\{(\psi,v,\phi_0)\in \mathbb{R} \times \mathbb{R} \times S^1 \big\}$. The values of $x_0$ for which the Melnikov function is zero, parameterize the discrete set of homoclinic intersection points on a given Poincare Section $\phi_0$. It is clear that the condition that must be satisfied for the existence of such intersections is 
\begin{equation}\label{eq:12}
\frac{\gamma}{A}<\frac{15\sqrt{\beta}\pi k^2}{32\beta^3\sinh{\frac{\pi k}{2\sqrt{\beta}}}} 
\end{equation}

\begin{figure}[pt]	
\begin{center}
 	\subfigure[]{\scalebox{\scl}{\includegraphics{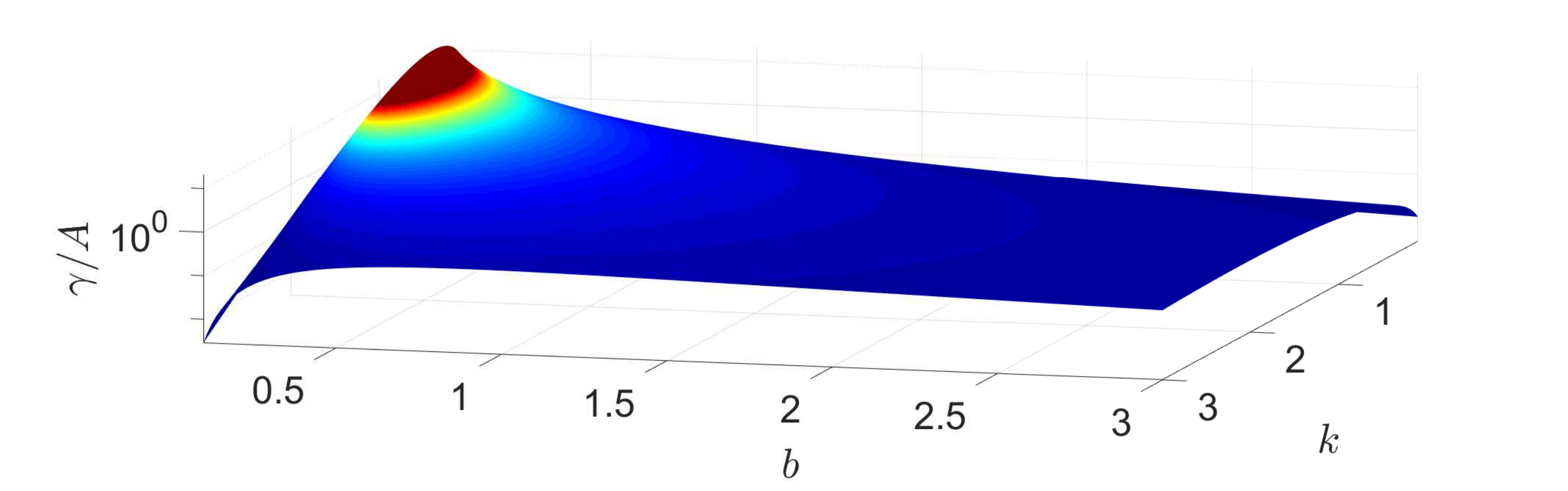}}}\\
  	\subfigure[]{\scalebox{\scl}{\includegraphics{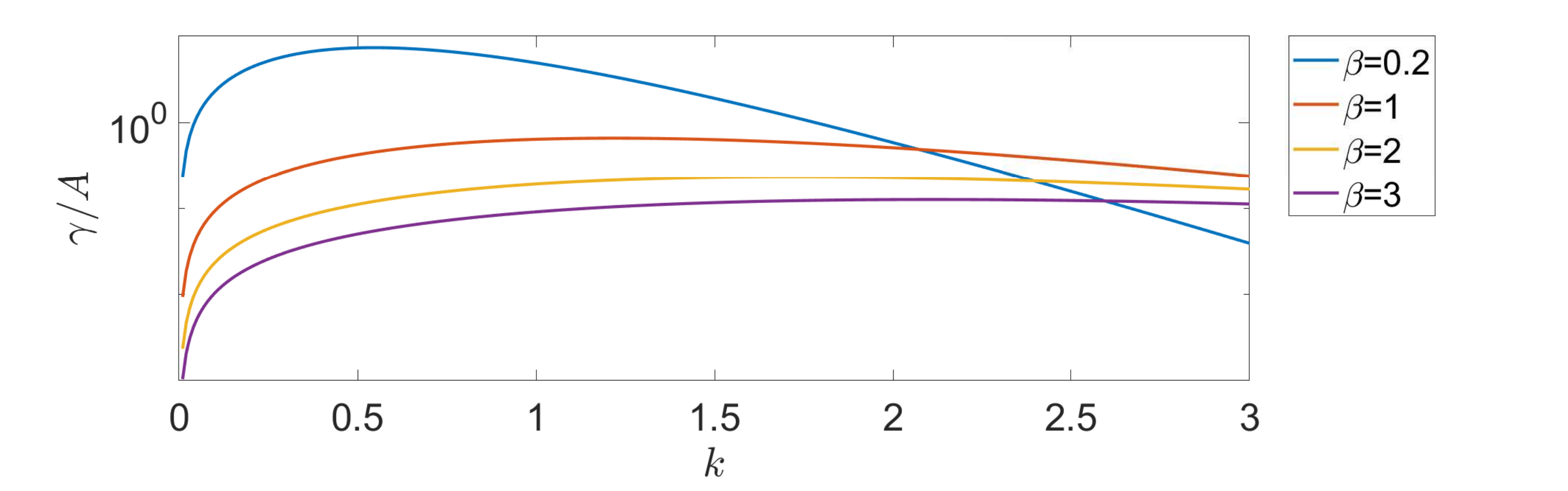}}}
  	\subfigure[]{\scalebox{\scl}{\includegraphics{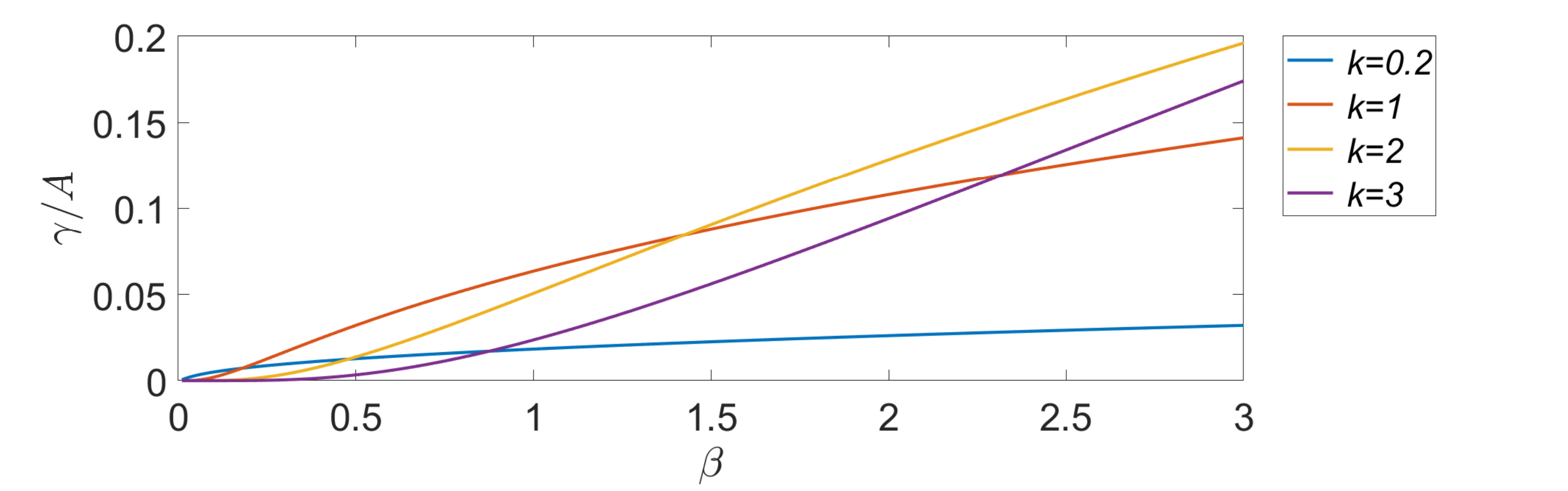}}}
  	\caption{Melnikov critical surface (a) and cross-sections of the surface for four different cases of propagation constants $\beta$ (b) and spatial frequencies of the inhomogeneity $k$ (c) respectively. The vertical axis depicting $\gamma/A$ in (a) and (b) is in logarithmic scale due to its exponential dependence on the ratio of the characteristic spatial scales, namely the period of the inhomogeneity $2\pi/k$ and the SW width $1/\sqrt{\beta}$.}
 	\label{Fig:1}
\end{center}
\end{figure}

For parameter values satisfying relation (\ref{eq:12}), the existence of homoclinic orbits for (\ref{eq:4}) and as a consequence the existence of SWs of Bright type for (\ref{eq:3}) is deduced. In order to visualize this relation, we illustrate the resulting critical surface in Fig.\ref{Fig:1}(a) along with cross-sections of the critical surface for four different cases of fixed propagation constants and wavenumbers in Fig.\ref{Fig:1}(b) and (c) respectively. It is clear that the condition for the formation of Bright SW depends strongly on both the parameters of the medium $(\gamma, k)$ and the propagation constant of the SW ($\beta$). In fact, the presence of the hyperbolic sine term in relation (\ref{eq:12}), implies an exponential dependence on the ratio of the characteristic spatial scales, namely the period of the inhomogeneity $2\pi/k$ and the SW width $1/\sqrt{\beta}$. Under this perspective, it turns out that through the Melnikov condition for homoclinic orbits, we can extract quantitative conclusions for the existence of different SW in a wide range of inhomogeneous nonlinear photonic structures with nonlocal response.

\section{\label{sec:level3}	RESULTS AND DISCUSSION}

In order to proceed to the determination of the exact forms of the SW profiles, the dynamics of the system (\ref{eq:6}) is going to be studied by utilizing a phase space analysis based on the Poincare surfaces of section, $\Sigma^{\phi_0}$, due to the periodicity of the $\phi$ dependence. We are interested in determining the invariant manifolds $W^{u,s}(\gamma_\epsilon(x))$ of the hyperbolic trajectory at the origin $\gamma_\epsilon(x)$. In order to obtain accurate SW solutions for each parametric subset, we utilized an analytical approximation of the local invariant manifolds of the hyperbolic trajectory restricted on a Poincare surface of section, $W_{loc}^{u,s}(\gamma_\epsilon(x))\cap\Sigma^{\phi_0}$. Due to the reduction of the dynamics on a constant, arbitrarily chosen, $\phi_0$ plane, we restrict our study on the first two parts of the system of eq.(\ref{eq:6}) 
\begin{equation} \label{eq:13}
\begin{gathered}
\dot{\psi} = v\equiv F_1(\psi,v)\\
\dot{v} = \beta\psi-2\psi^3-4\gamma\psi^2v+Asin(\phi_0)\psi \equiv F_2(\psi,v)\\
\end{gathered}
\end{equation}
In the above reduced system, we mention that $F_1(0,0)=F_2(0,0)=0$ as a consequence of the fact that the intersection of the hyperbolic trajectory $\gamma(x)$ with the $\Sigma^{\phi_0}$ plane is the point $(0,0,\phi_0)$ on the extended phase space of the system. By applying a linear transformation to transform the coordinates of the reduced nonlinear vector field in the appropriate form, we make use of the stable, unstable and center manifolds theorem for vector fields \cite{wiggins2003}. The exploitation of the initial as well as the tangency conditions for the hyperbolic trajectory's local invariant manifolds implied by the latter theorem, allows for the calculation of a Taylor expansion representing the restriction of the manifolds on a Poincare surface of section.  Following this approximation, we apply the inverse linear transformation to obtain the respective analytical expressions in the $(\psi,v)$ system of coordinates. The tracing of the global invariant manifolds on $\Sigma^{\phi_0}$ are given through the following relation
\begin{equation} \label{eq:14}
W^{u,s}(\gamma(x))\cap\Sigma^{\phi_0}=\bigcup_{\substack{x_u=n\frac{2\pi}{k}, x_s=-n\frac{2\pi}{k}, \\n\in\mathbb{N}}}q_{x_{u,s}} (W_{loc}^{u,s}(\gamma(x))\cap\Sigma^{\phi_0}), \qquad x \in \mathbb{R}
\end{equation}
where $q_x$ is the solution of the dynamical system's (\ref{eq:6}) initial value problem. 

\subsection{\label{ssec:level1} BRIGHT SOLITARY WAVES}
Bright solitary waves correspond to asymptotic states with zero values at infinity. As a consequence, the respective orbits determining their profile are those homoclinic to the hyperbolic trajectory or to the Poincare map's saddle fixed point on $\Sigma^{\phi_0}$, and their existence is valid for parameter sets satisfying Melnikov condition. By fixing the wave's propagation constant to the value $\beta=1$, we begin with the examination of SW formation dynamics for two quite different cases of photonic structures, in terms of their characteristic parameters. In Fig. 2(a) and 2(b) the intersection of the stable and unstable manifolds of the saddle point with the Poincare surface of section is shown for the parameter sets satisfying Melnikov's condition $(\gamma=0.00025,A=0.008,k=1.8)$ and $(\gamma=0.0075,A=0.8,k=2.6)$ respectively. Transverse intersections of the respective invariant curves correspond to different Bright SW profiles. 

Regarding the first case, it is shown that a relatively small amplitude of the spatial inhomogeneity combined with a relatively small magnitude of nonlocality, results in a 
small degree of deformation of the stable and unstable manifolds with respect to their unperturbed form. Fundamental SW profiles corresponding to the first two intersection points (1)-(2) of Fig. 2(a), have simple profiles with single maxima located close to the maxima and minima of the refractive index as shown in Fig. 3(a),(b). More complex profiles having more than one maxima correspond to intersection points (3)-(6) of Fig. 2(a) as shown in Fig. 3(c)-(f). The latter can be intuitively considered as bound states consisting of in-phase and out-of-phase combinations of the simple single-maximum fundamental states.  Despite of the profile similarity of the fundamental SW of Figs. 3(a) and 3(b), the relative position of their centers with respect to the underlying inhomogeneity plays a crucial role on their propagation dynamics. As shown in Figs. 4(a) and 4(b), oscillatory propagation takes place for SW centered arround a local minimum of the refractive index whereas a purely stable propagation takes place for SW centered around local maxima. SW profiles having the form of bound states undergo more complex propagation dynamics, including oscillatory instabilities such as in Fig. 4(e) as well as breaking of the bound states to simpler SW propagating and undergoing mutual attracting or repelling interactions, as shown in Figs. 4(c), (d), (f). \

For the case of a photonic structure with a parameter set corresponding to stronger perturbation, as shown in Fig. 2(b), the stable and unstable manifolds present large excursions from their unperturbed form and intensive multiple foldings forming a complex structure of homoclinic intersection points which preludes a more complex morphology of homoclinic orbits and as a consequence the existence of more complex SW profiles. Indeed, it is evident from Figs. 5 that in addition to simple SW profiles like those shown in Figs. 5(a) and 5(b), more complex ones are also possible as in Figs. 5(c)-(f). The latter have the form of strongly bounded states which cannot be considered as consisting of combinations of fundamental SW and they are characterized by a high degree of asymmetry with no clear correspondence of their profiles to the extrema of the underlying inhomogeneity of the refractive index. Inspite of the strong perturbation, fundamental SW centered around the local maxima of the refractive index and relatively simple bound states, propagate in a stable or oscillatory fashion as shown in Figs. 6(a), (b)

The drastic exponential dependence of the Melnikov condition (\ref{eq:12}) on the propagation constant $\beta$ suggests that even in the same photonic structure significantly different dynamics of SW formation and propagation can take place. For the case of a parameter set  $(\gamma=0.0025,A=0.1,k=1.6)$ and $\beta=1, 0.2$ with both values of $\beta$ satisfying the Melnikov's condition, we can have qualitatively different forms of the stable and unstable manifolds and the corresponding sets of homoclinic points as shown in Fig. 7(a) and (b). In fact the form of the Melnikov function, measuring the distance between the two manifolds suggests that the effective perturbation strength and the deviation of the manifolds from their unperturbed form depends strongly on $\beta$. Characteristic SW profiles along with their propagation dynamics are depicted in Figs. 8 and 9, for $\beta=1, 0.2$ respectively. It is clear that, since a smaller propagation constant also results to a wider spatial profile of the unperturbed soliton (\ref{eq:8}), the SW profiles with $\beta=0.2$ extend to larger number of periods of the underlying inhomogeneity of the refractive index in comparison to the case of $\beta=1$.

\subsection{\label{ssec:level2} KINK SOLITARY WAVES}
For the case of parameter sets where the Melnikov condition is not fulfilled, the stable and unstable manifolds of the saddle point at the origin do not intersect and therefore the existence of Bright SW is not supported. However, it is possible that one of the invariant manifolds of the saddle can be either smoothly connected or transversely intersected with an invariant manifold of opposite stability of another closed hyperbolic trajectory. This case corresponds to a SW profile characterized by the spatially localized transition between two different asymptotic states, that is known as a Kink SW. In fact, in our case the unstable manifold of the saddle point at the origin is joint smoothly with the stable manifold of a periodic orbit as shown in the Poincare surface of section of the system in Fig. 10(a)-(d) for the indicative parametric sets $(\beta=0.2,\gamma=0.175,A=0.02,k=0.5)$, $(\beta=0.2,\gamma=0.425,A=0.001,k=0.5)$, $(\beta=1,\gamma=0.3,A=0.001,k=3)$ and $(\beta=0.2, \gamma=0.4, Α=0.05, k= 0.5)$ respectively. The smooth connection of the two manifolds results in the existence of a continuous family of Kink SW each member of which corresponds to a different point of this connection. This is in sharp contrast to the case where Bright SW exist, where SW can be formed in a discrete set of positions corresponding to the intersections of stable and unstable manifolds and manifestates the role of a stronger nonlocality.  It is worth mentioning that the periodic orbits, corresponding to spatially periodic stationary solutions, result from resonances between the frequencies of the periodic orbits of the unperturbed system, under no inhomogeneity and nonlocality, with the frequencies of the inhomogeneity. Their existence, similarly to the case of homoclinic orbits, depends on the interplay of the inhomogeneity and the nonlocality and can be analytically predicted in terms of the sub-harmonic Melnikov method \cite{wiggins2003}.\

The corresponding profiles of the Kink SW along with their propagation dynamics are depicted in Figs. 11 and 12 for different parameters. It is shown that both the parameters of the underlying inhomogeneous structure and the SW characteristic width $\beta$, determine the SW profile in terms of the amplitude and period of the nonzero asymptotic state as well as the spatial extent of the transition between the two states. It is worth emphasizing that the SW profiles shown in Fig. 12 are indicative members of a continuous family of SW having the same propagation constant $\beta$. Moreover, it is shown that the Kink SW can be quite robust under propagation, whereas cases of complex propagation dynamics are also possible, such as the one depicted in Fig. 11(f) where the SW profile decomposes to secondary beams propagating along a parabolic trajectory.

\section{\label{sec:level4}SUMMARY AND CONCLUSIONS}
Self-localization dynamics and Solitary Wave formation have been studied for a photonic structure characterized by a transversely inhomogeneous linear refractive index and an asymmetric nonlocal nonlinear response, by utilizing a standard NLSE model. SW formation dynamics has been investigated under a phase space approach and specific conditions for the formation of SW of either Bright or Kink type have been obtained analytically in terms of Melnikov's method. The conditions involve the parameters of the photonic structure (nonlocality parameter, magnitude and period of the inhomogeneity) and the wave (propagation constant, spatial width) and provide an analytical criterion for the result of the antagonistic interplay between the effects of the inhomogeneity and the nonlocality, tending to discretize and smooth the medium response respectively. It is shown that either discrete families of Bright SW located at specific positions or continuous families of Kink SW can be formed, depending of the relative strength of these effects. The latter is expressed in terms of the three characteristic spatial scales of the system, that is the SW width, the period of the inhomogeneity and the spatial range of nonlocality. The exact SW profiles have been obtained via the calculation of the location of homoclinic points in a Poincare surface of section and their propagation dynamics have been studied numerically. Cases of stable and oscillatory propagation where shown as well as cases where transformations between different SW profiles occur along propagation.\

It is shown that different types and families of SW can be formed in the same photonic structure depending on their power (also related to their spatial width) suggesting an advanced functionality of such photonic structures in terms of power discrimination and selectivity that is quite promising for applications. Cases of self-defocusing nonlinearities supporting the existence of Dark and anti-Dark solitons as well as two-dimensional inhomogeneous nonlocal configurations can also be studied under similar approaches.

\newpage

\begin{figure}[pt]	
\begin{center}
 	\subfigure[]{\scalebox{\scl}{\includegraphics{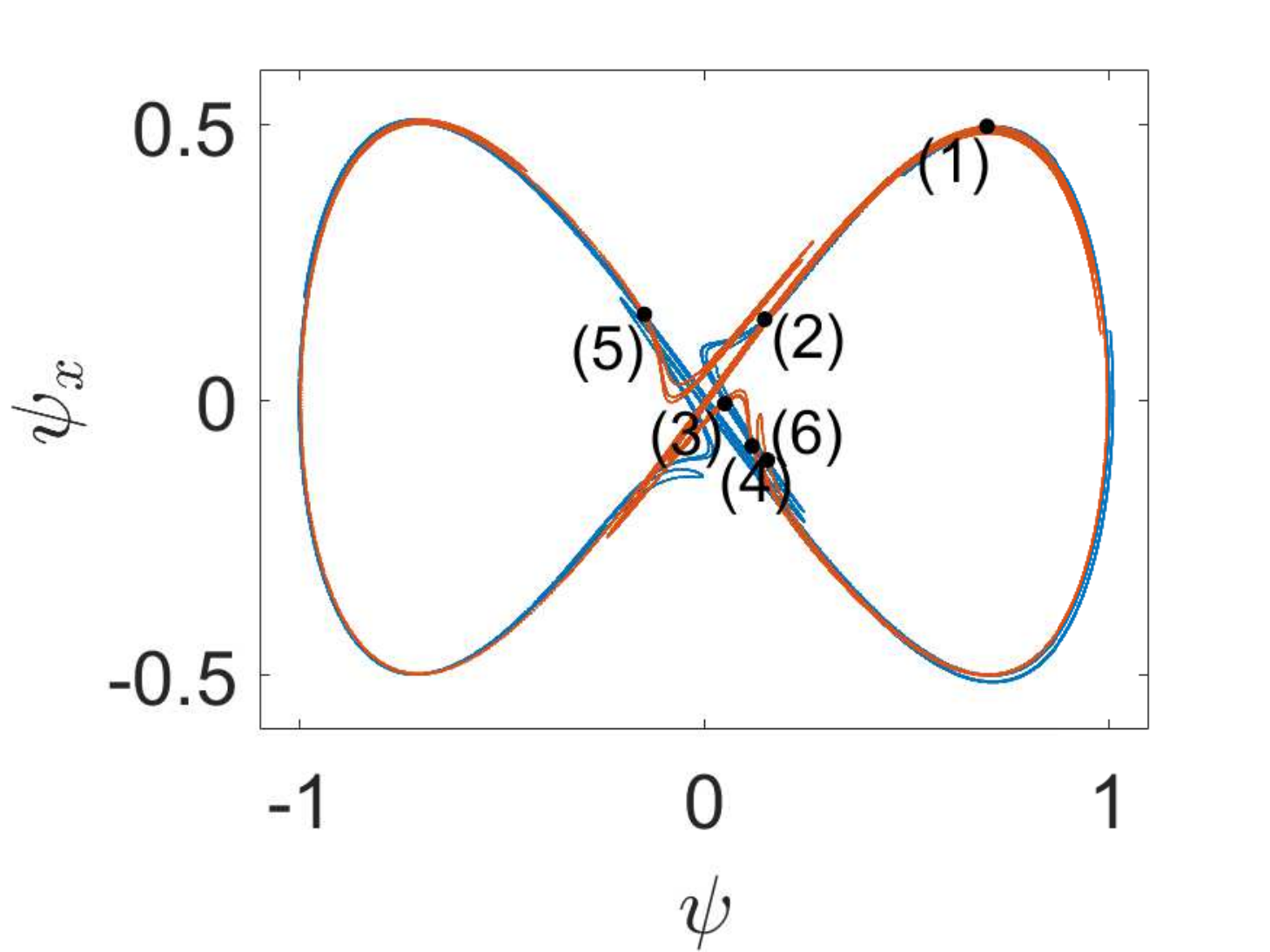}}}
 	\subfigure[]{\scalebox{\scl}{\includegraphics{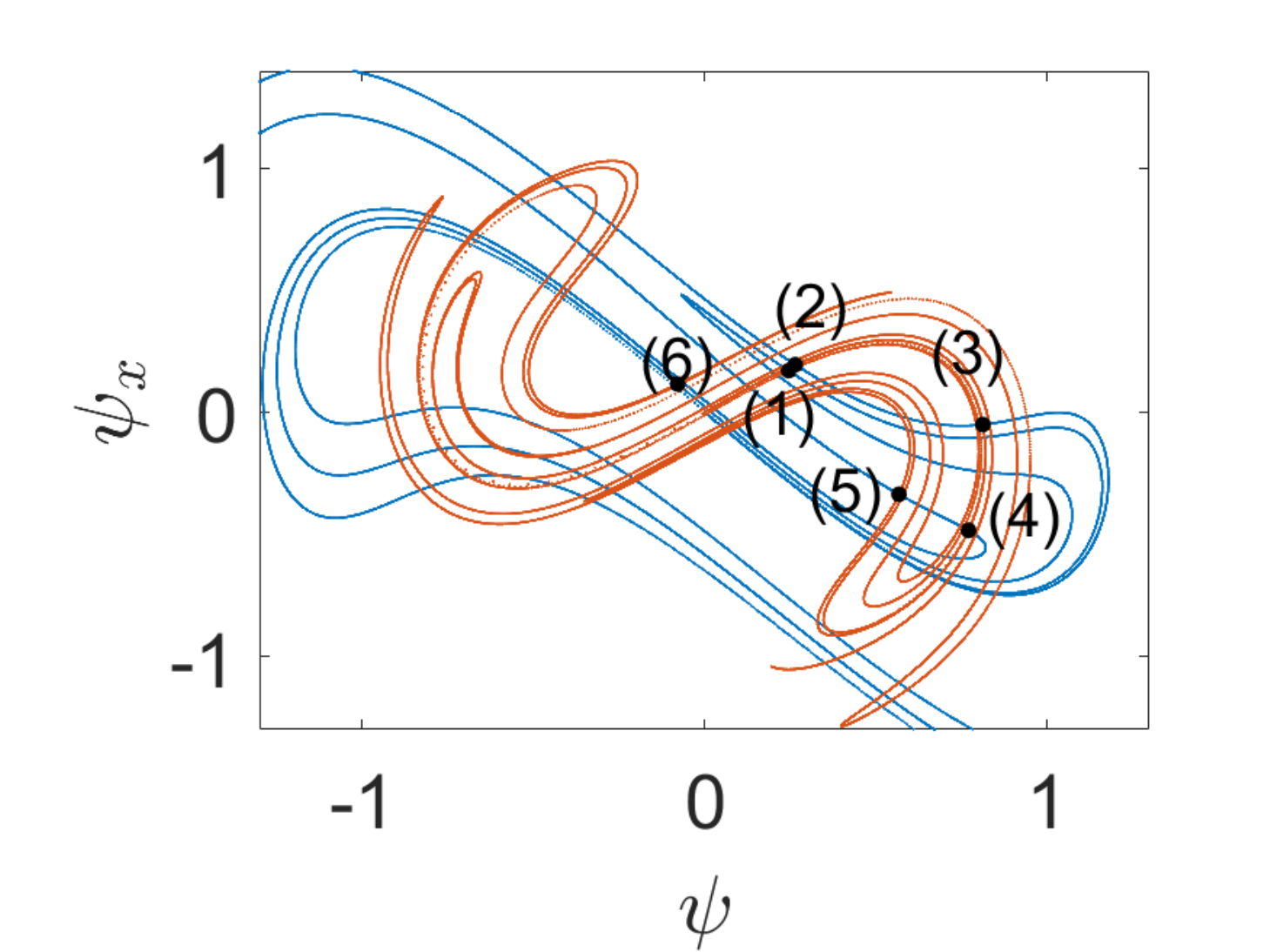}}}
  	\caption{Branches of the stale (blue) and unstable (red) manifolds of the hyperbolic trajectory in the Poincare surface of section $\Sigma^{0}$, for two different case of photonic structures namely, (a) $\beta=1,\gamma=0.00025,A=0.008,k=1.8$ and (b) $\beta=1,\gamma=0.0075,A=0.8,k=2.6$. Black dots denote the intersections of the manifolds for which SW profiles were studied. Both parameter sets correspond to cases where the Melnikov condition for the existence of homoclinic orbits is satisfied.}
 	\label{Fig:2}
\end{center}
\end{figure}

\begin{figure}[pt]
\begin{center}
 	\subfigure[]{\scalebox{\scl}{\includegraphics{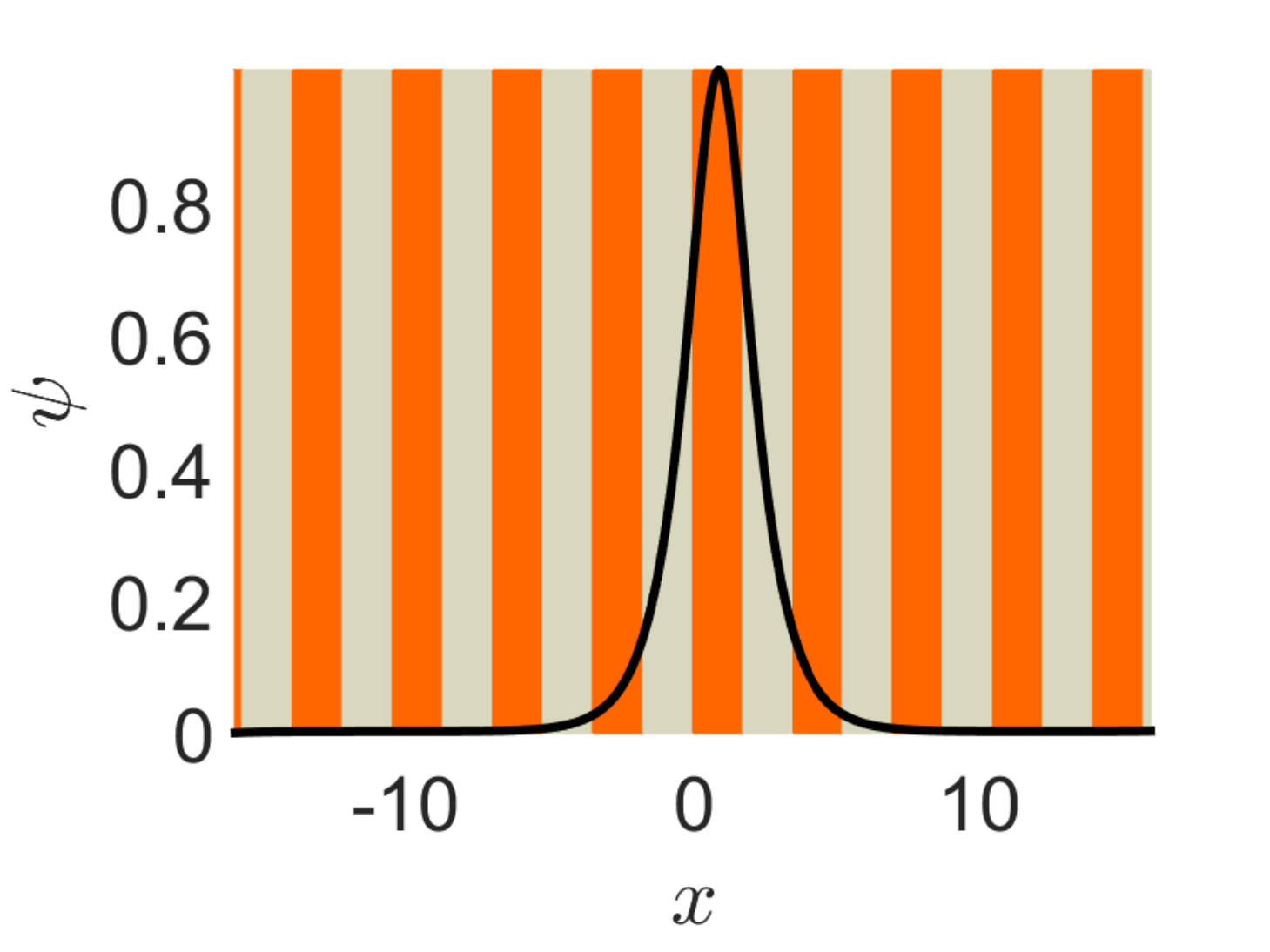}}}
  	\subfigure[]{\scalebox{\scl}{\includegraphics{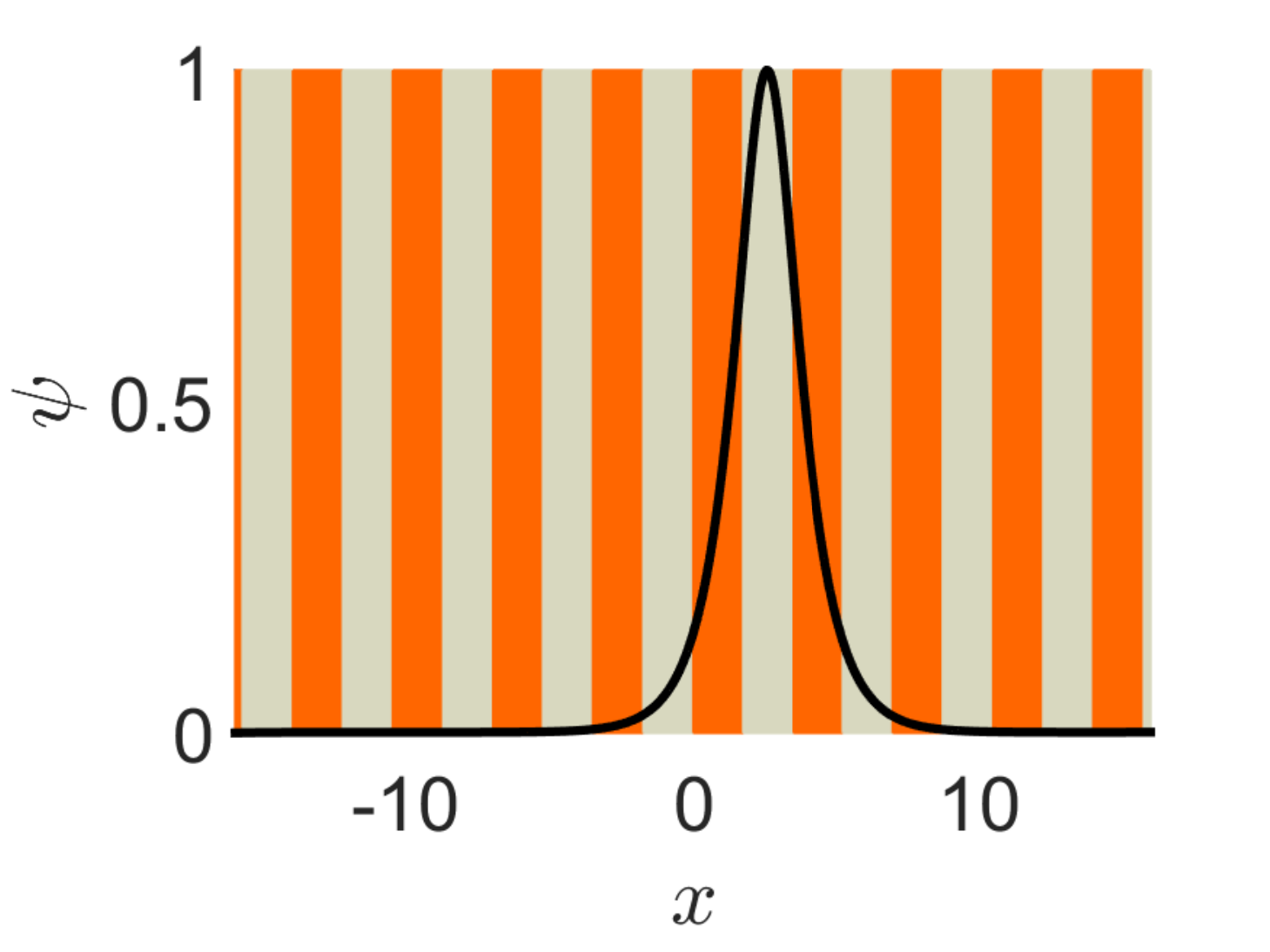}}}\\
  	\subfigure[]{\scalebox{\scl}{\includegraphics{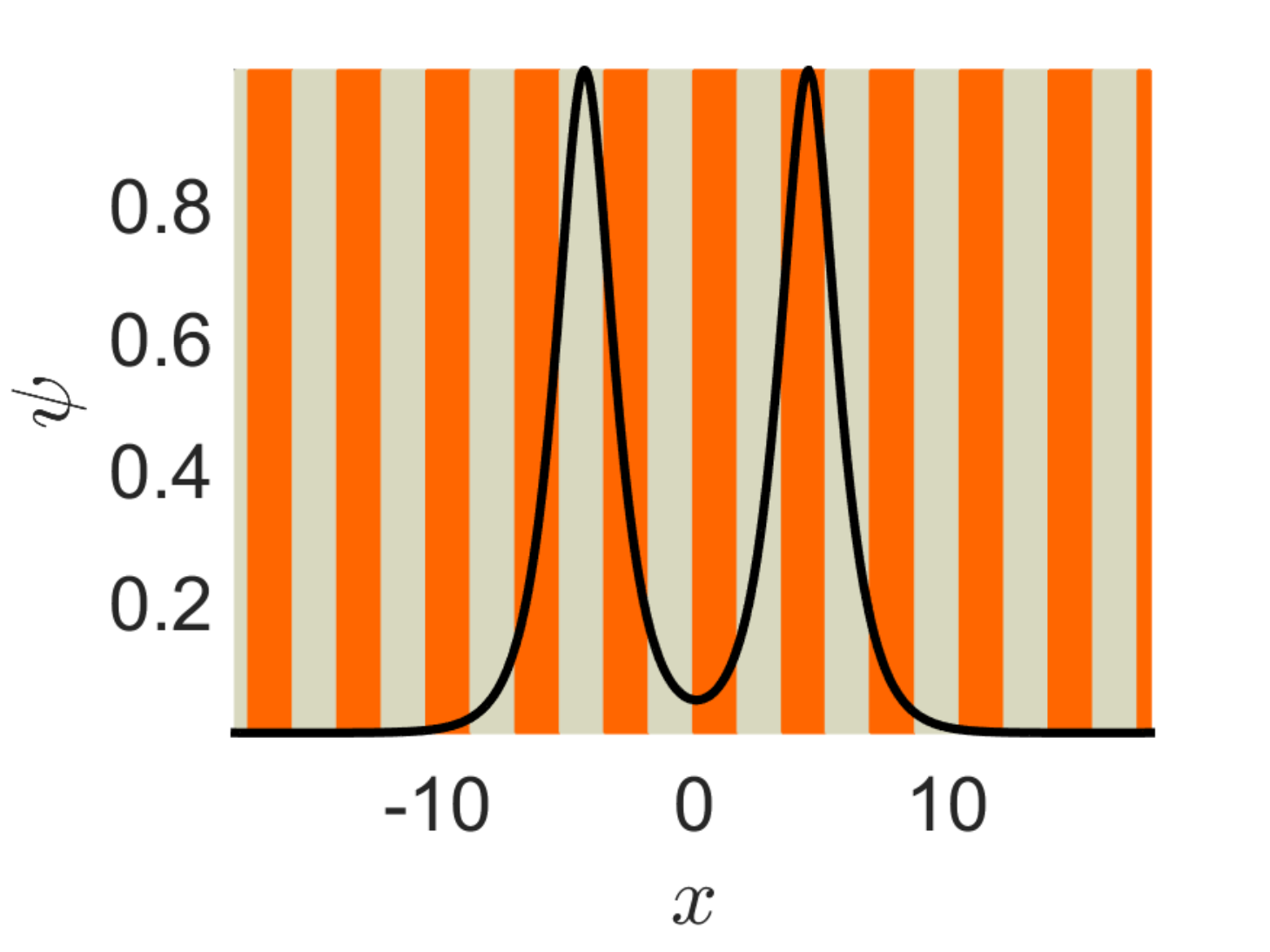}}}
  	\subfigure[]{\scalebox{\scl}{\includegraphics{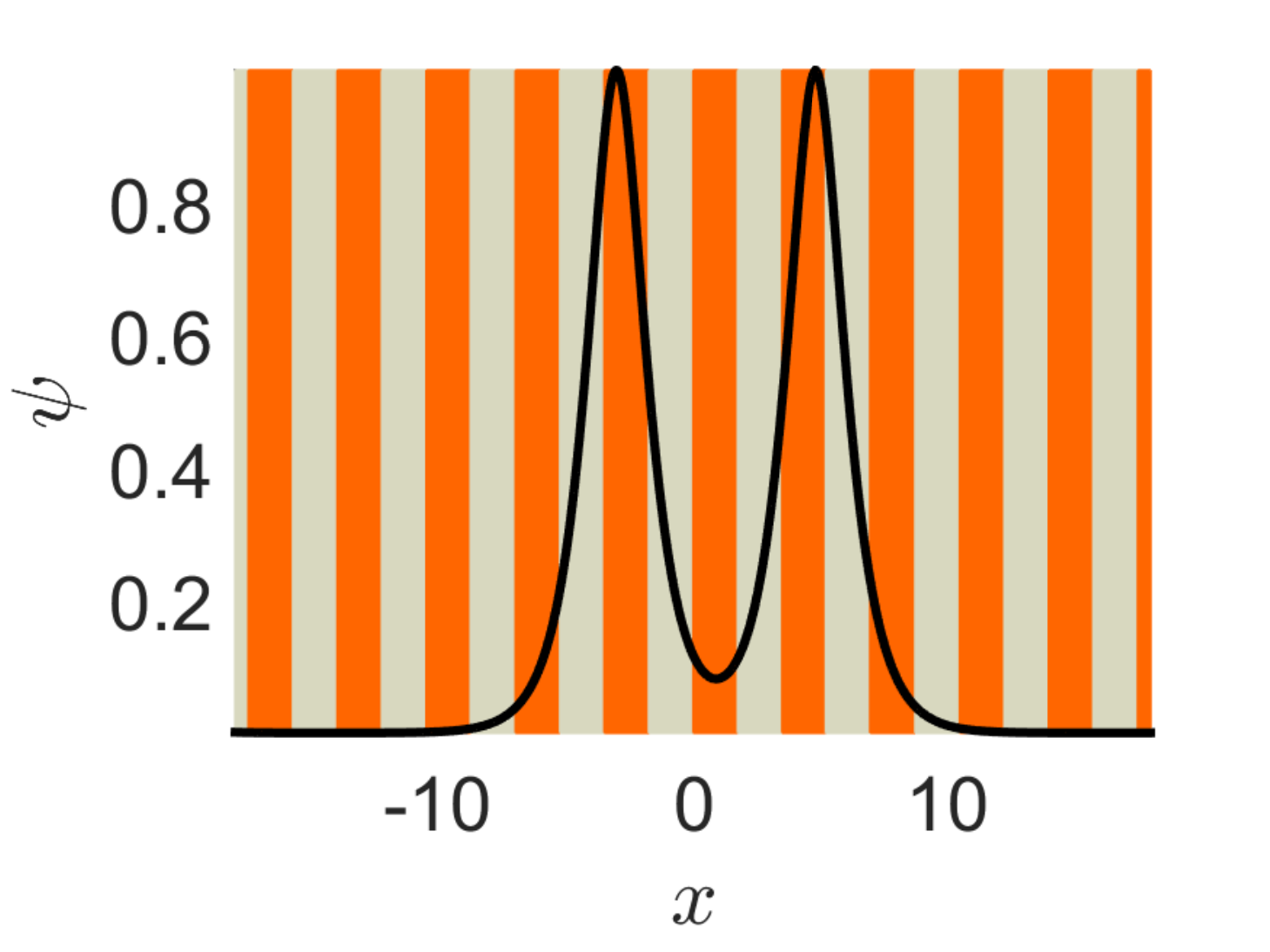}}}\\
  	\subfigure[]{\scalebox{\scl}{\includegraphics{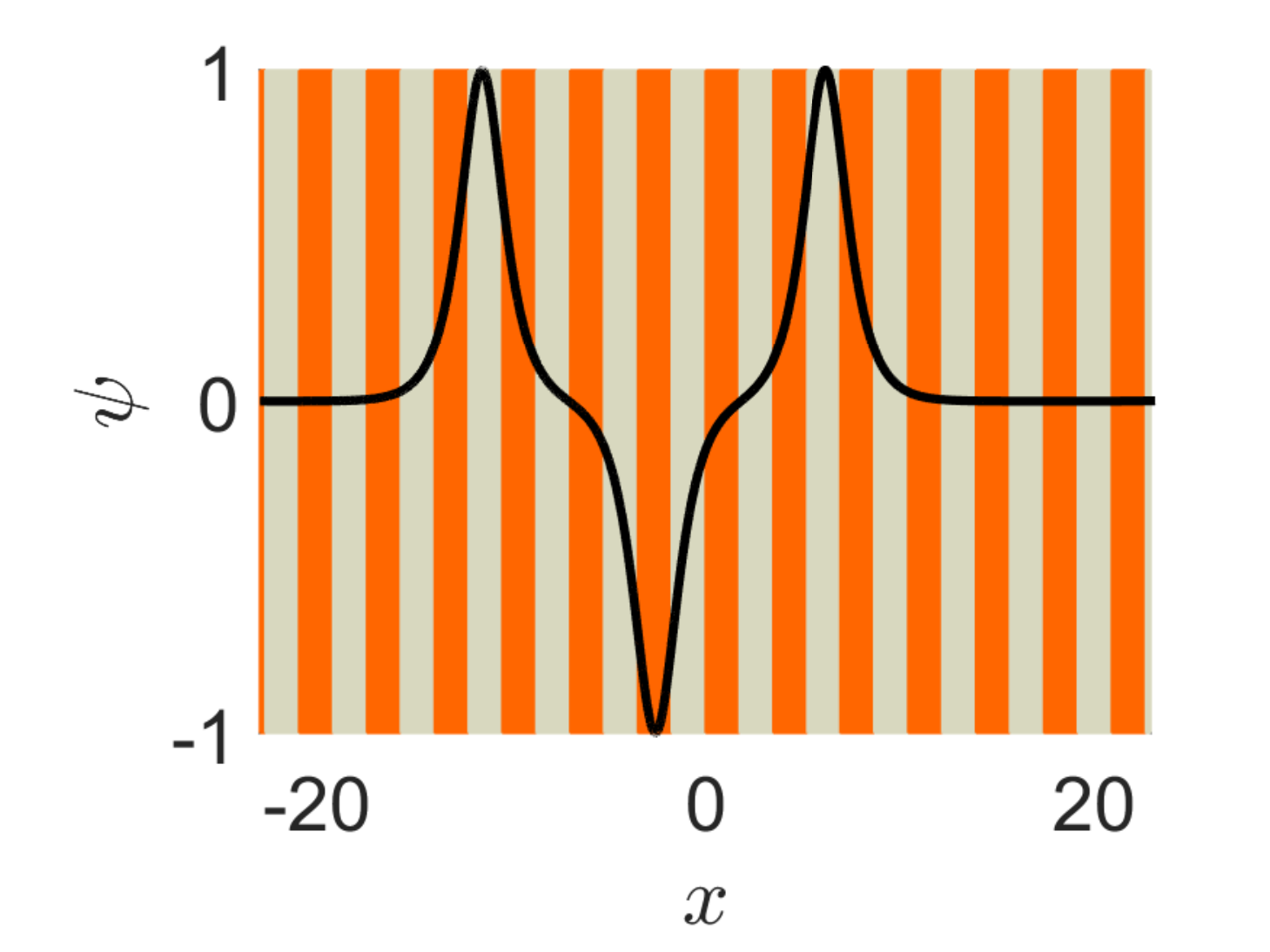}}}
  	\subfigure[]{\scalebox{\scl}{\includegraphics{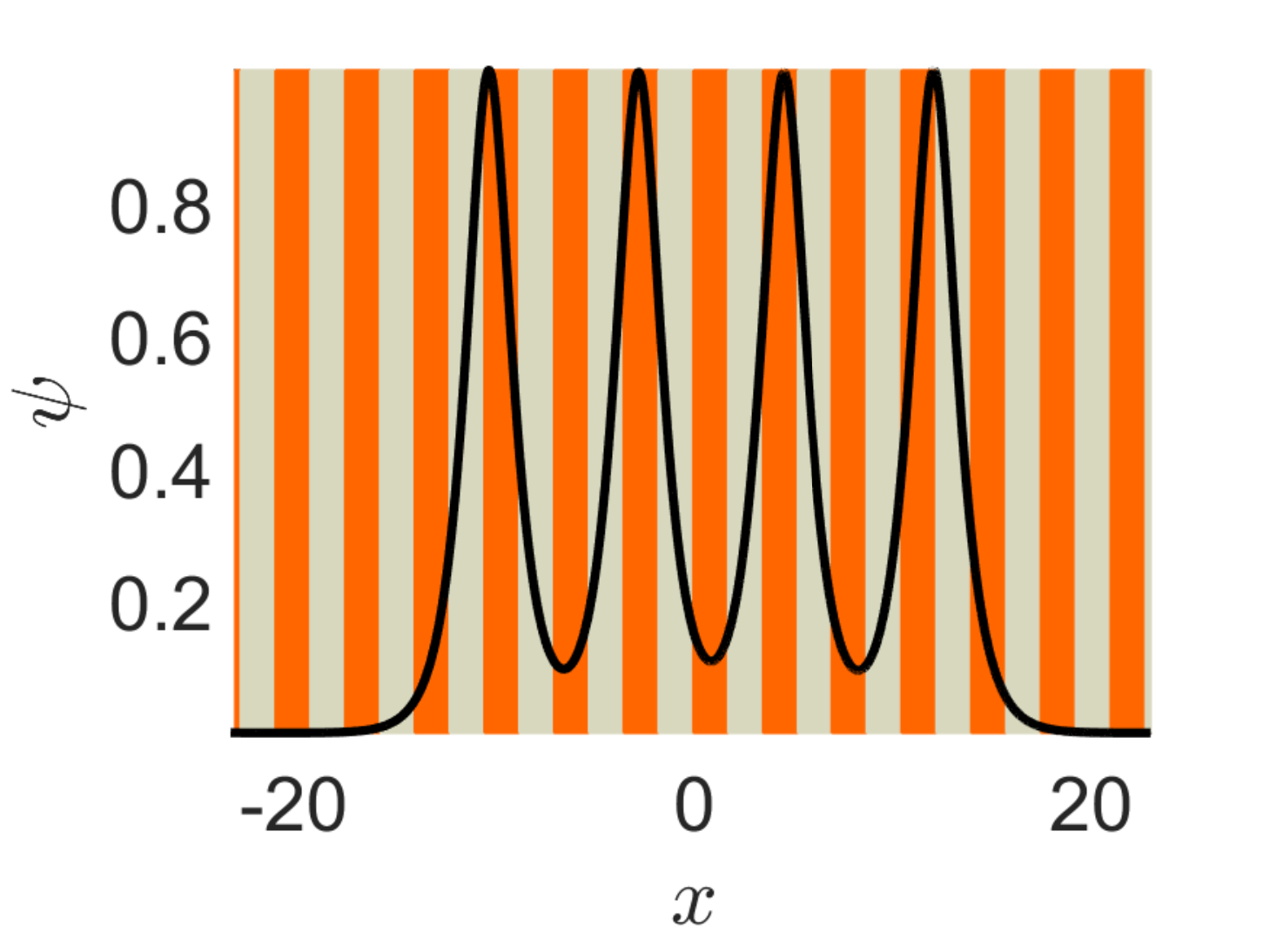}}}
  	\caption{Bright SW profiles corresponding to intersection points (a)-(1), (b)-(2), (c)-(3), (d)-(4), (e)-(5) and (f)-(6) of Fig. \ref{Fig:2}(a). Simple profiles, shown in (a) and (b), correspond the first two intersection points and describe fundamental SW with single maxima. More complex profiles depicted in (c)-(f) correspond to subsequent intersections and that can be intuitively considered as bound states consisting of in phase or out of phase combinations of fundamental SW. Dark orange and light grey regions denote the positions of minima and maxima of the refractive index variation respectively. }
 	\label{Fig:3}
\end{center}
\end{figure}

\begin{figure}[pt]
\begin{center}
 	\subfigure[]{\scalebox{\scl}{\includegraphics{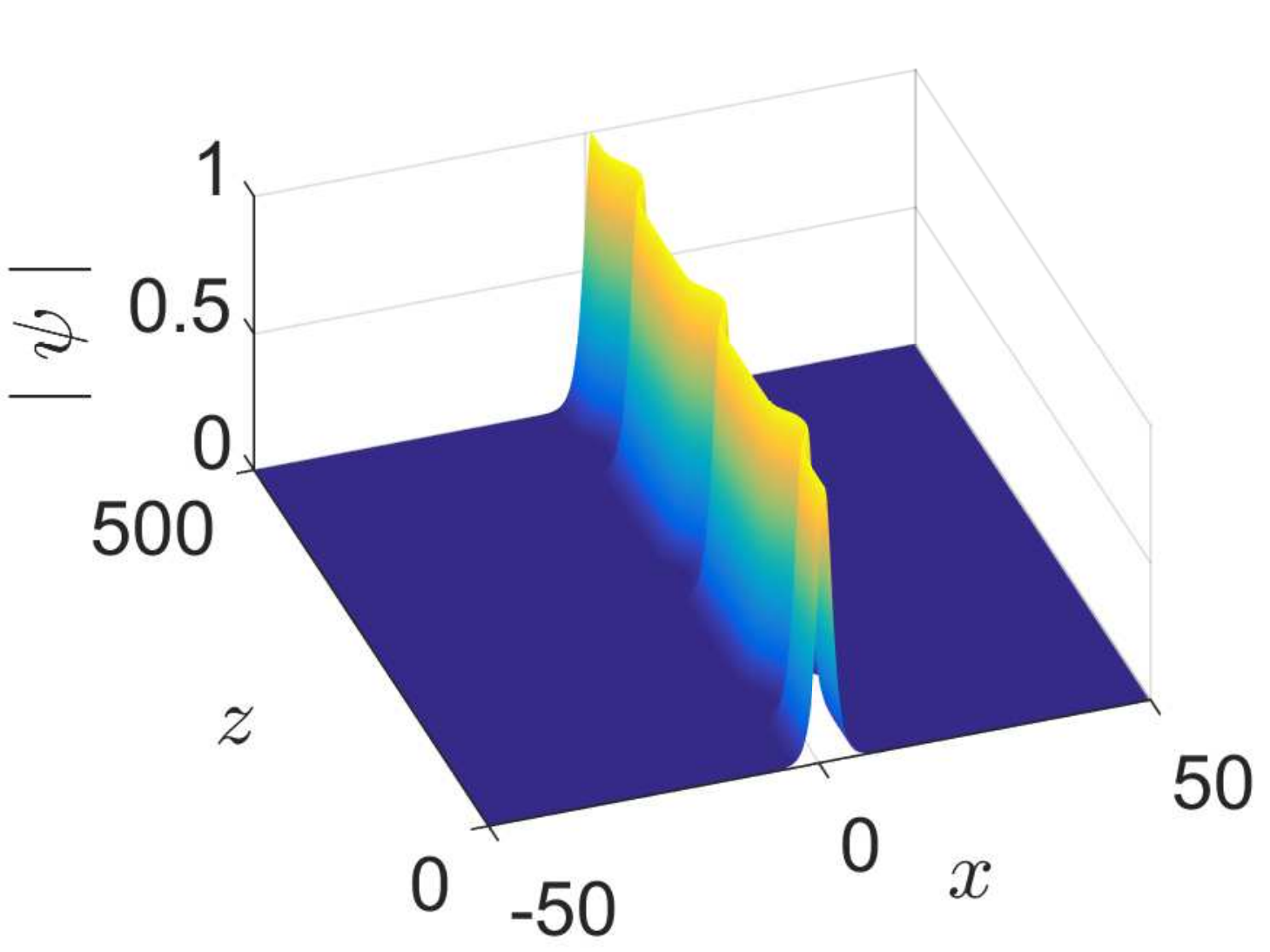}}}
  	\subfigure[]{\scalebox{\scl}{\includegraphics{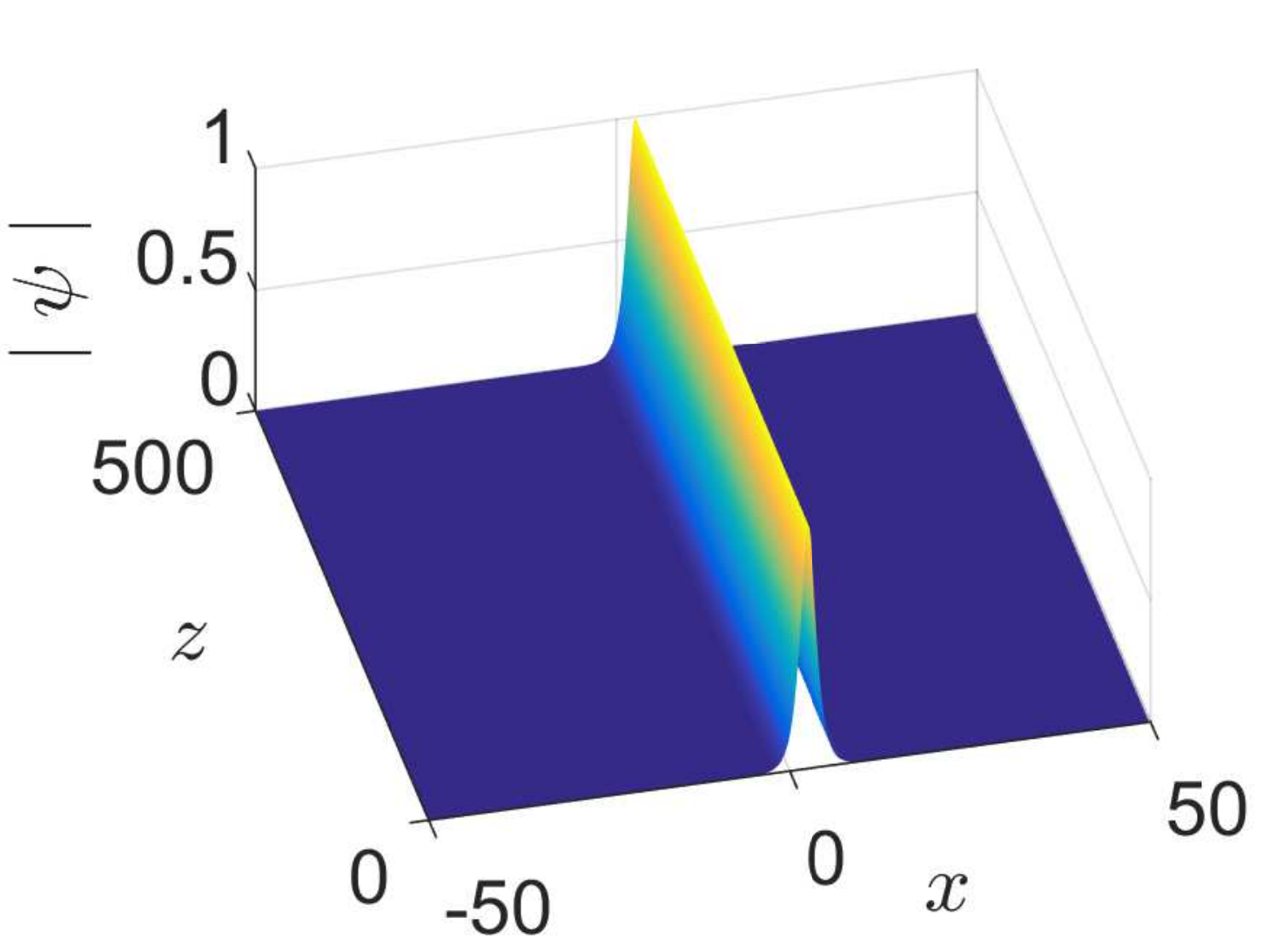}}}\\
  	\subfigure[]{\scalebox{\scl}{\includegraphics{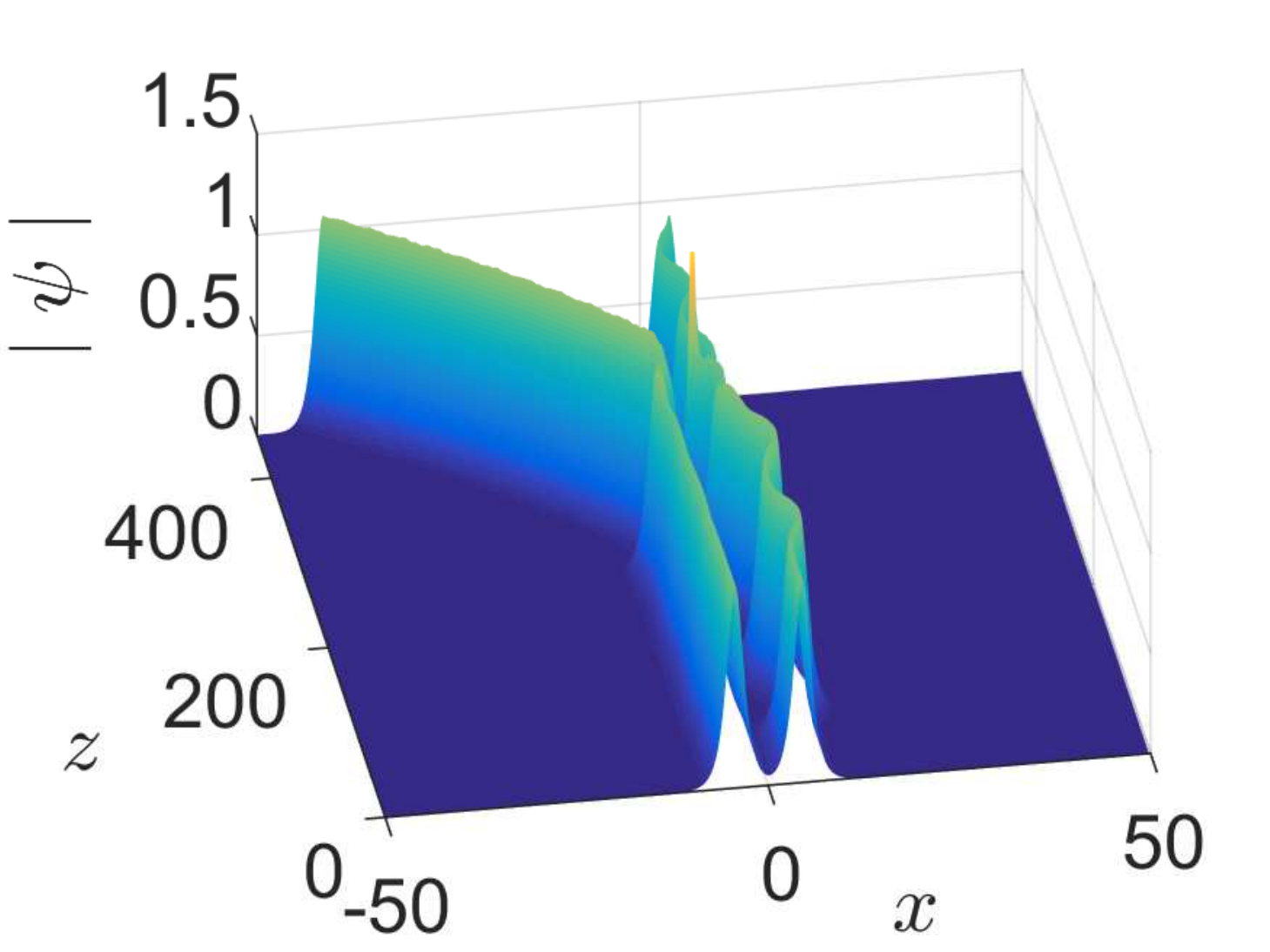}}}
  	\subfigure[]{\scalebox{\scl}{\includegraphics{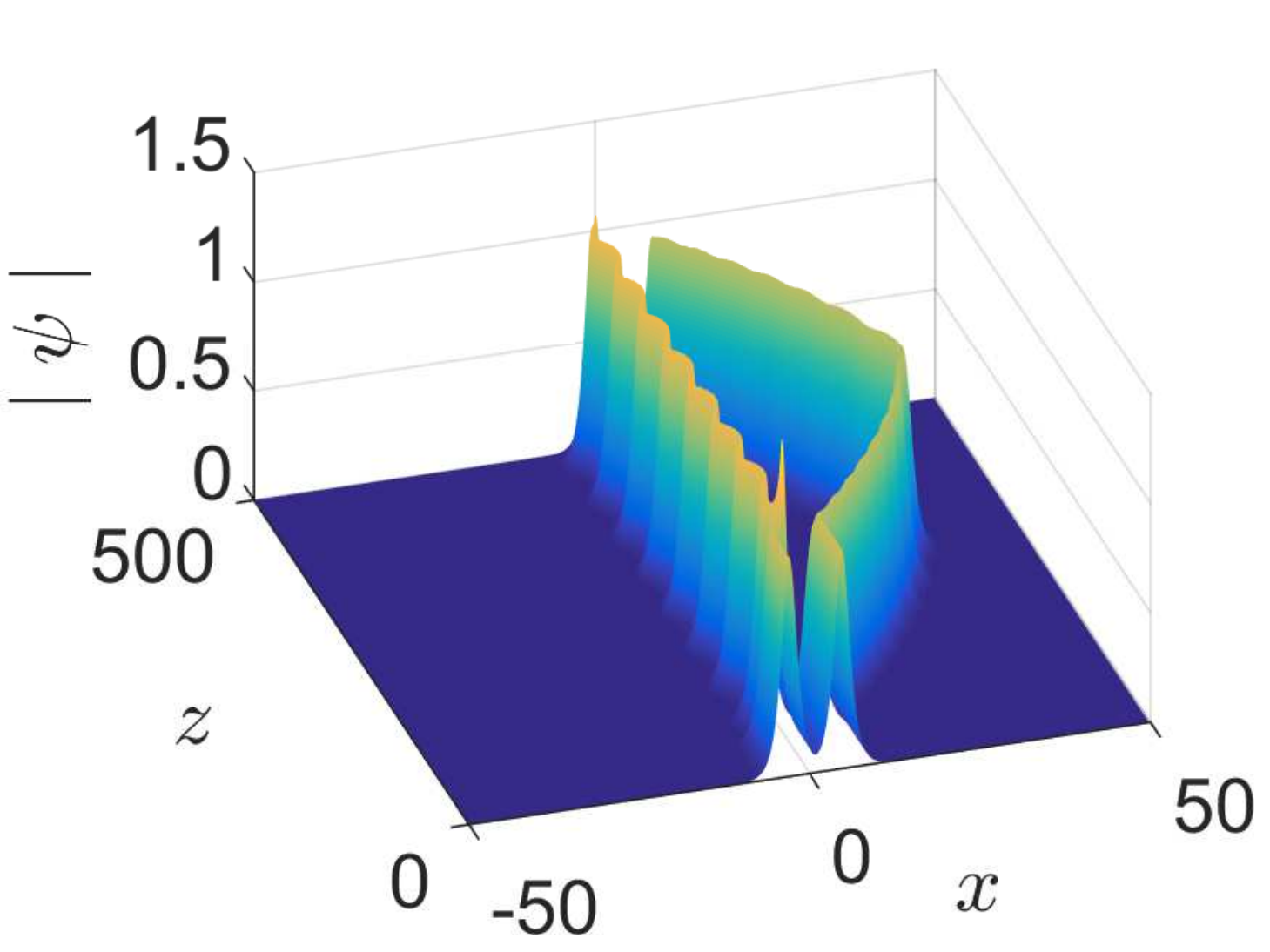}}}\\
  	\subfigure[]{\scalebox{\scl}{\includegraphics{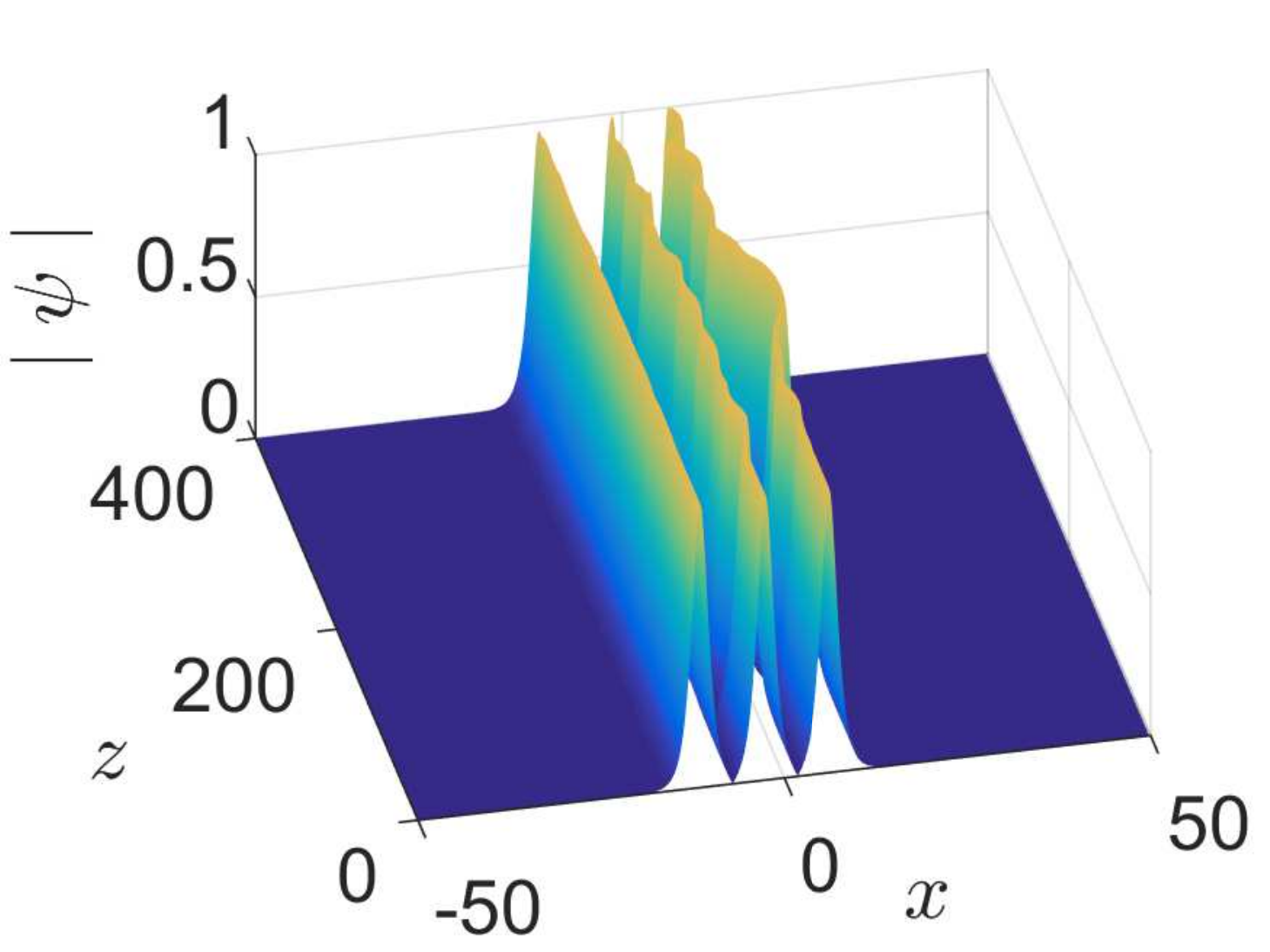}}}
  	\subfigure[]{\scalebox{\scl}{\includegraphics{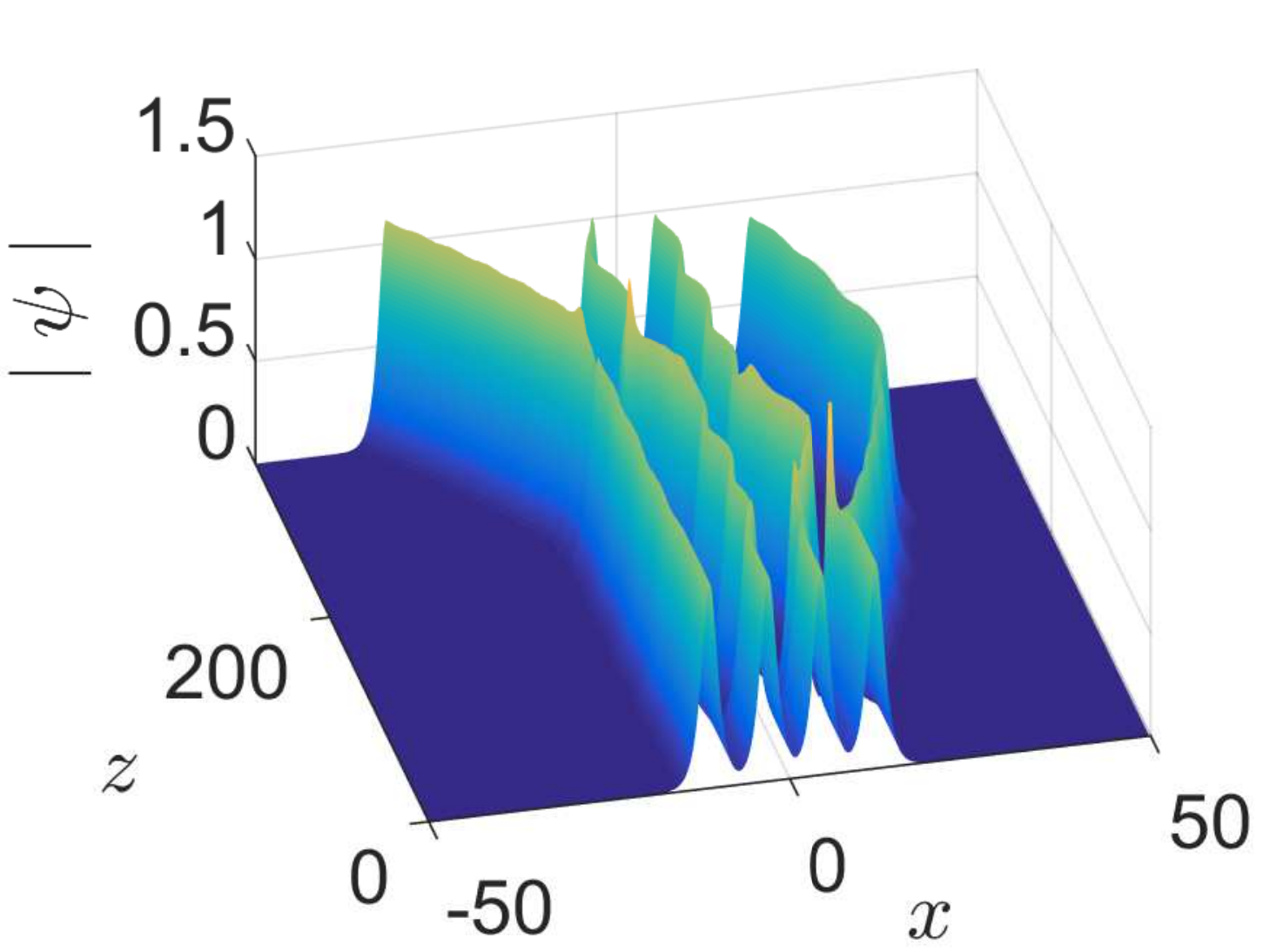}}}
  	\caption{Propagation of bright SW with the corresponding profiles depicted in Fig. \ref{Fig:3}. Propagation of simple SW profiles centered around a local minimum (a) and a local maximum (b) of the refractive index undergo oscillatory and stable propagation respectively. Bound states in (c)-(f) are shown to undergo more complex propagation dynamics exhibiting oscillatory instabilities or breaking of bound states to simpler SW undergoing mutual attractive or repelling interactions.}
 	\label{Fig:4}
\end{center}
\end{figure}

\begin{figure}[pt]
\begin{center}
 	\subfigure[]{\scalebox{\scl}{\includegraphics{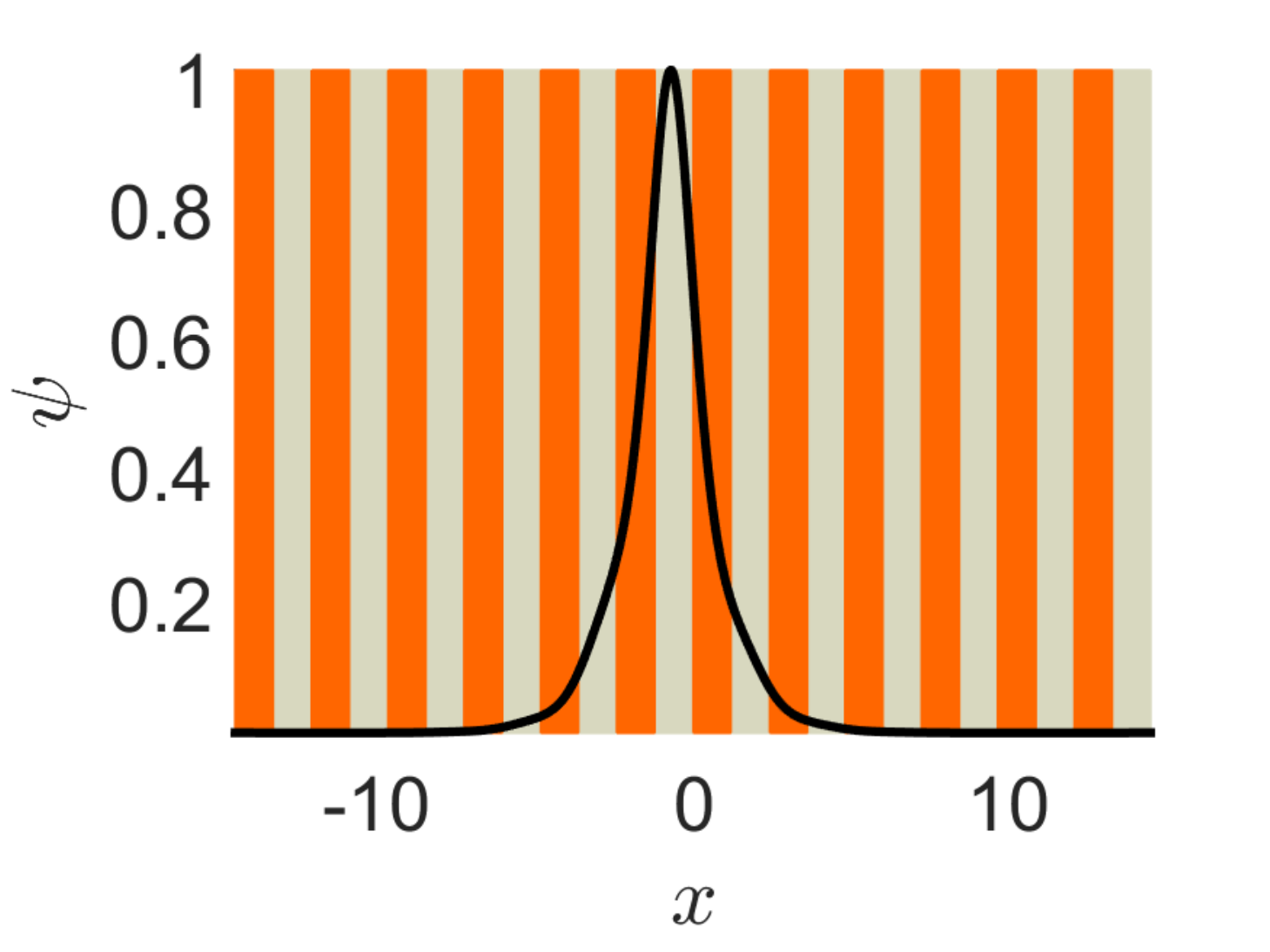}}}
  	\subfigure[]{\scalebox{\scl}{\includegraphics{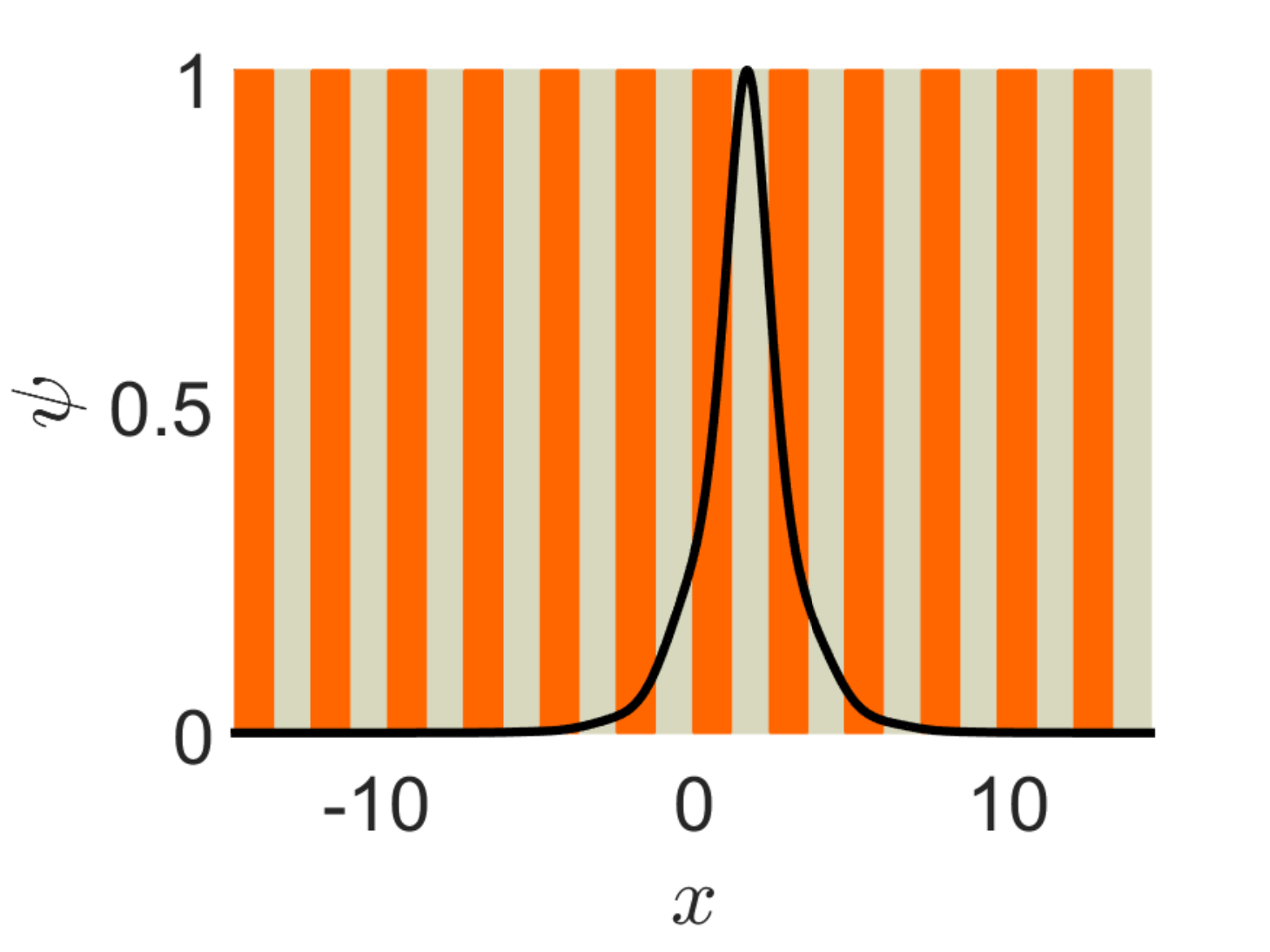}}}\\
  	\subfigure[]{\scalebox{\scl}{\includegraphics{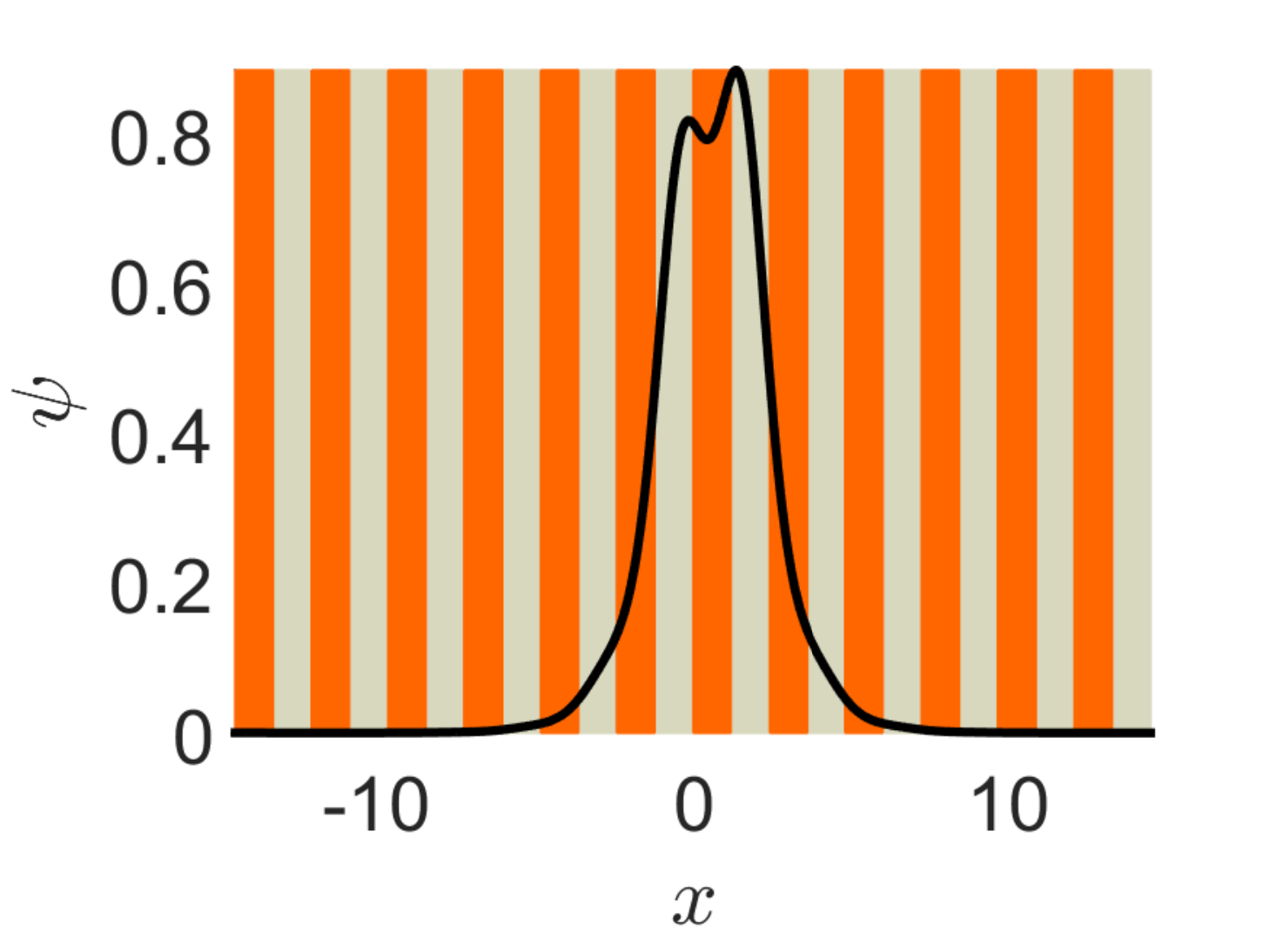}}}
  	\subfigure[]{\scalebox{\scl}{\includegraphics{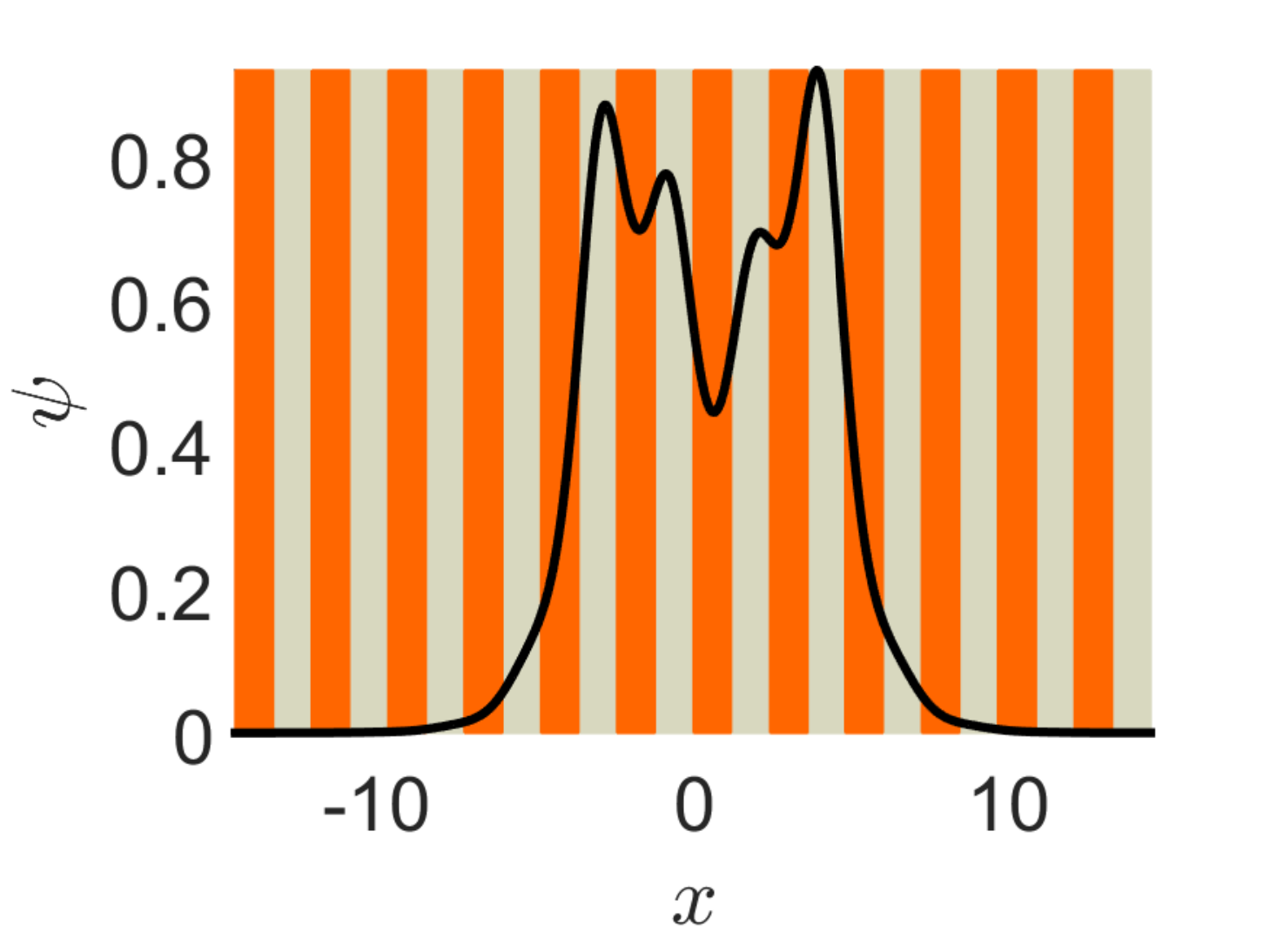}}}\\
  	\subfigure[]{\scalebox{\scl}{\includegraphics{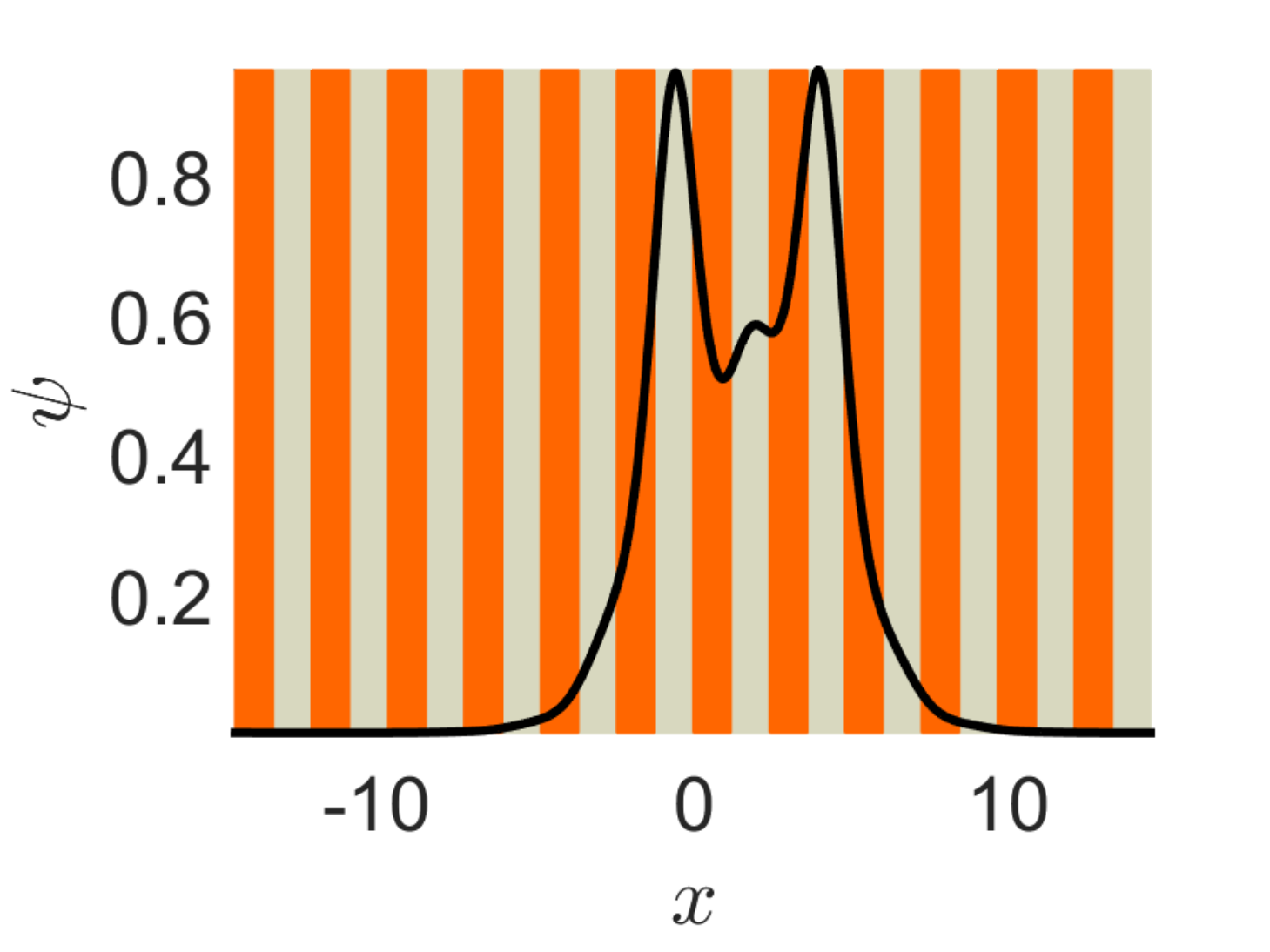}}}
  	\subfigure[]{\scalebox{\scl}{\includegraphics{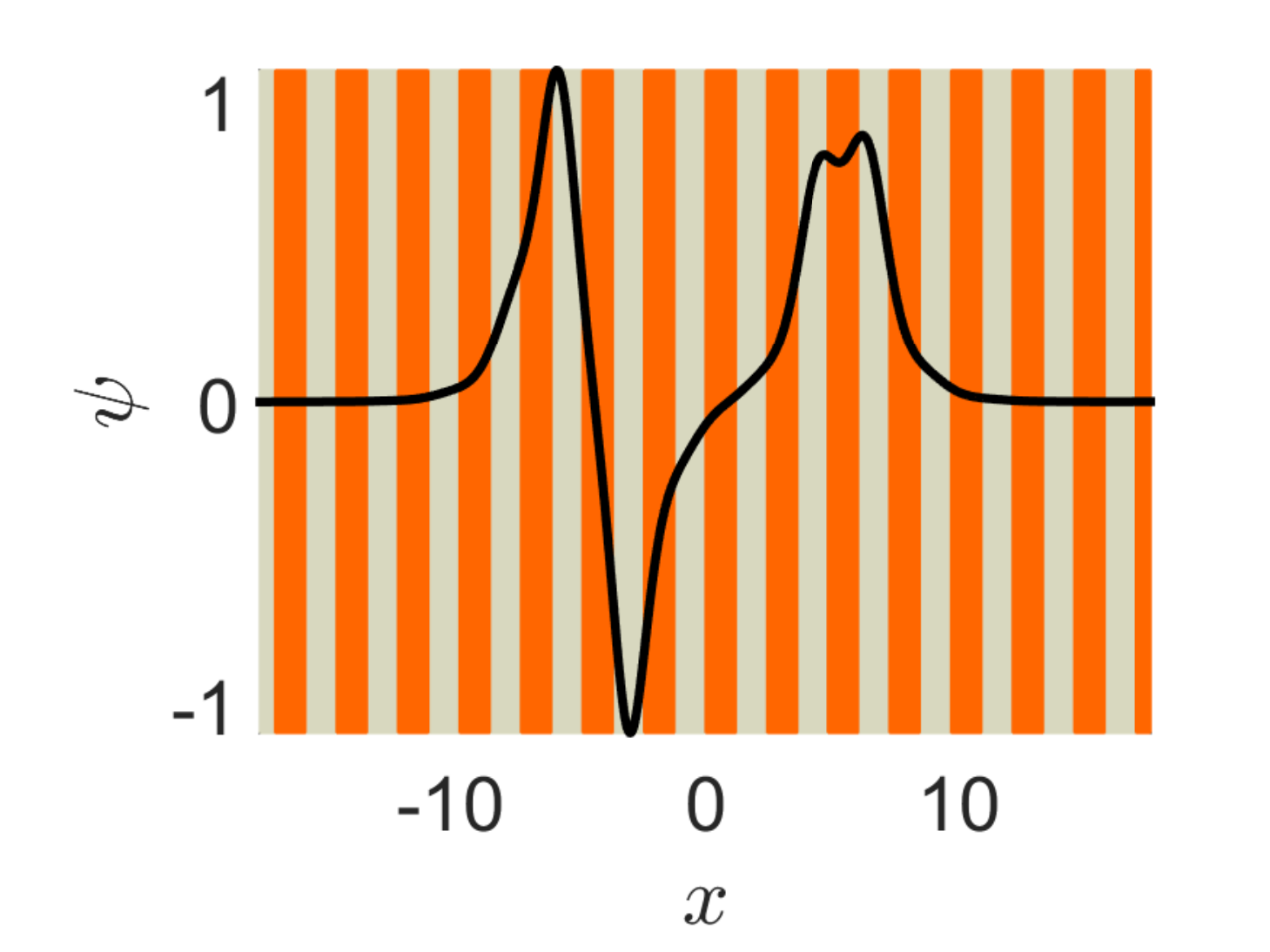}}}
  	\caption{Bright SW profiles corresponding to intersection points (a)-(1), (b)-(2), (c)-(3), (d)-(4), (e)-(5) and (f)-(6) of Fig. \ref{Fig:2}(b). Simple SW profiles, shown in (a) and (b), correspond to first intersections of the stable and unstable manifolds of the saddle, whereas more complex profiles in (c)-(f) have a high degree of asymmetry. Dark orange and light grey regions denote the positions of minima and maxima of the refractive index variation respectively.}
 	\label{Fig:5}
\end{center}
\end{figure}

\begin{figure}[pt]
\begin{center}
 	\subfigure[]{\scalebox{\scl}{\includegraphics{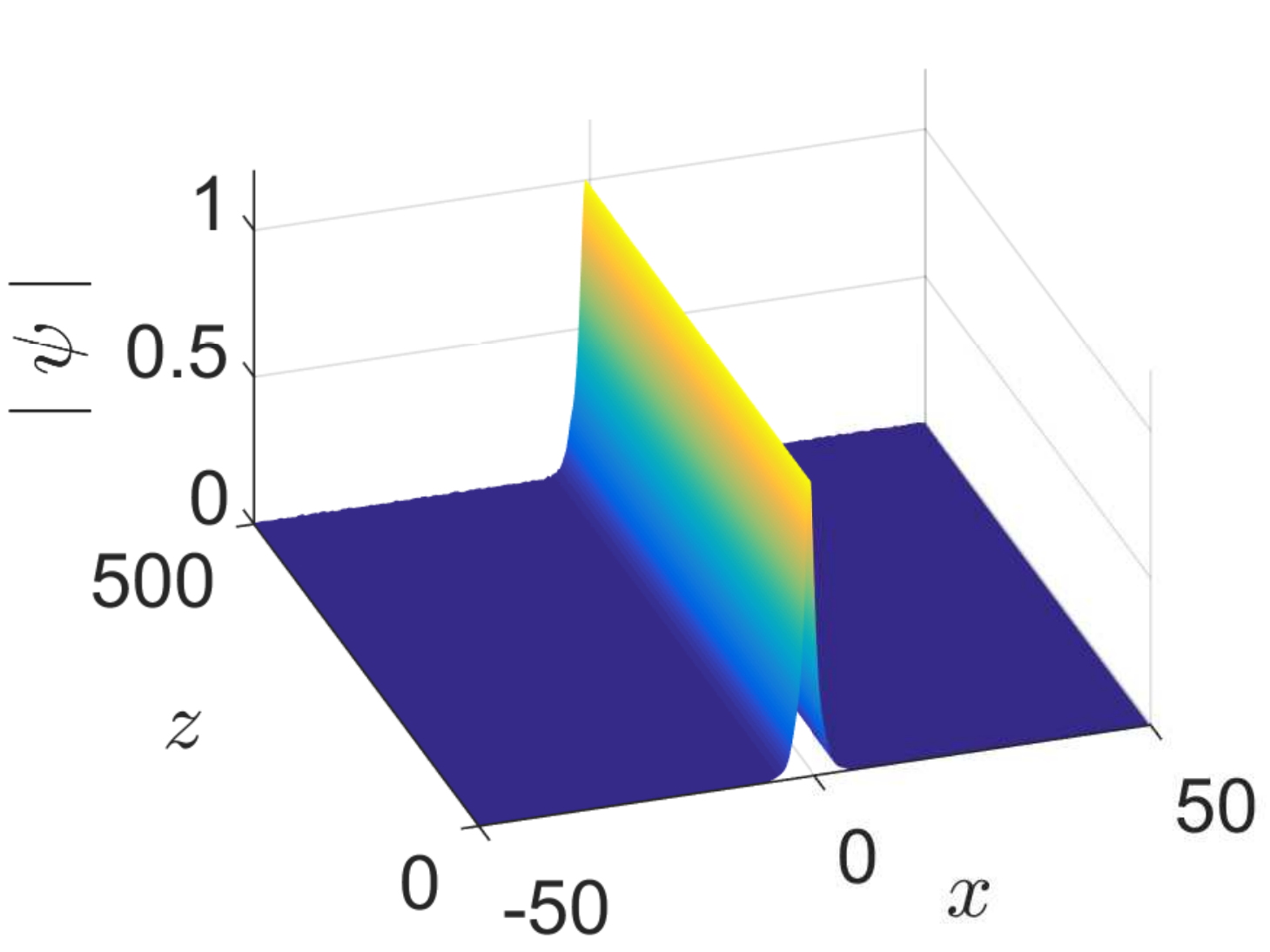}}}
  	\subfigure[]{\scalebox{\scl}{\includegraphics{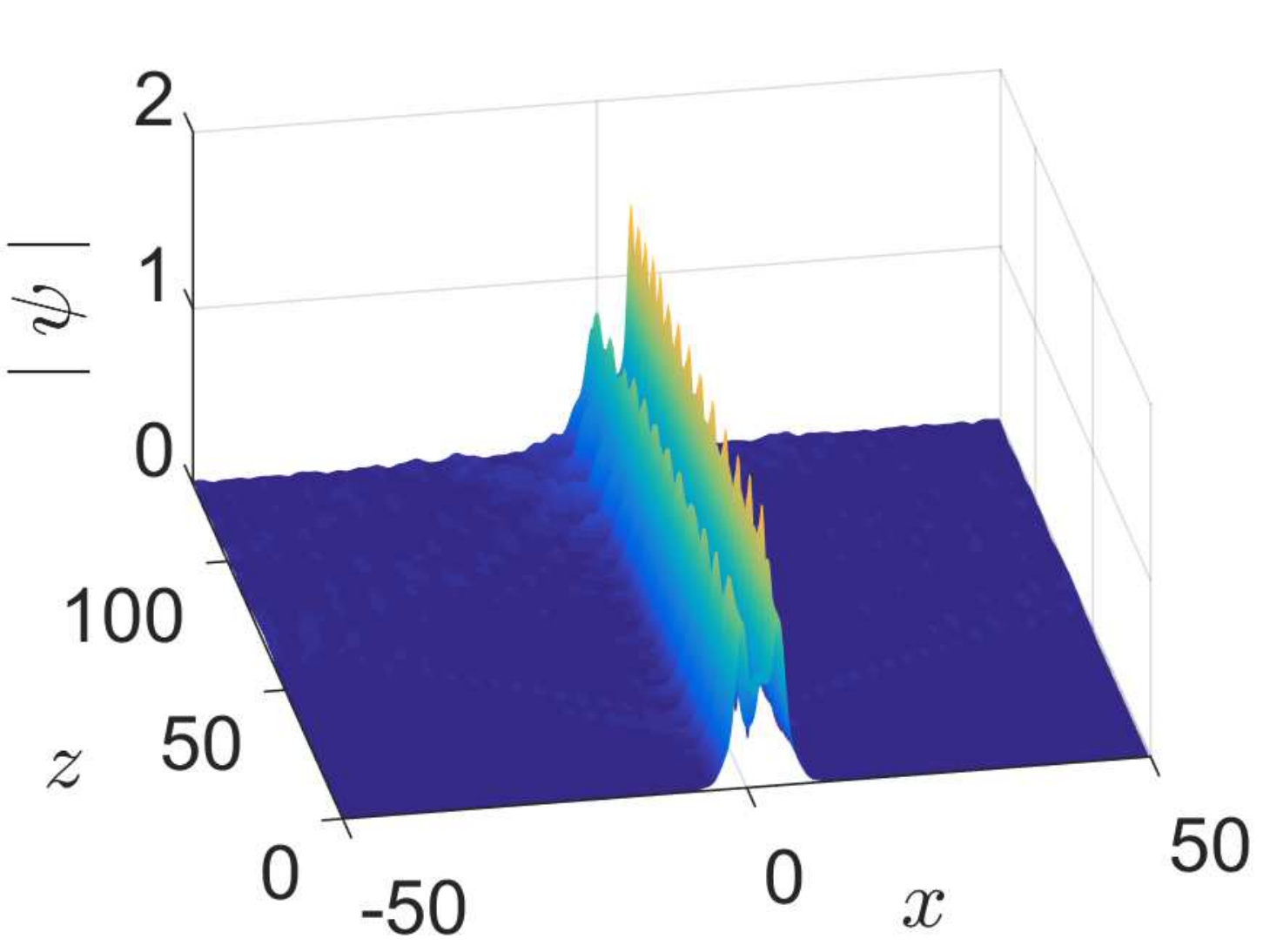}}}
  	\caption{Propagation dynamics of (a) the bright SW of Fig. \ref{Fig:5}(a) and (b) the bright SW of Fig. \ref{Fig:5}(e). Both the fundamental SW (a) and the relatively simple bound state (b) centered around local maxima of the refractive index, propagate in a stable and oscillatory fashion respectively.  }
 	\label{Fig:6}
\end{center}
\end{figure}

\begin{figure}	[pt]
\begin{center}
 	\subfigure[]{\scalebox{\scl}{\includegraphics{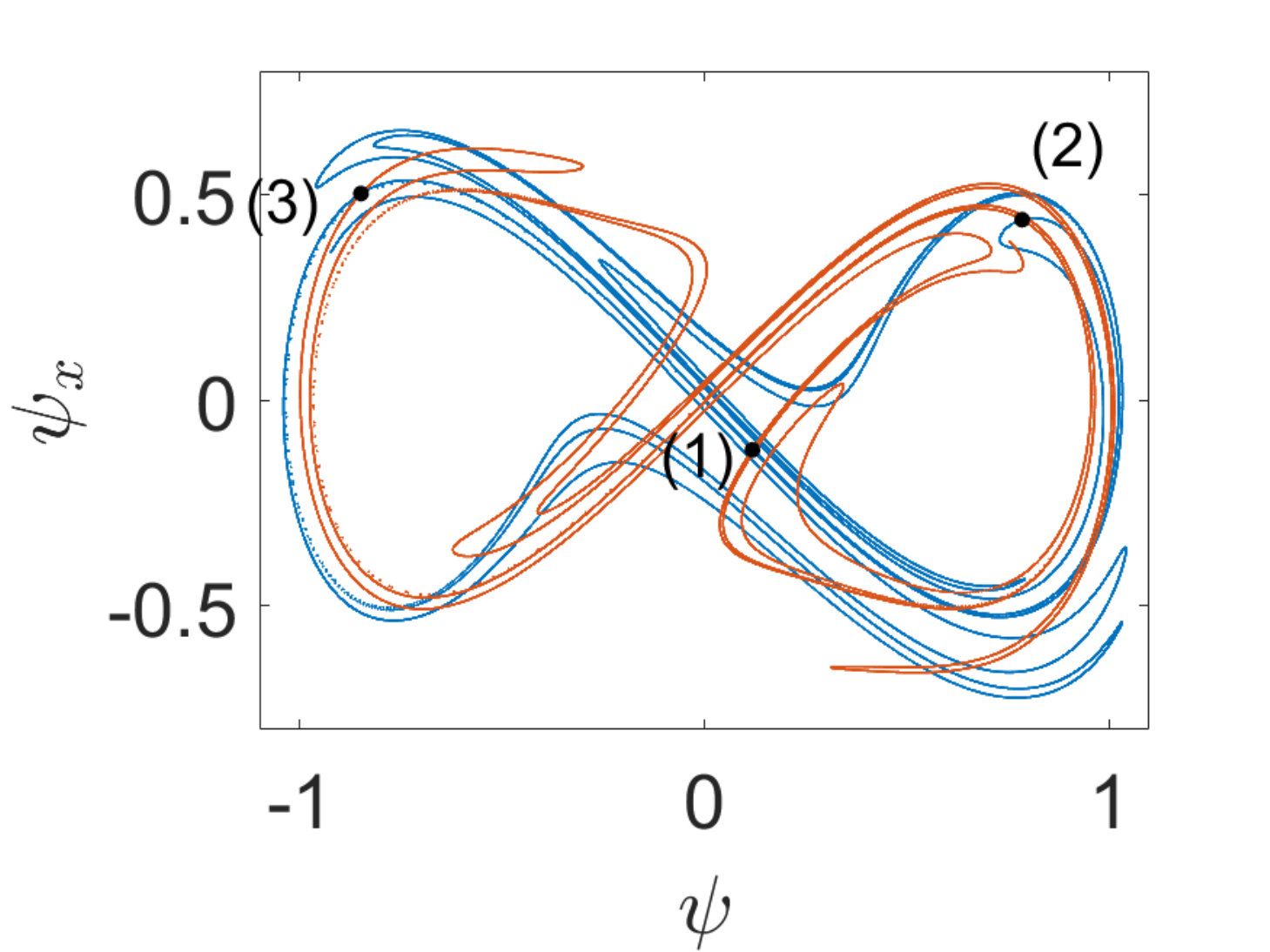}}}
 	\subfigure[]{\scalebox{\scl}{\includegraphics{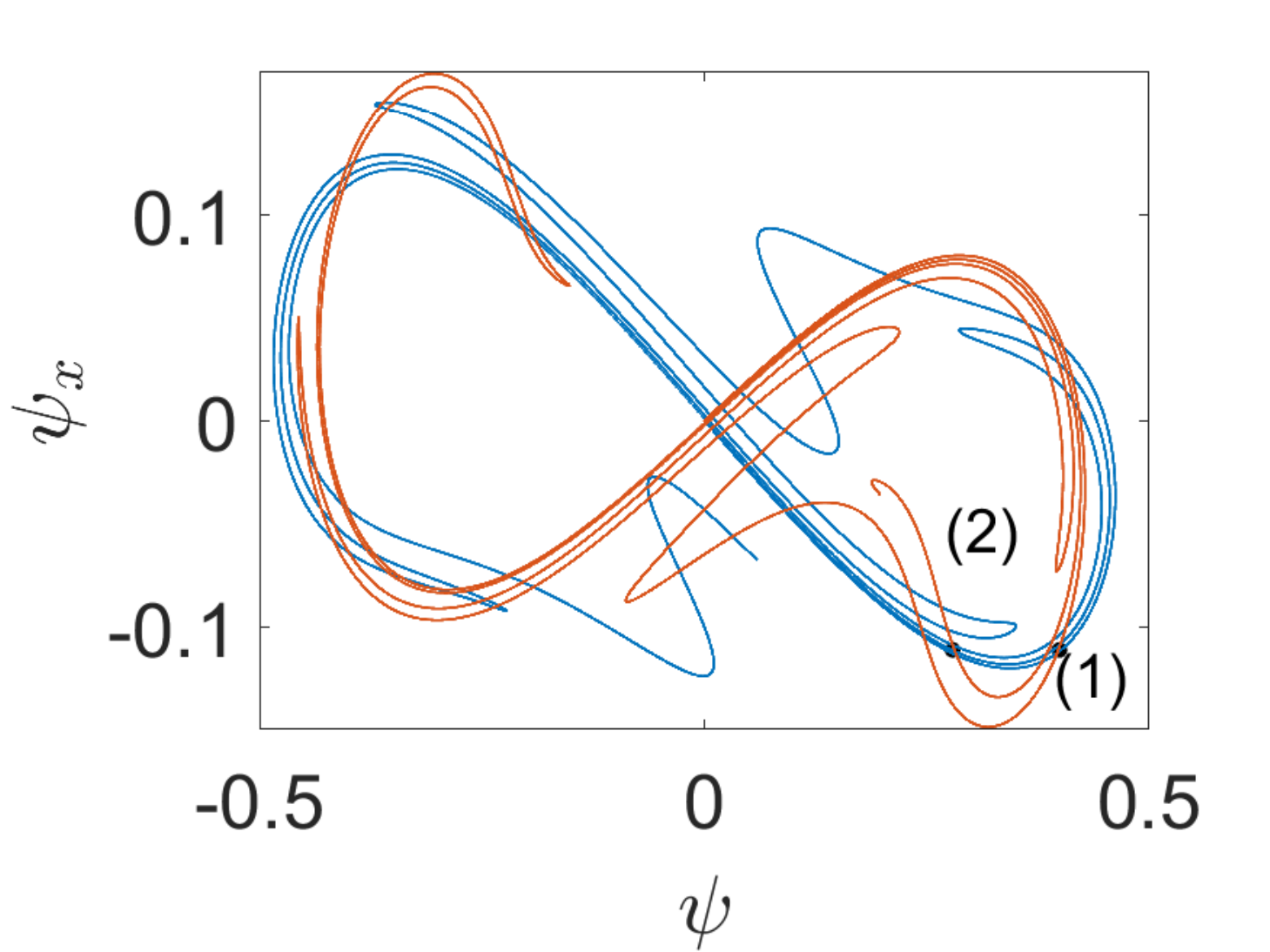}}}
  	\caption{Branches of the stale (blue) and unstable (red) manifolds of the hyperbolic trajectory in a Poincare surface of section for parameter sets corresponding to the same photonic structure $\gamma=0.0025,A=0.1,k=1.6$ for two cases of different propagation constants satisfying Melnikov's condition (a) $\beta=1$ and (b) $\beta=0.2$. Black dots denote the intersections of the manifolds for which solitary waves profiles were studied.  The qualitatively different forms of the stable and unstable manifolds in the same photonic structure signifies a drastic dependence of the SW formation dynamics on the propagation constant.}
 	\label{Fig:7}
\end{center}
\end{figure}

\begin{figure}[pt]
\begin{center}
 	\subfigure[]{\scalebox{\scl}{\includegraphics{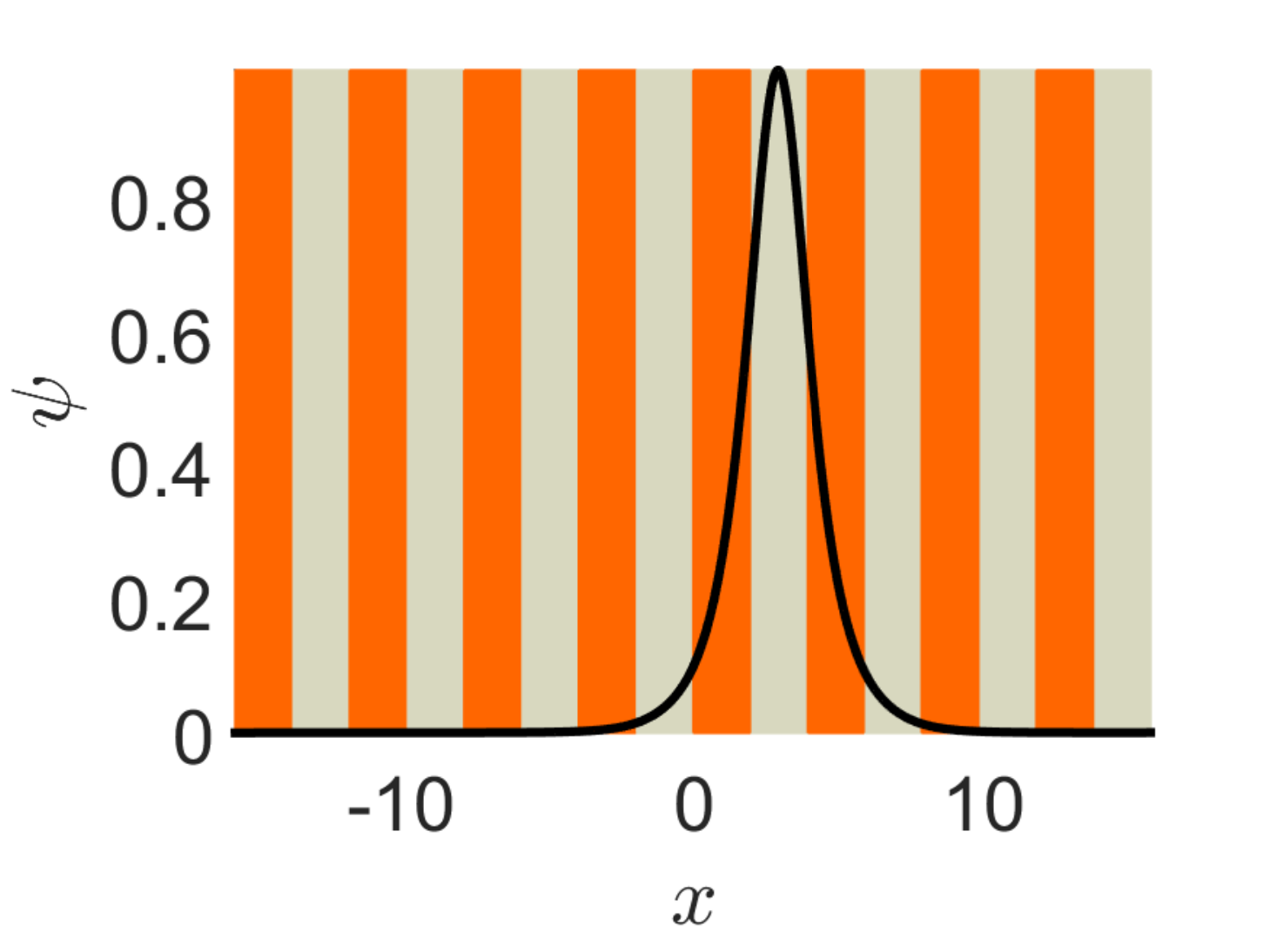}}}
 	\subfigure[]{\scalebox{\scl}{\includegraphics{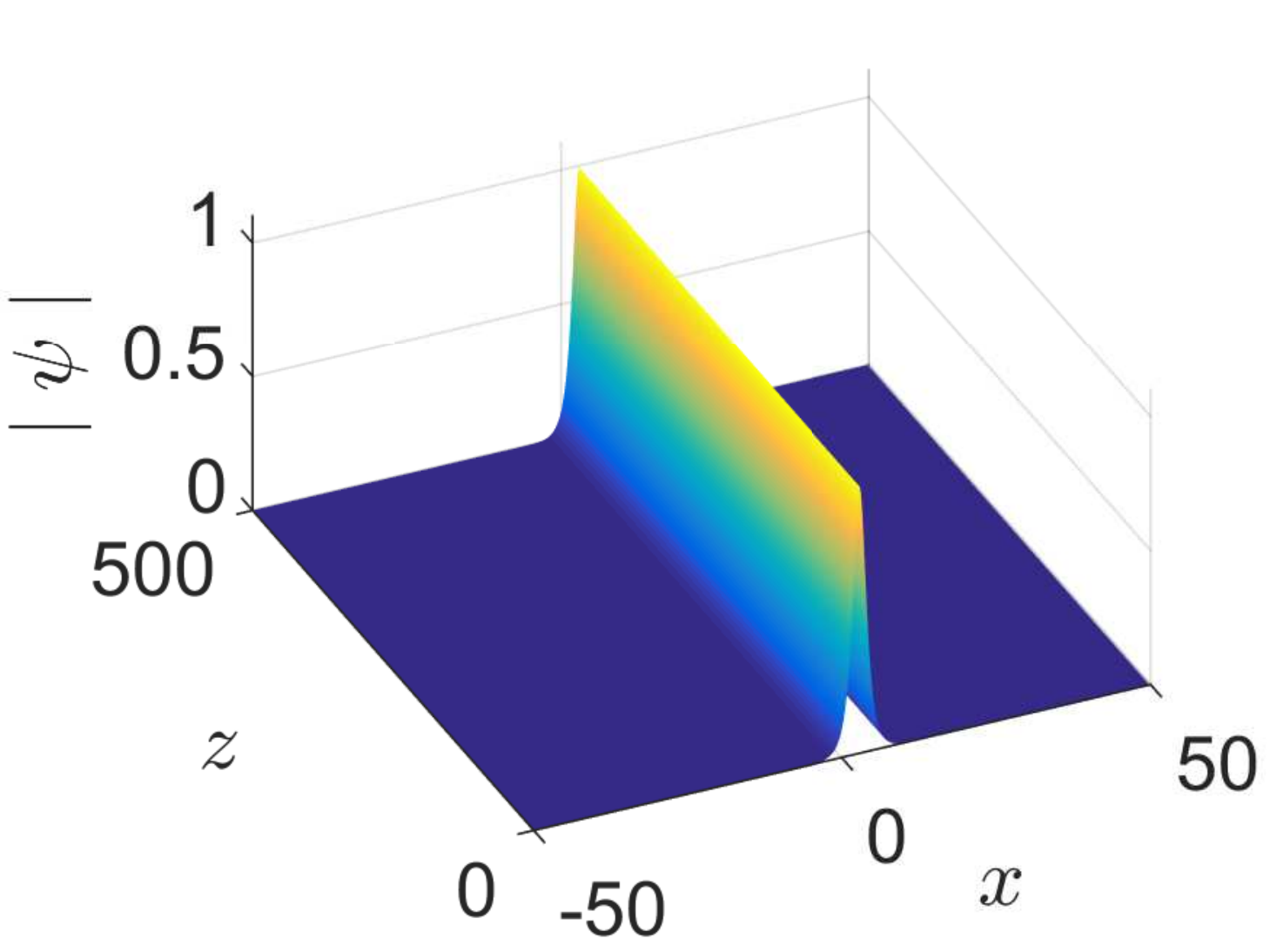}}}\\
 	\subfigure[]{\scalebox{\scl}{\includegraphics{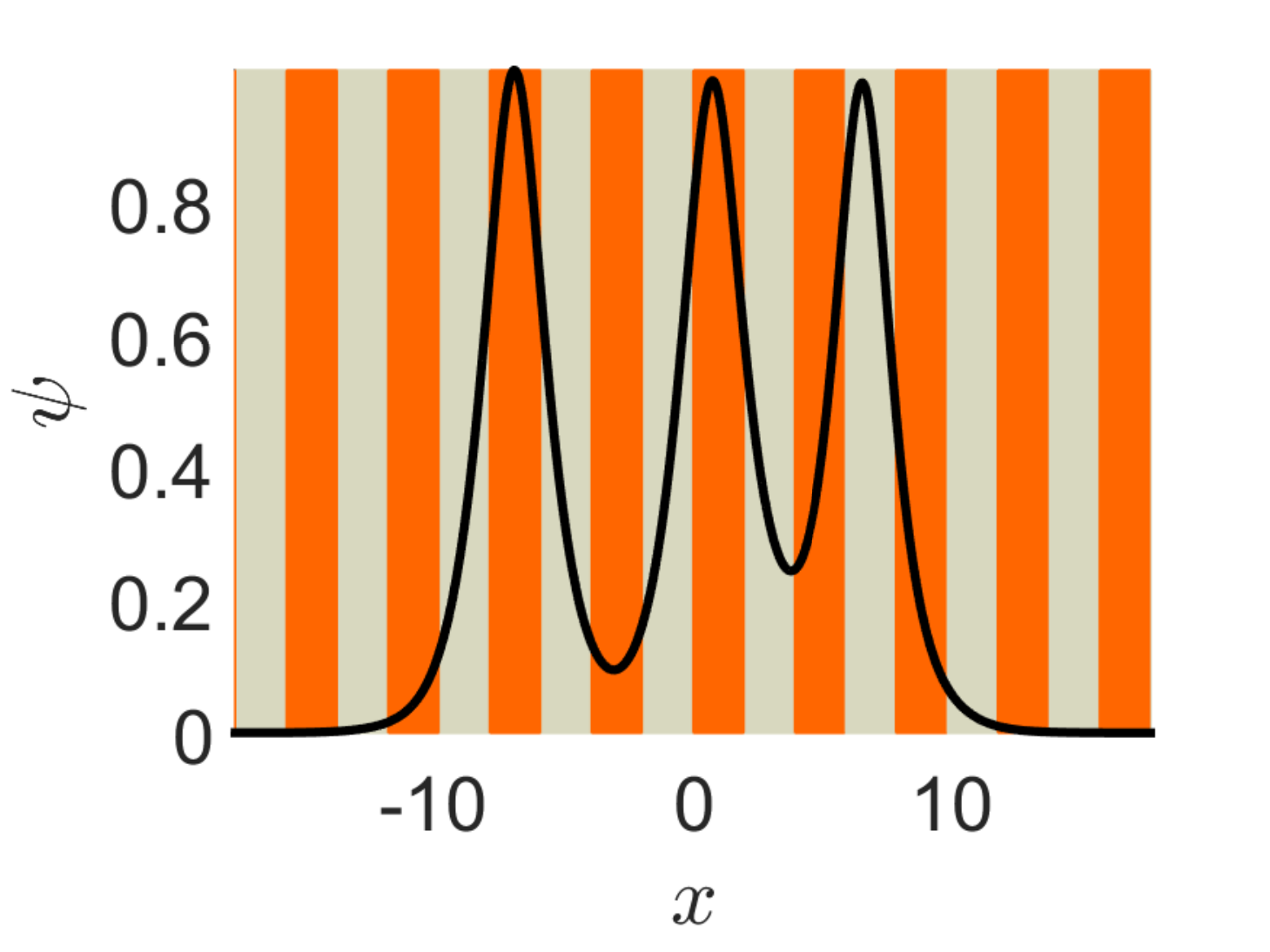}}}
  	\subfigure[]{\scalebox{\scl}{\includegraphics{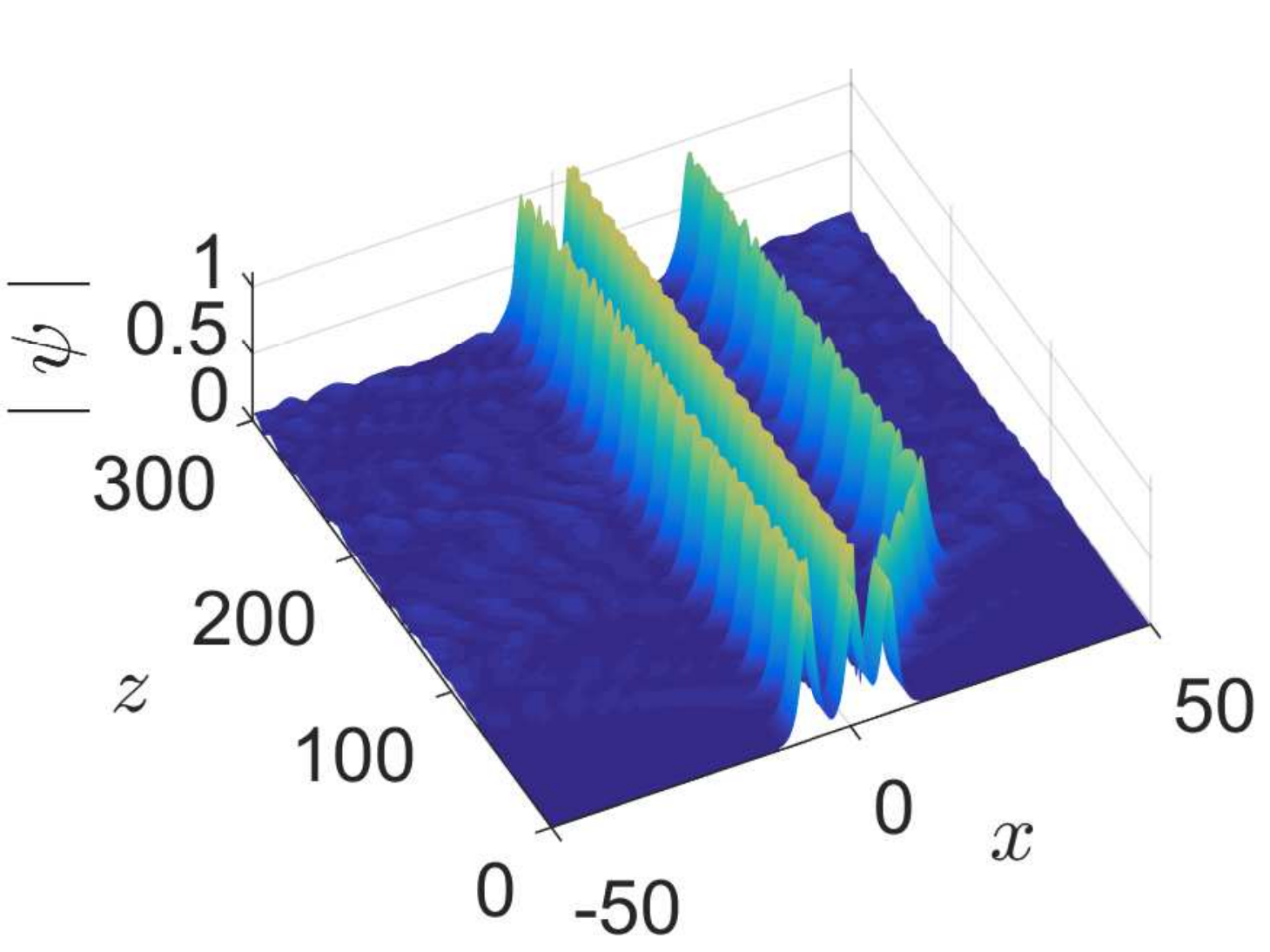}}}\\
  	\subfigure[]{\scalebox{\scl}{\includegraphics{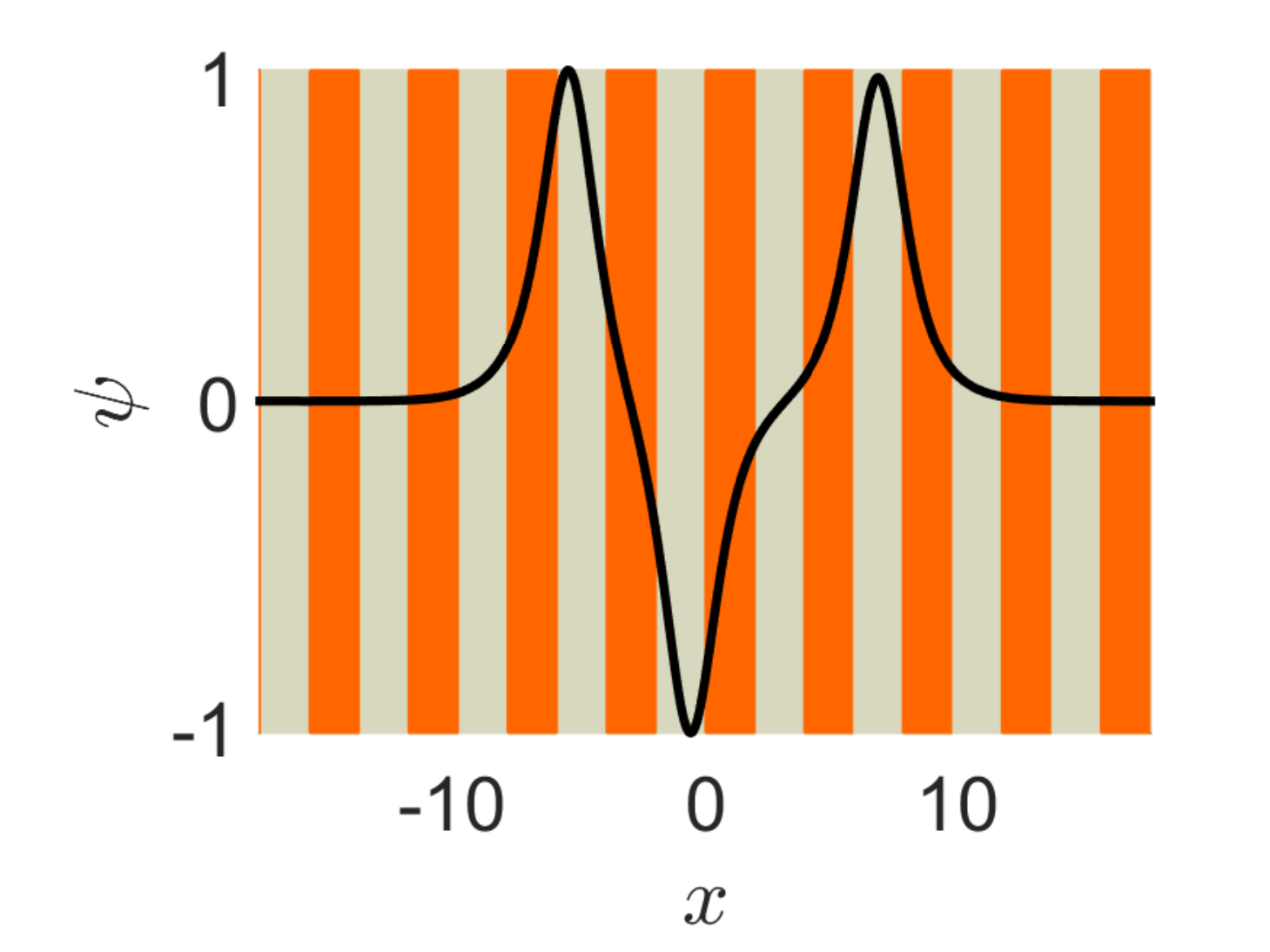}}}
  	\subfigure[]{\scalebox{\scl}{\includegraphics{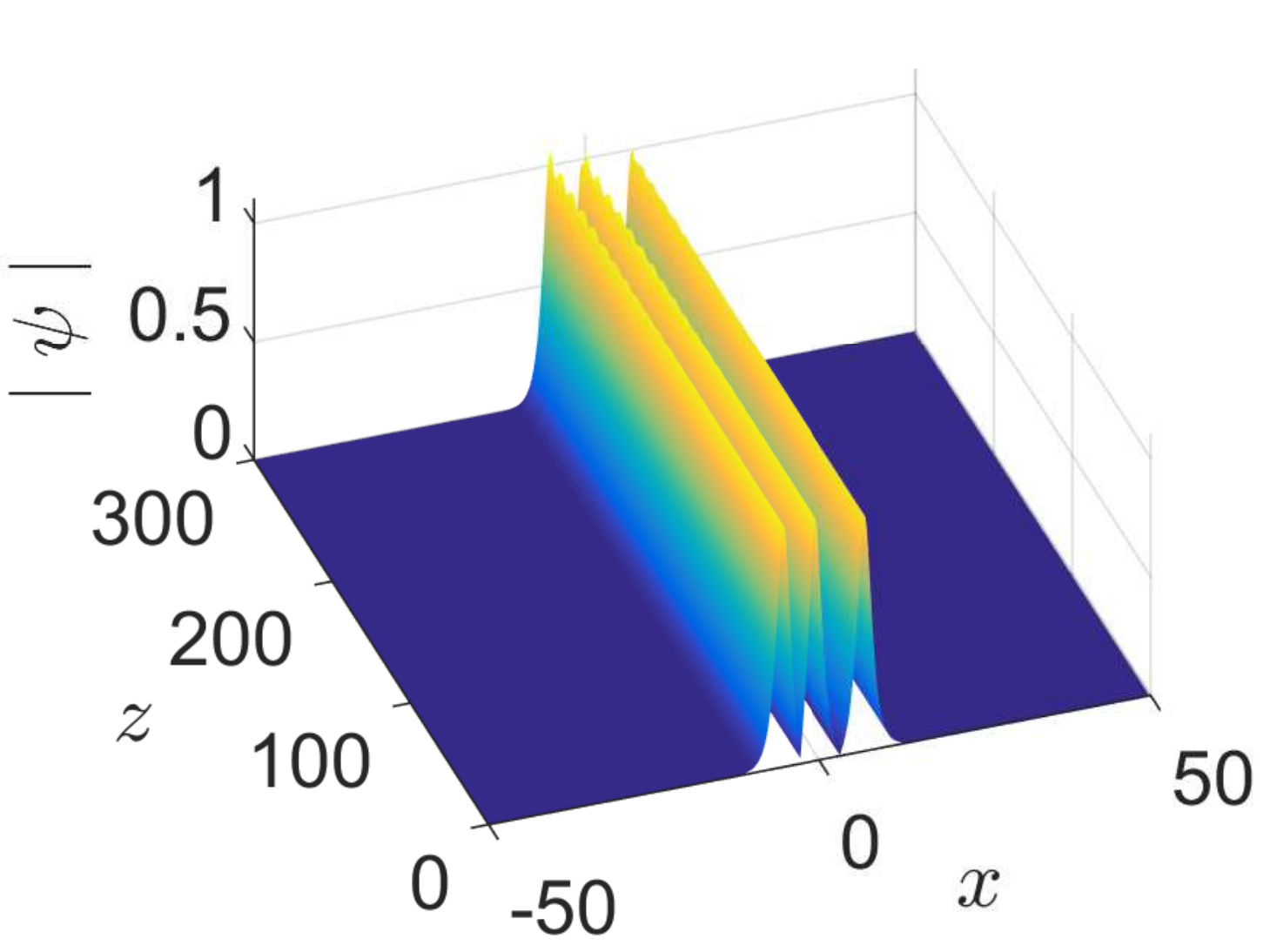}}}
  	\caption{ Bright SW profiles (a), (c), (e) corresponding to intersection points (1), (2) and (3) of Fig. \ref{Fig:7}(a), along with their propagation dynamics (b), (d) and (f) respectively. Dark orange and light grey regions denote the positions of minima and maxima of the refractive index variation respectively. }
 	\label{Fig:8}
\end{center}
\end{figure}

\begin{figure}[pt]
\begin{center}
 	\subfigure[]{\scalebox{\scl}{\includegraphics{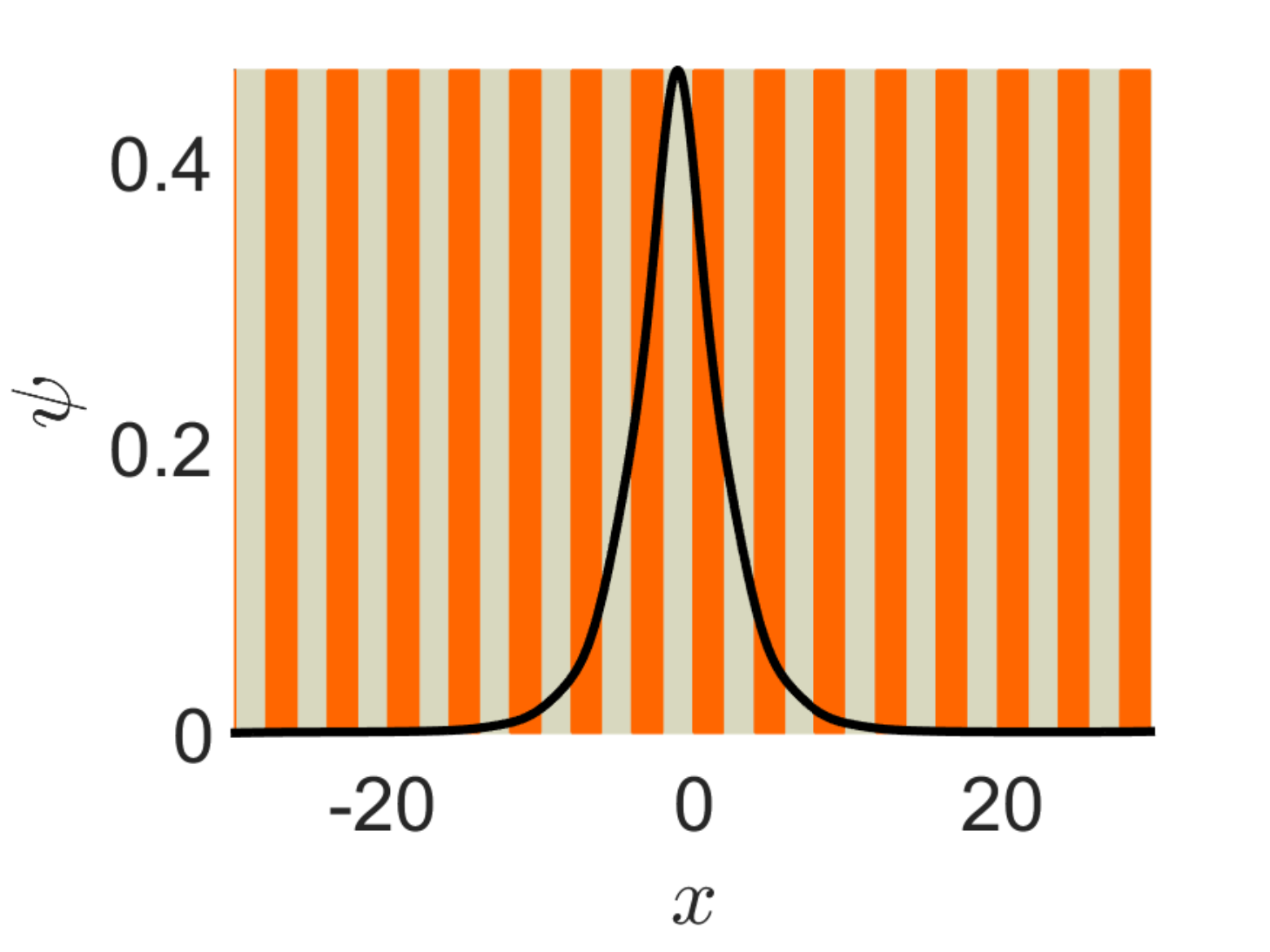}}}
 	\subfigure[]{\scalebox{\scl}{\includegraphics{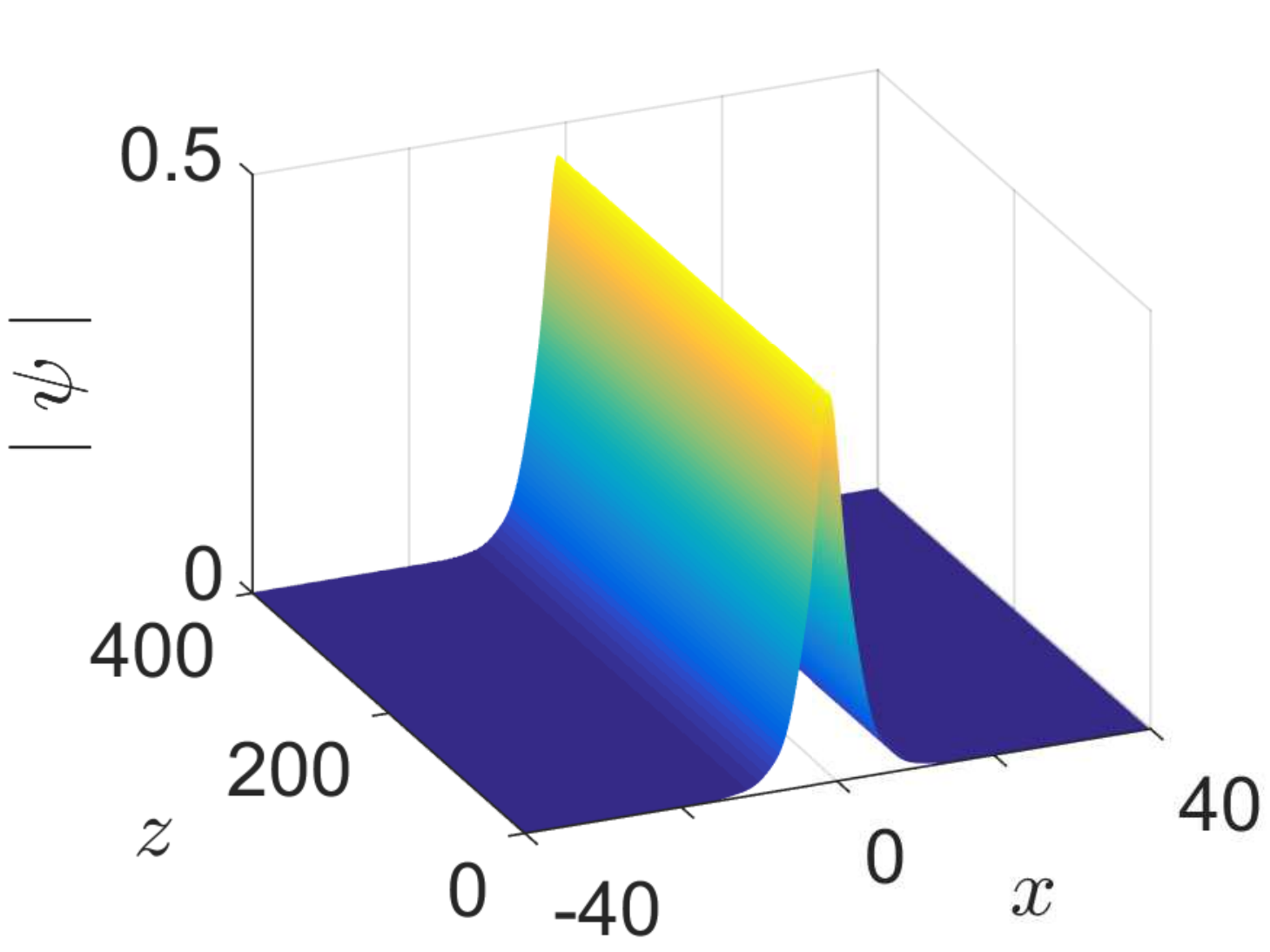}}}\\
 	\subfigure[]{\scalebox{\scl}{\includegraphics{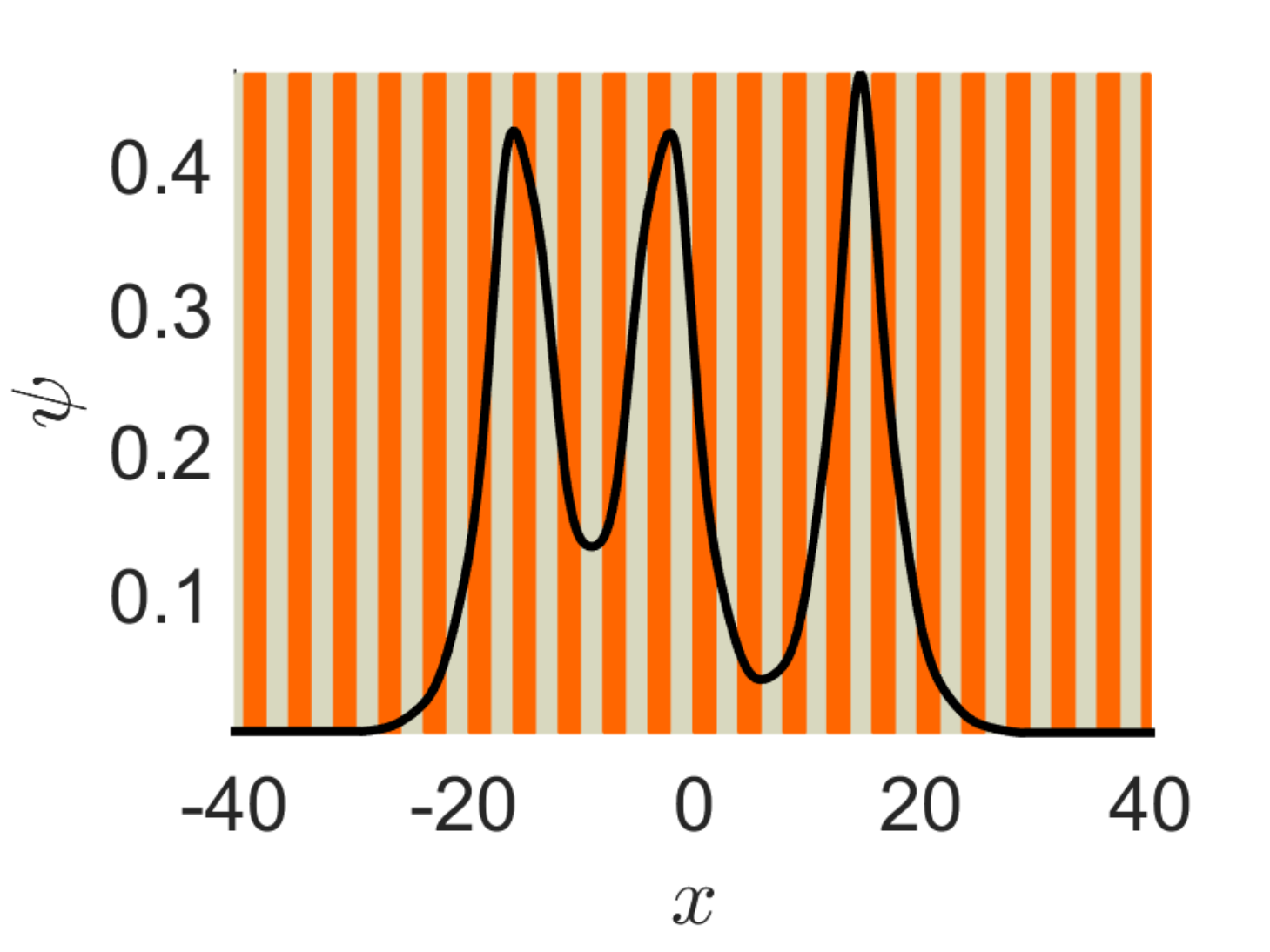}}}
  	\subfigure[]{\scalebox{\scl}{\includegraphics{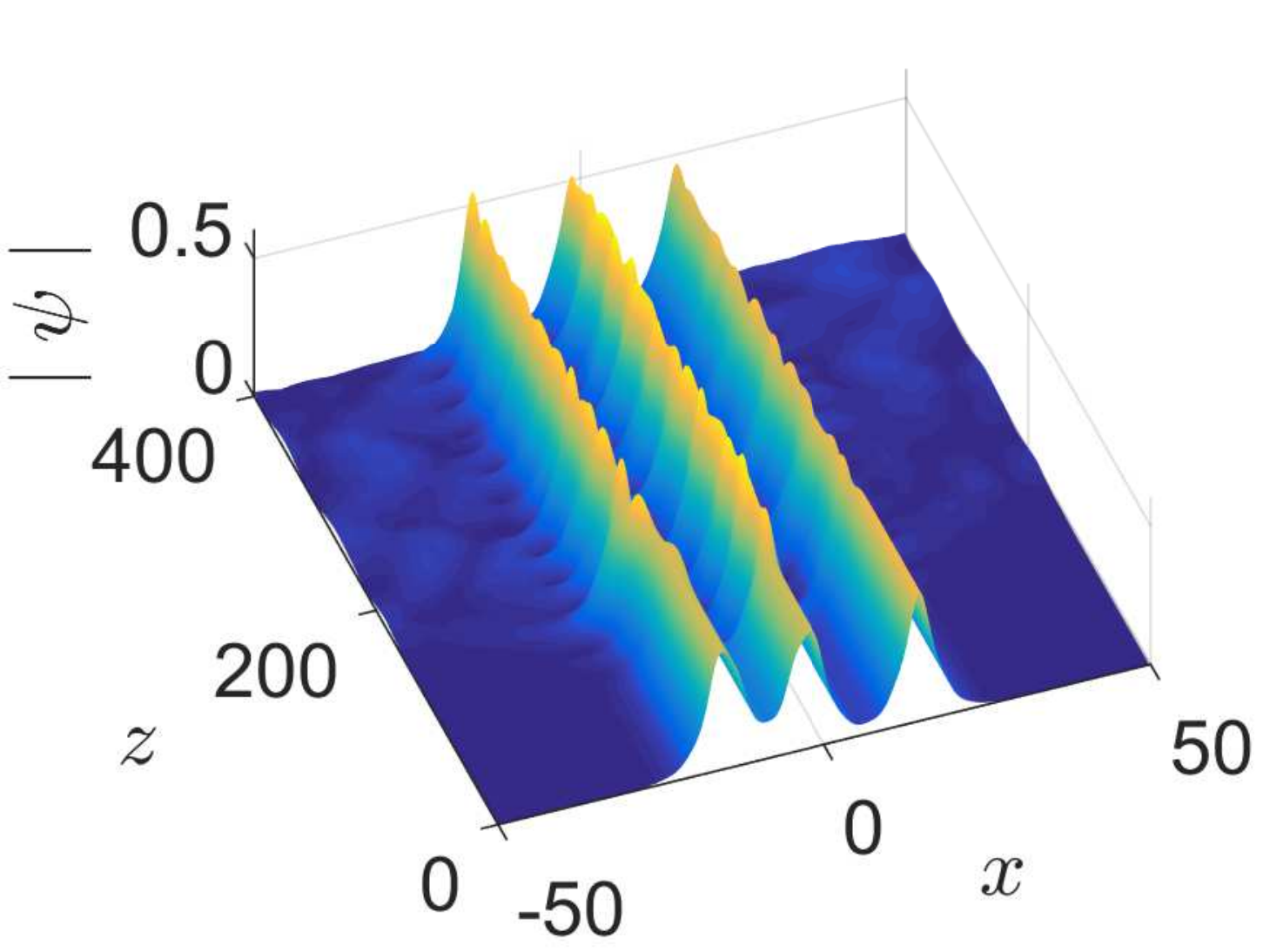}}}
  	\caption{Bright SW profiles (a), (c) corresponding to intersection points (1) and (2) of Fig. \ref{Fig:7}(b), along with their propagation dynamics (b) and (d) respectively. Dark orange and light grey regions denote the positions of minima and maxima of the refractive index variation respectively. }
 	\label{Fig:9}
\end{center}
\end{figure}

\begin{figure}[pt]
\begin{center}
 	\subfigure[]{\scalebox{\scl}{\includegraphics{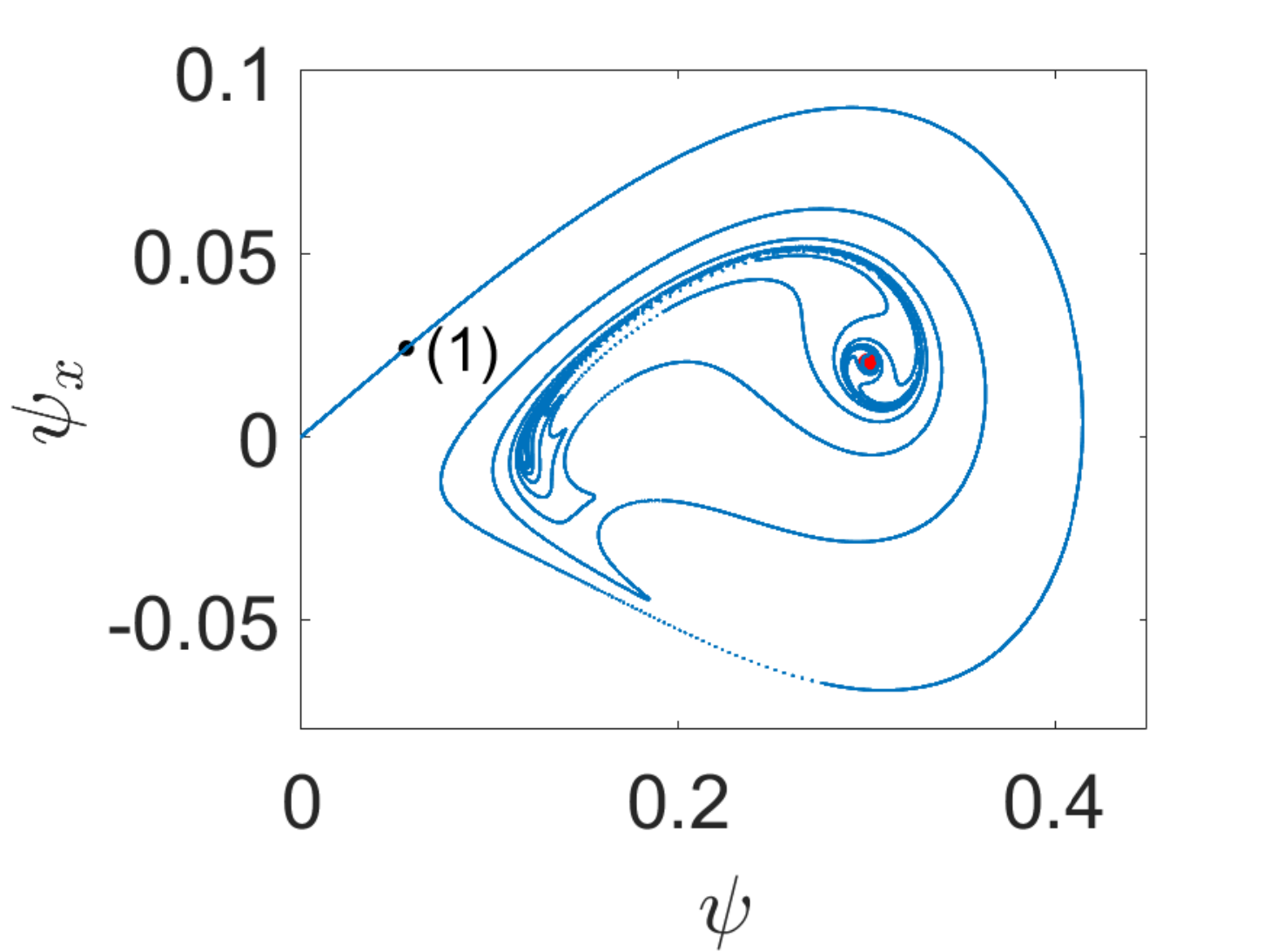}}}
 	\subfigure[]{\scalebox{\scl}{\includegraphics{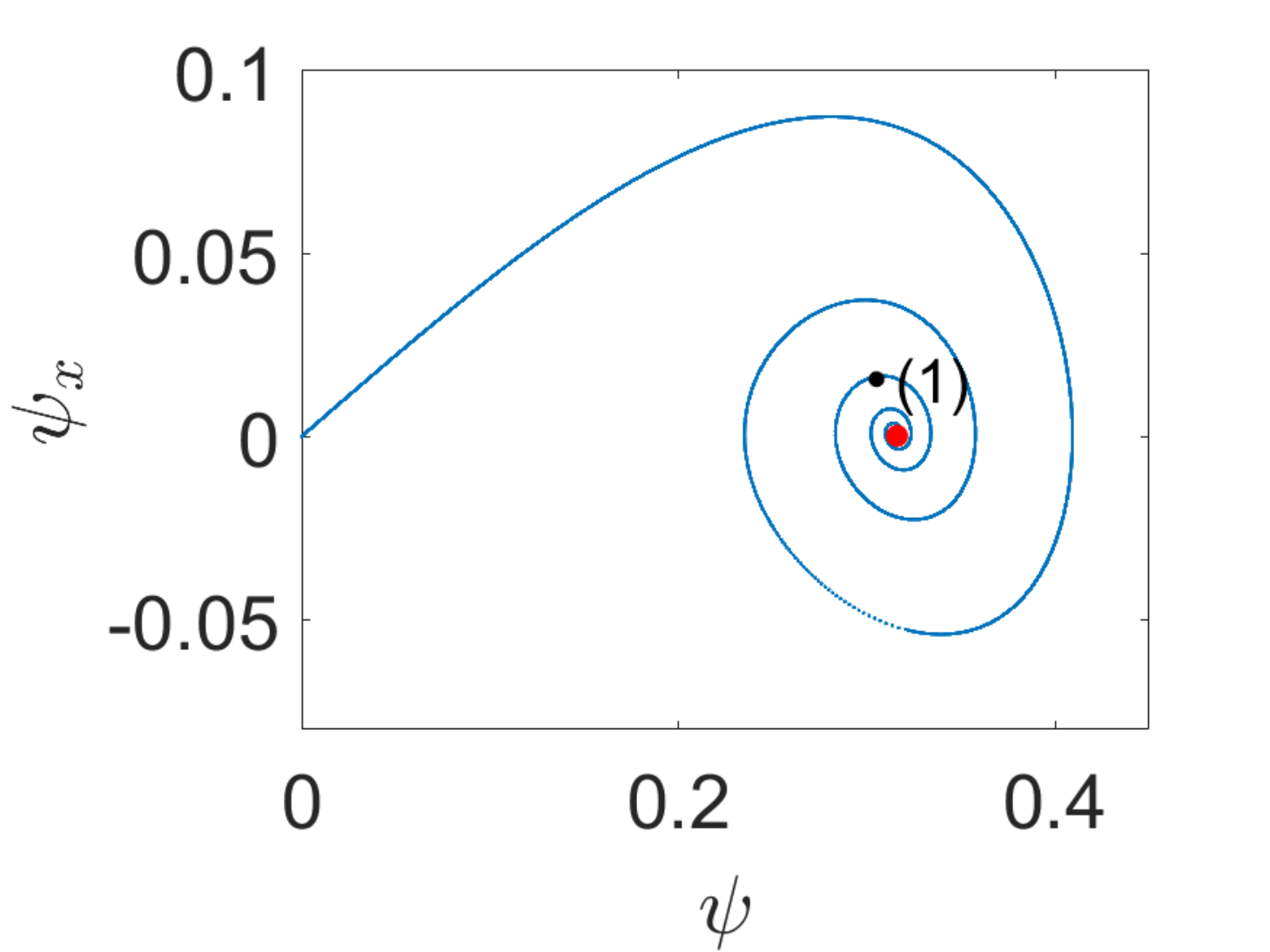}}}\\
 	\subfigure[]{\scalebox{\scl}{\includegraphics{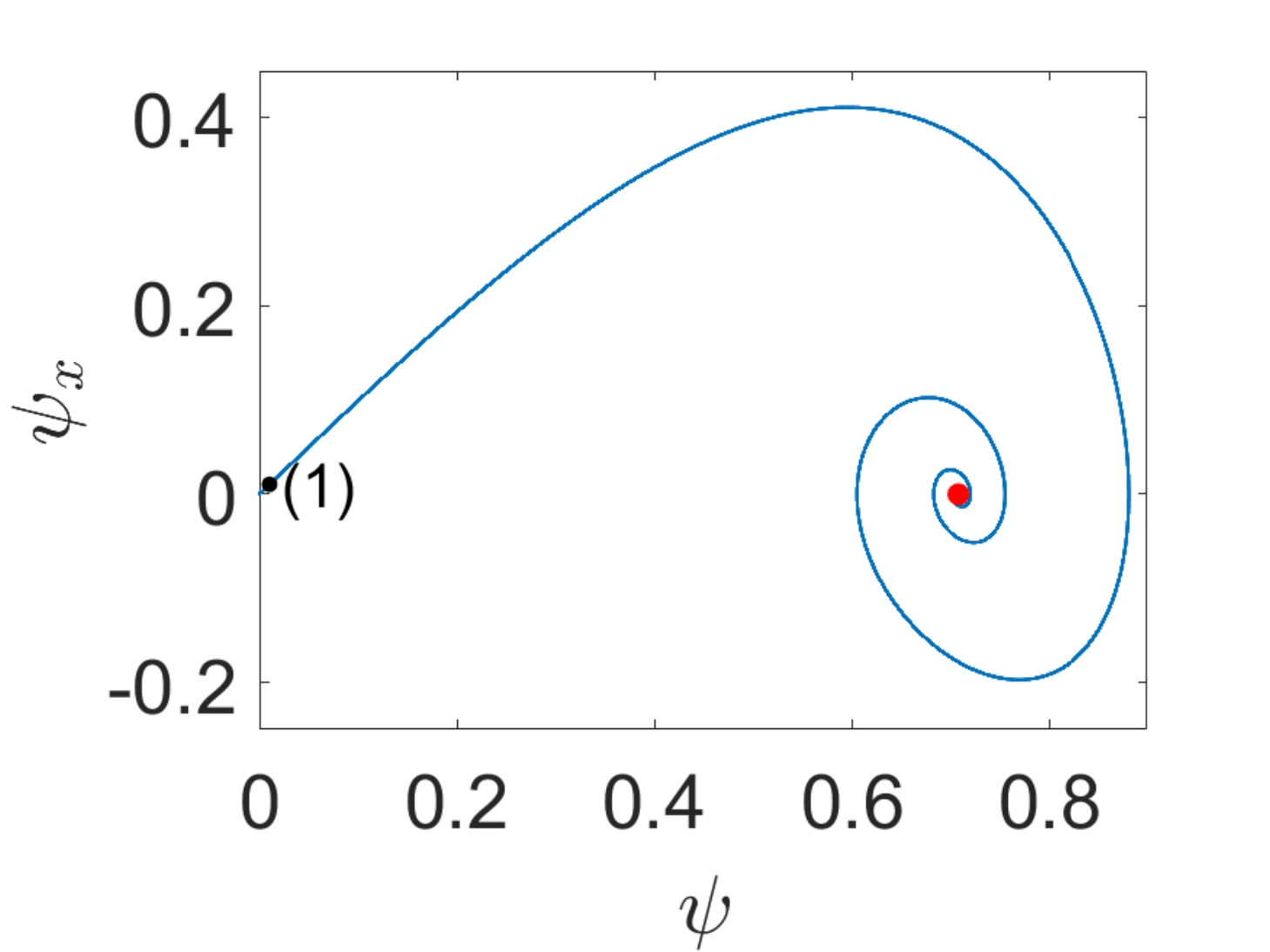}}}
  	\subfigure[]{\scalebox{\scl}{\includegraphics{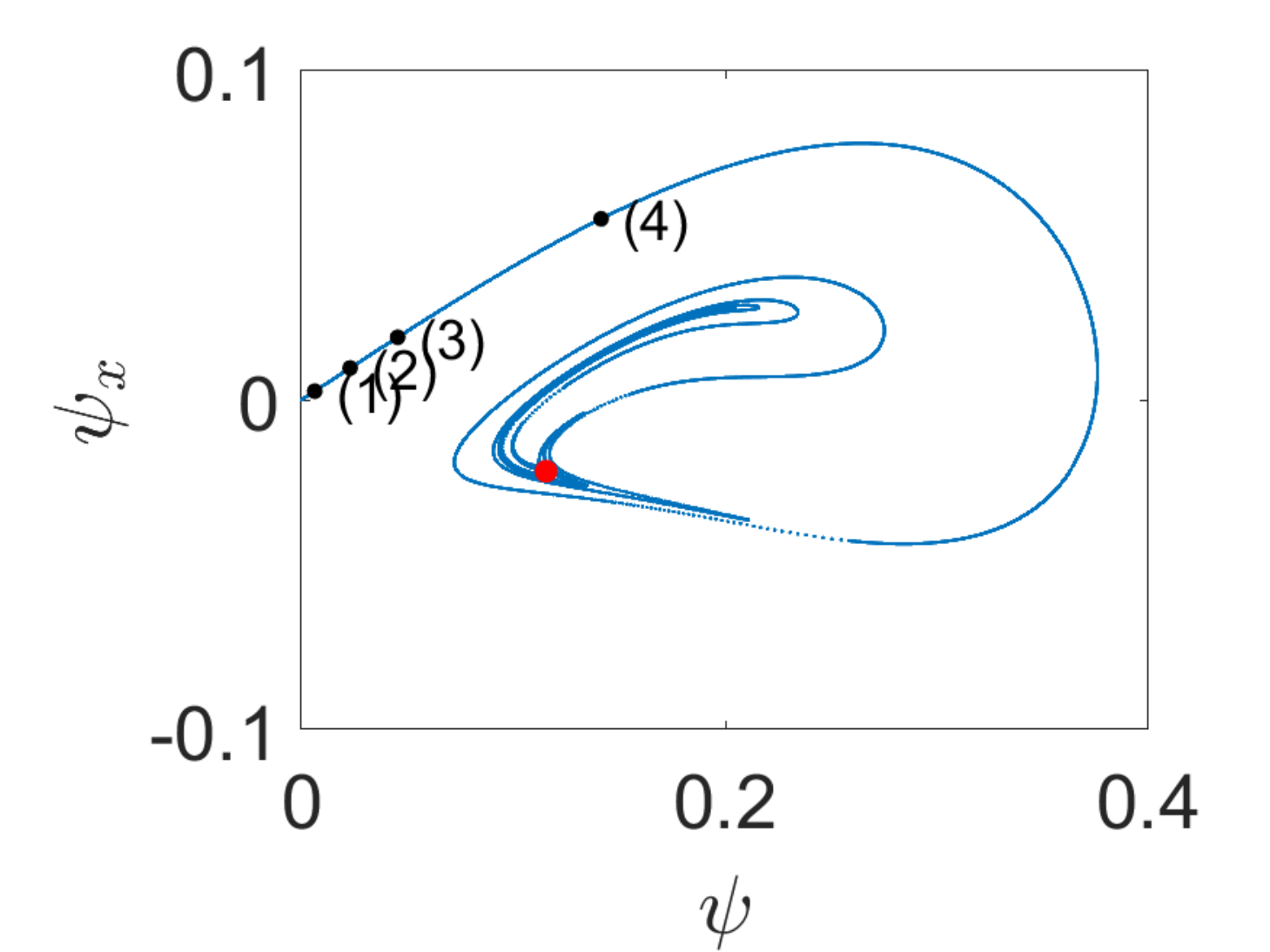}}}
  	\caption{Smooth join of the unstable manifold of the saddle at the origin with the stable manifold of a hyperbolic resonant periodic orbit depicted on a Poincare Section for the parameter sets $(\beta=0.2,\gamma=0.175, A=0.02,k=0.5)$ (a), $(\beta=0.2,\gamma=0.425,A=0.001,k=0.5)$ (b), $(\beta=1,\gamma=0.3,A=0.001,k=3)$ (c) and $(\beta=0.2, \gamma=0.4, Α=0.05, k= 0.5)$ (d).}
 	\label{Fig:10}
\end{center}
\end{figure} 

\begin{figure}[pt]
\begin{center}
 	\subfigure[]{\scalebox{\scl}{\includegraphics{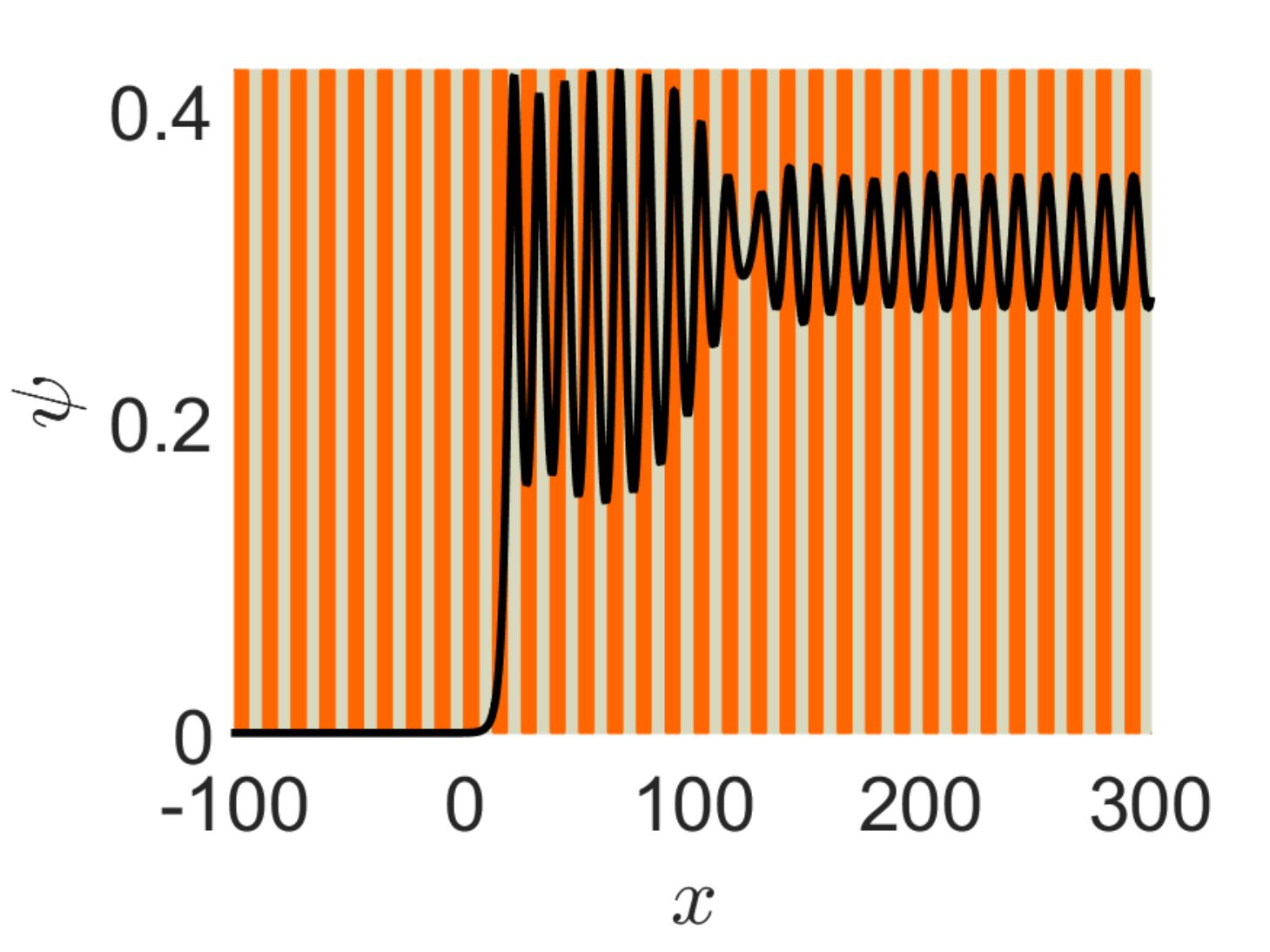}}}
 	\subfigure[]{\scalebox{\scl}{\includegraphics{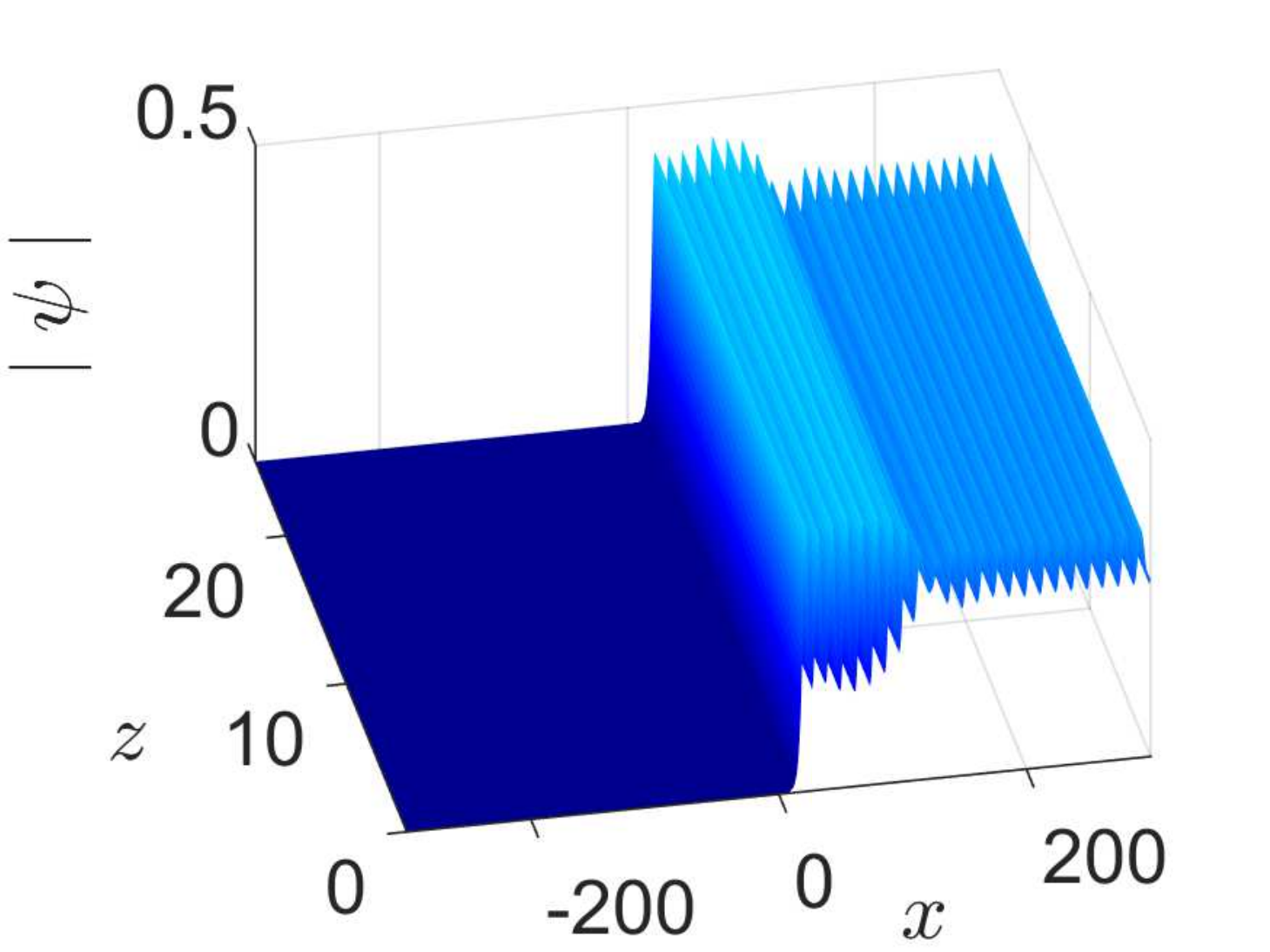}}}\\
 	\subfigure[]{\scalebox{\scl}{\includegraphics{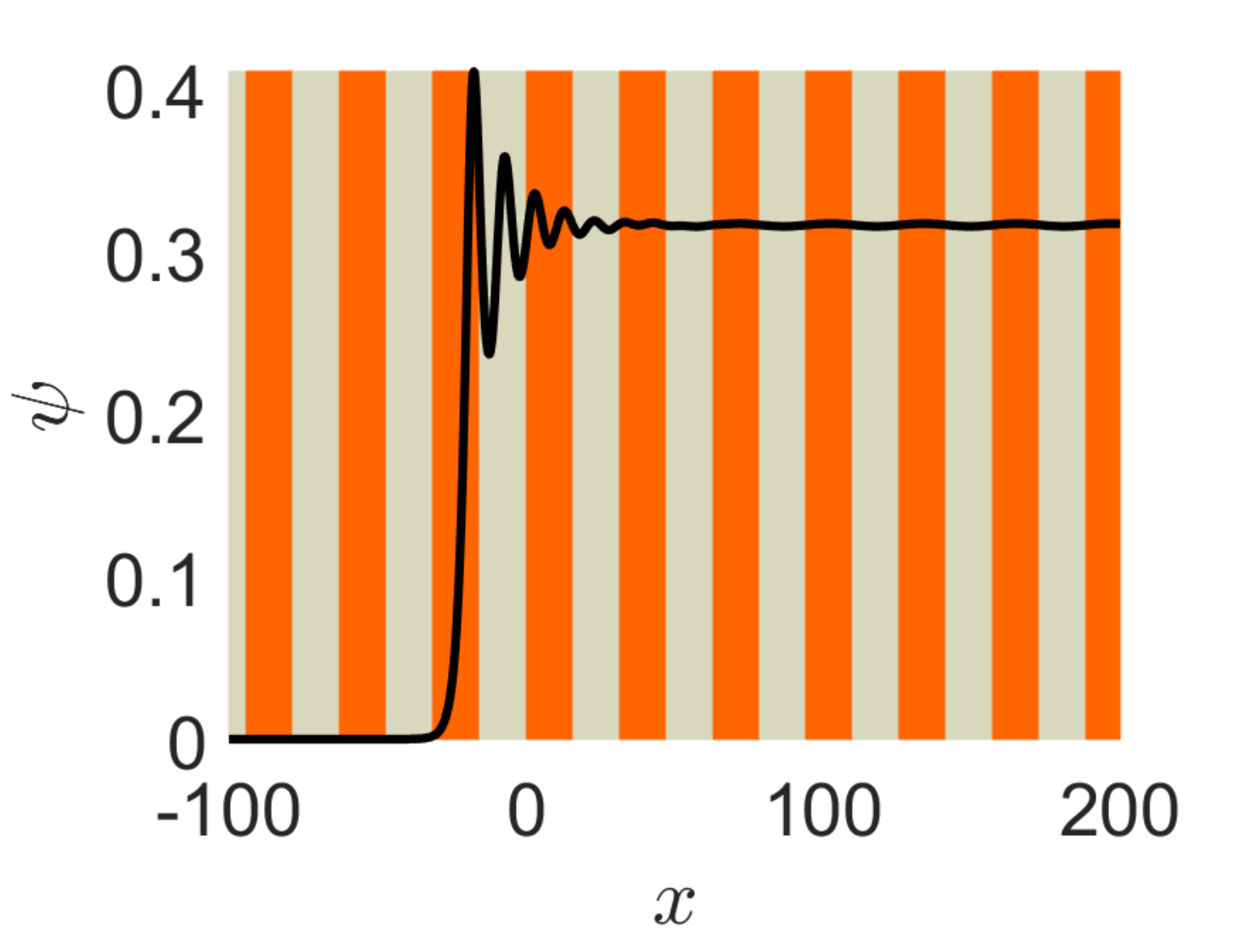}}}
  	\subfigure[]{\scalebox{\scl}{\includegraphics{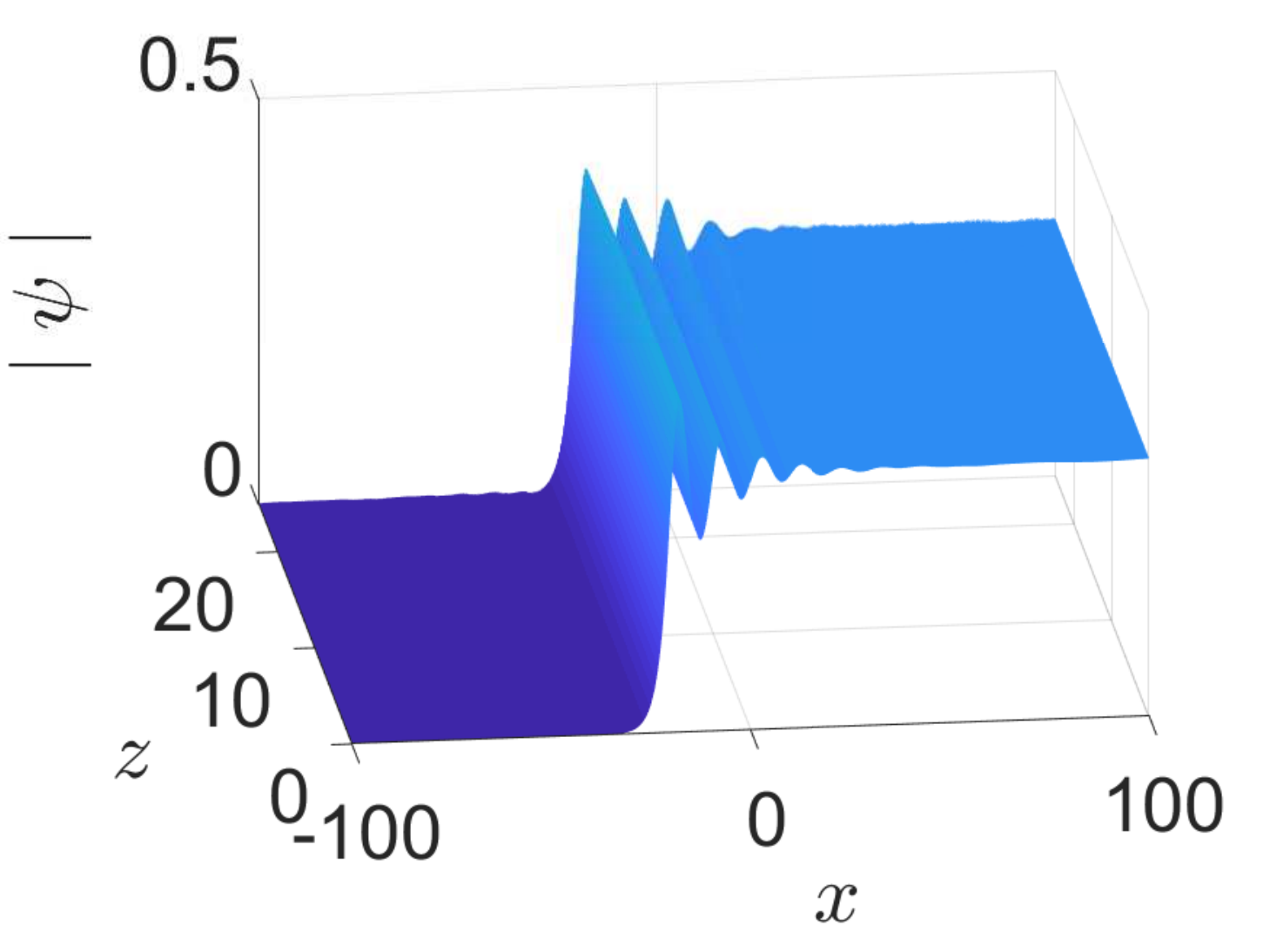}}}\\
  	\subfigure[]{\scalebox{\scl}{\includegraphics{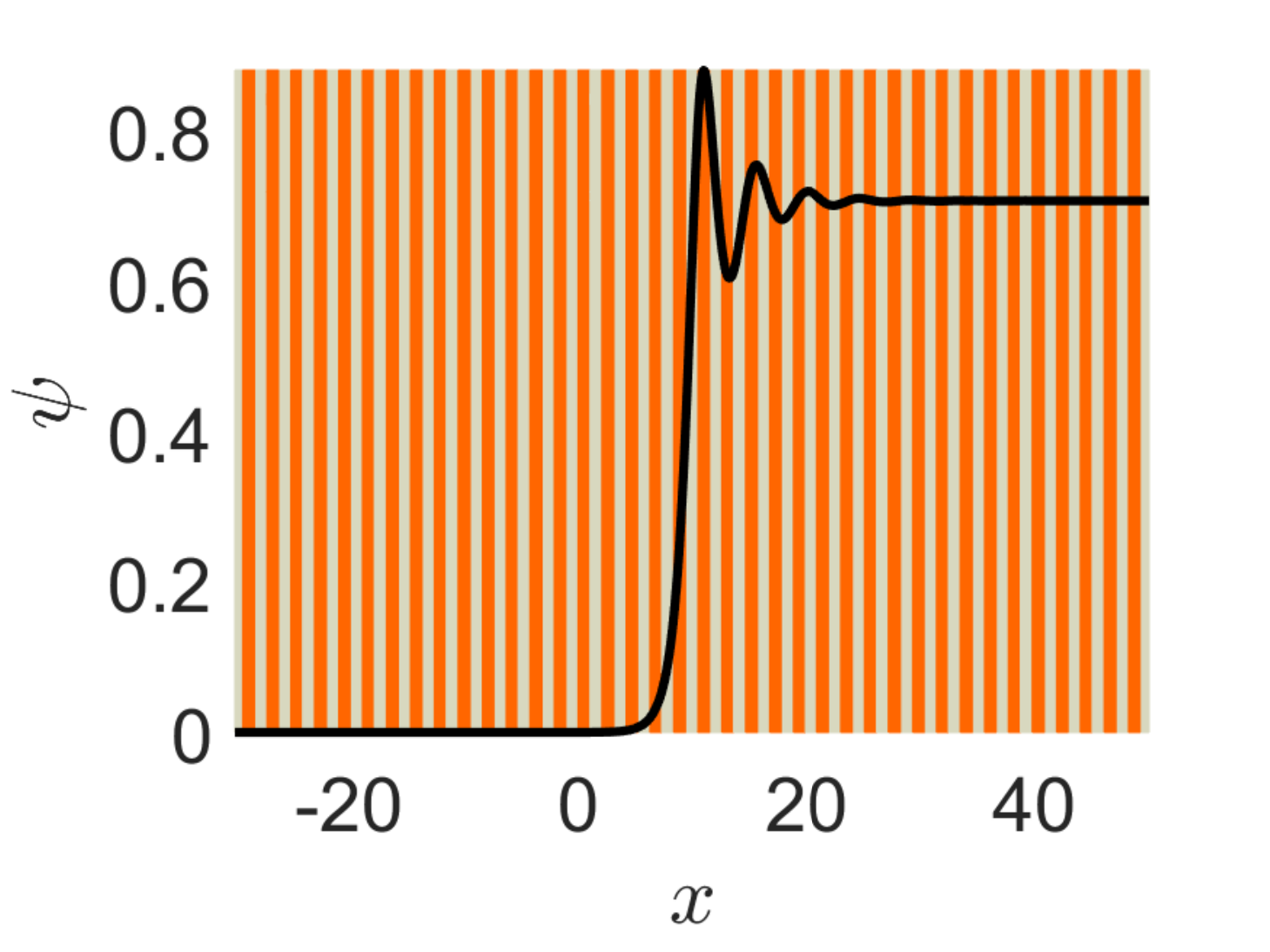}}}
  	\subfigure[]{\scalebox{\scl}{\includegraphics{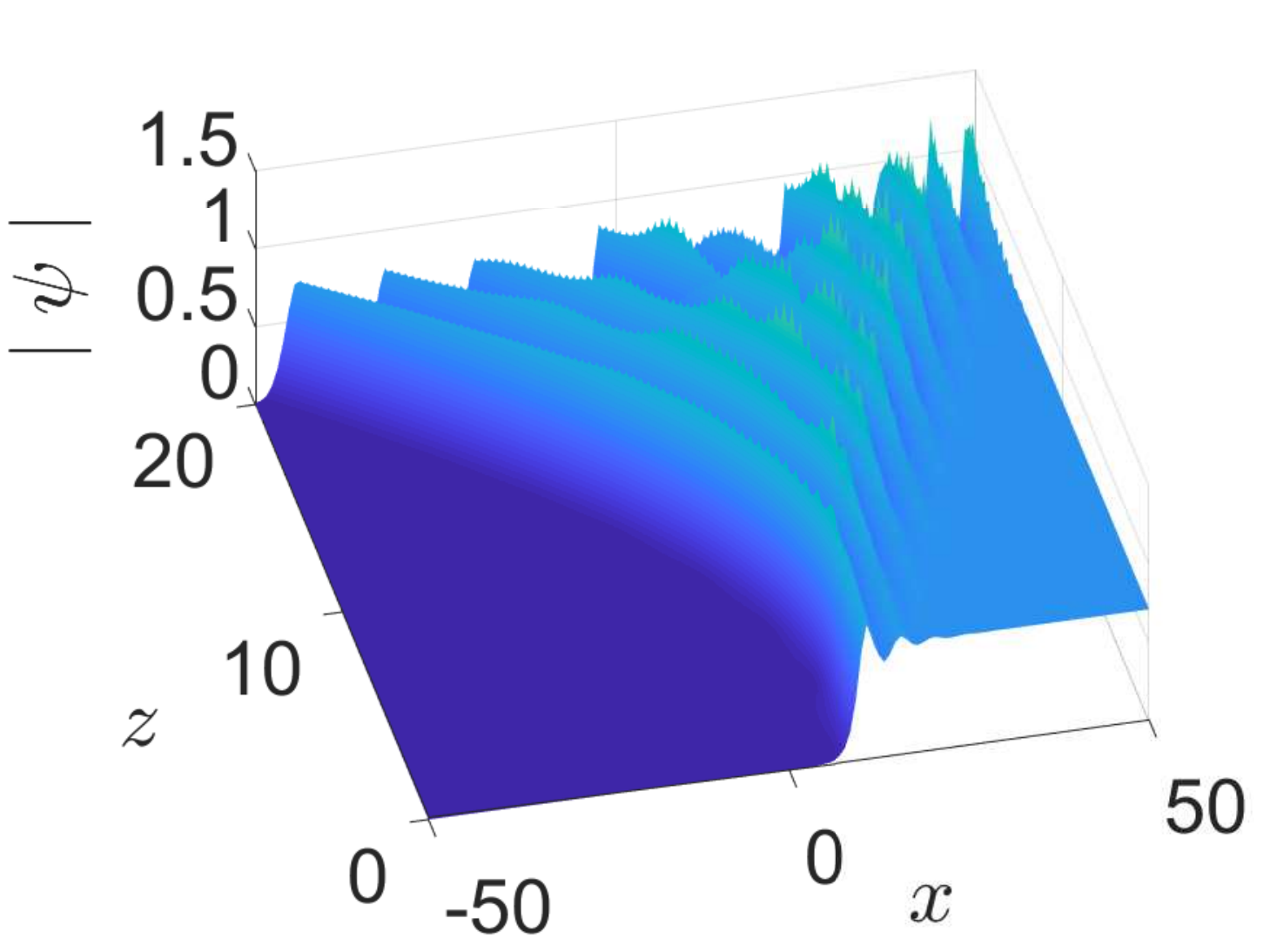}}}
  	\caption{Kink solitary wave profiles corresponding to the parameter sets $(\beta=0.2,\gamma=0.175,A=0.02,k=0.5)$ (a), $(\beta=0.2,\gamma=0.425,A=0.001,k=0.5)$ (c) and $(\beta=1,\gamma=0.3,A=0.001,k=3)$ (e), along with their propagation dynamics (b), (d) and (f). Dark orange and light grey regions denote the positions of minima and maxima of the refractive index variation respectively. }
 	\label{Fig:11}
\end{center}
\end{figure} 

\begin{figure}[pt]
\begin{center}
 	\subfigure[]{\scalebox{\scl}{\includegraphics{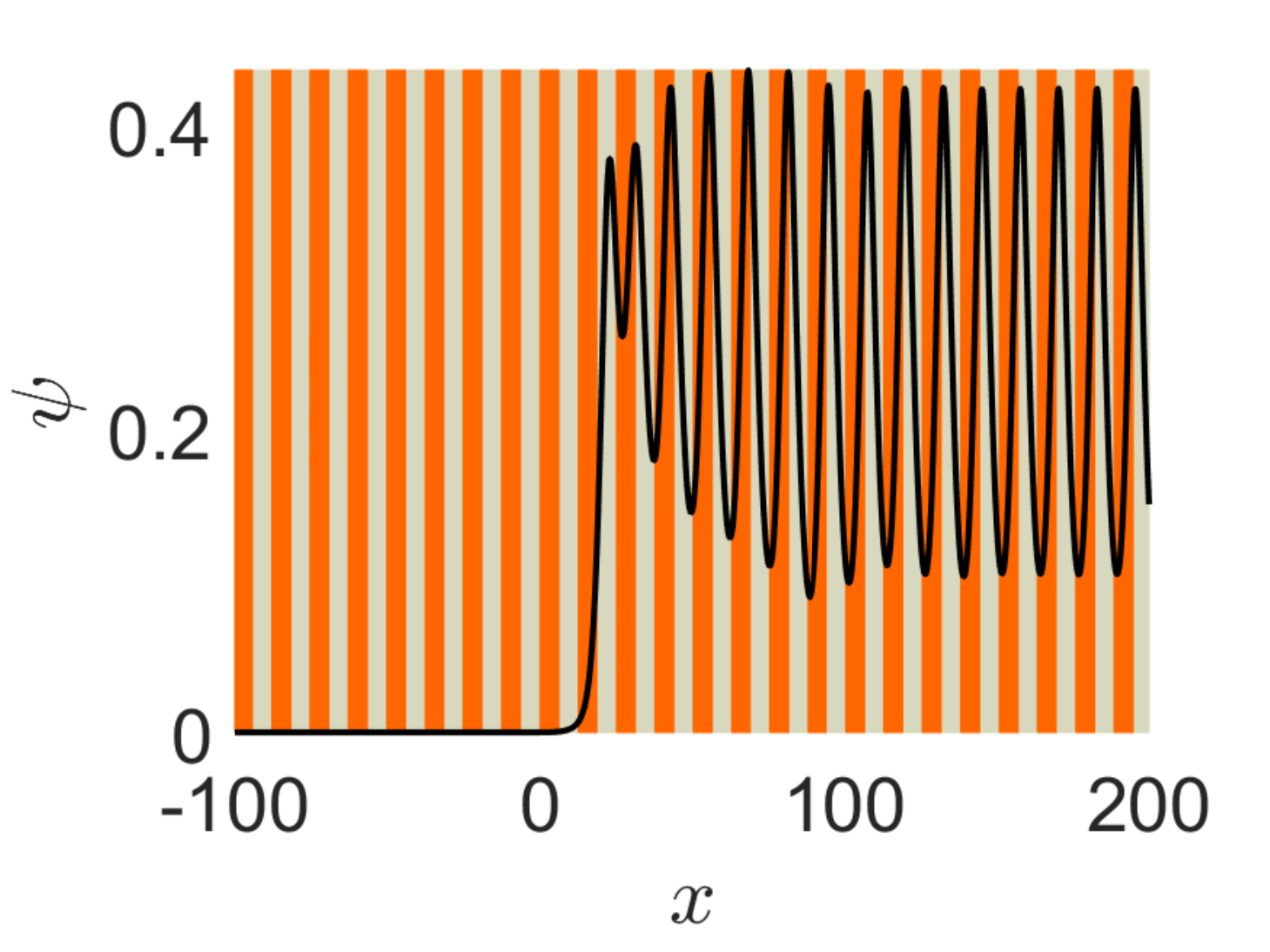}}}
 	\subfigure[]{\scalebox{\scl}{\includegraphics{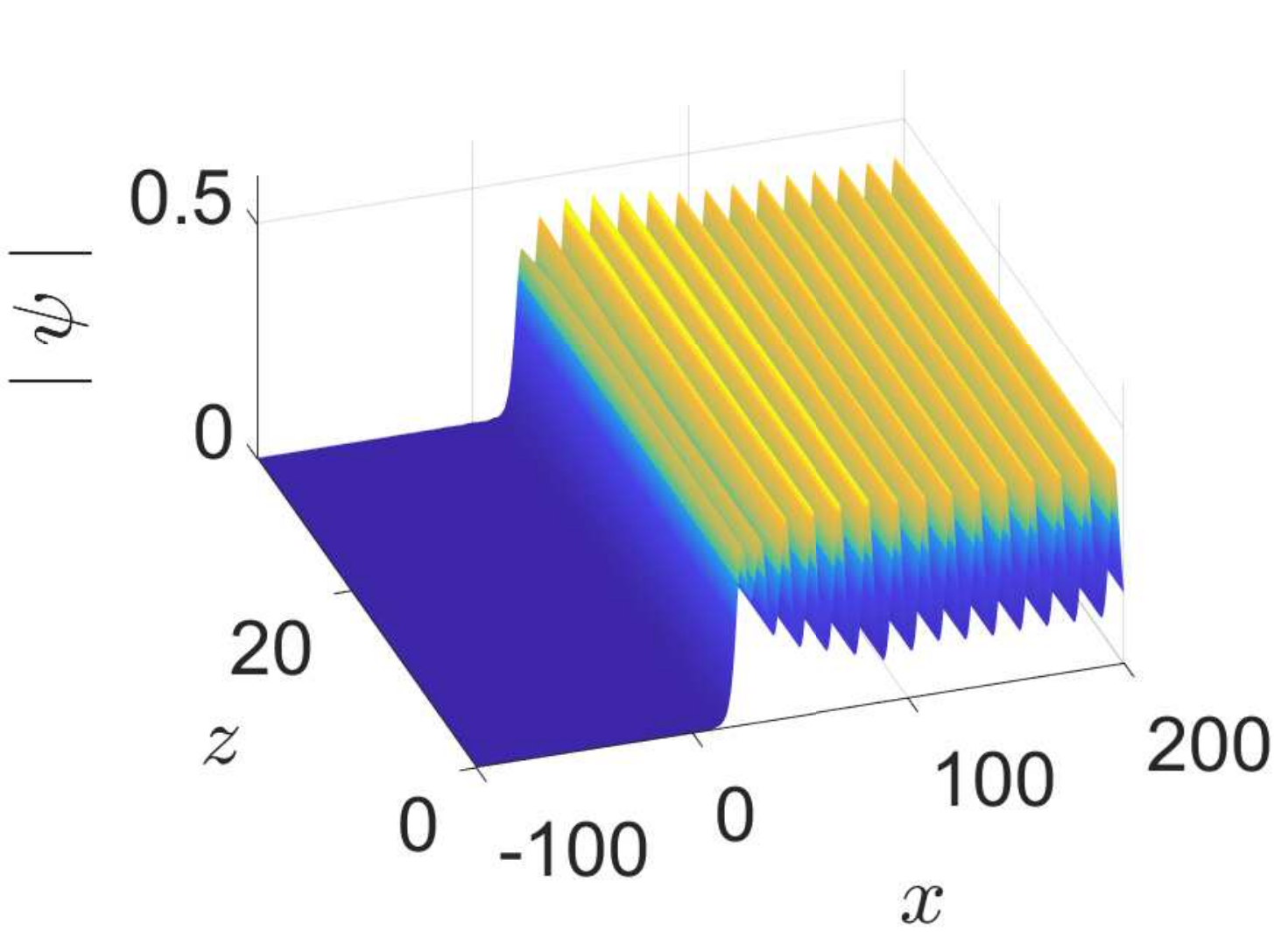}}}\\
 	\subfigure[]{\scalebox{\scl}{\includegraphics{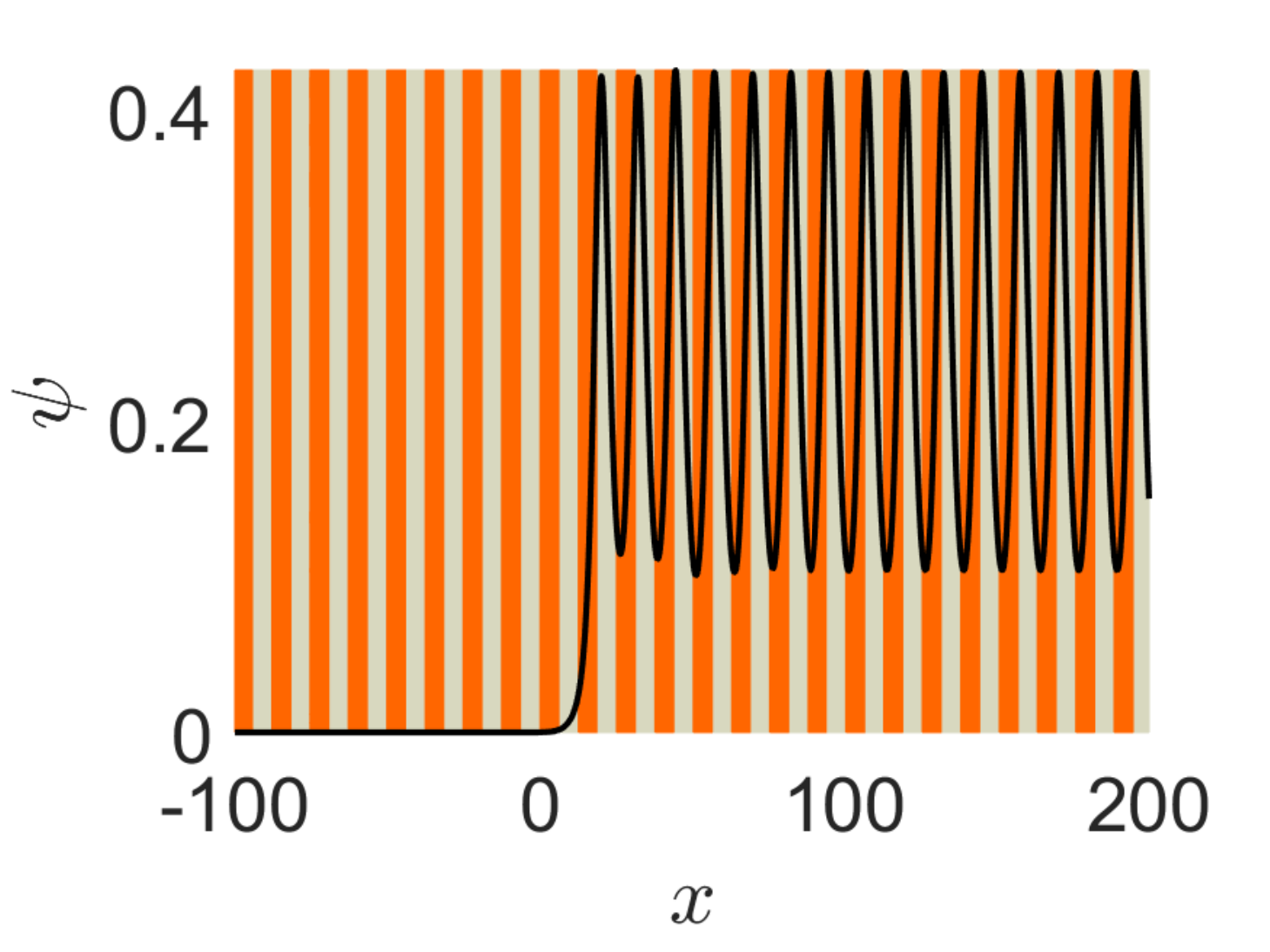}}}
  	\subfigure[]{\scalebox{\scl}{\includegraphics{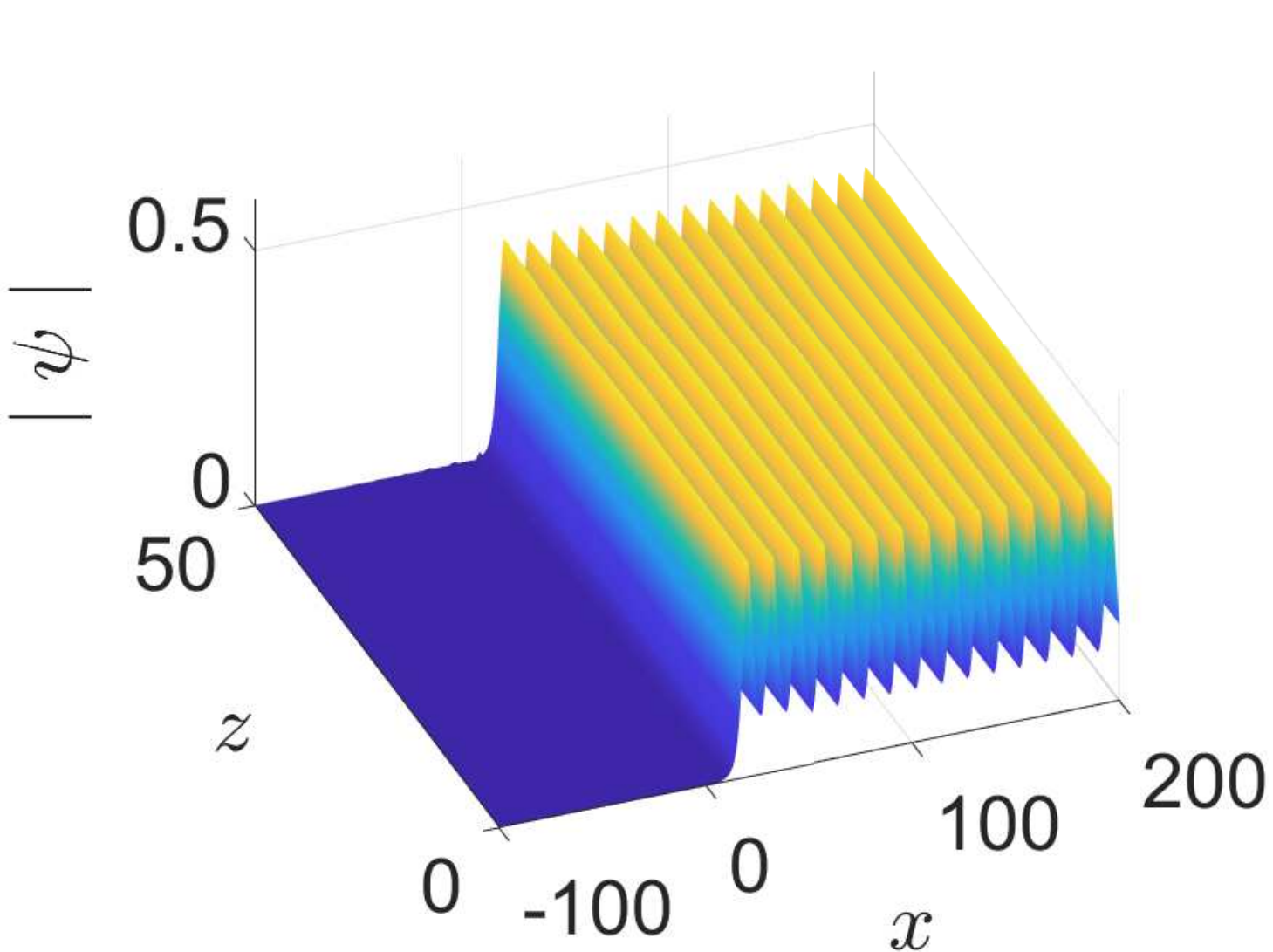}}}\\
  	\subfigure[]{\scalebox{\scl}{\includegraphics{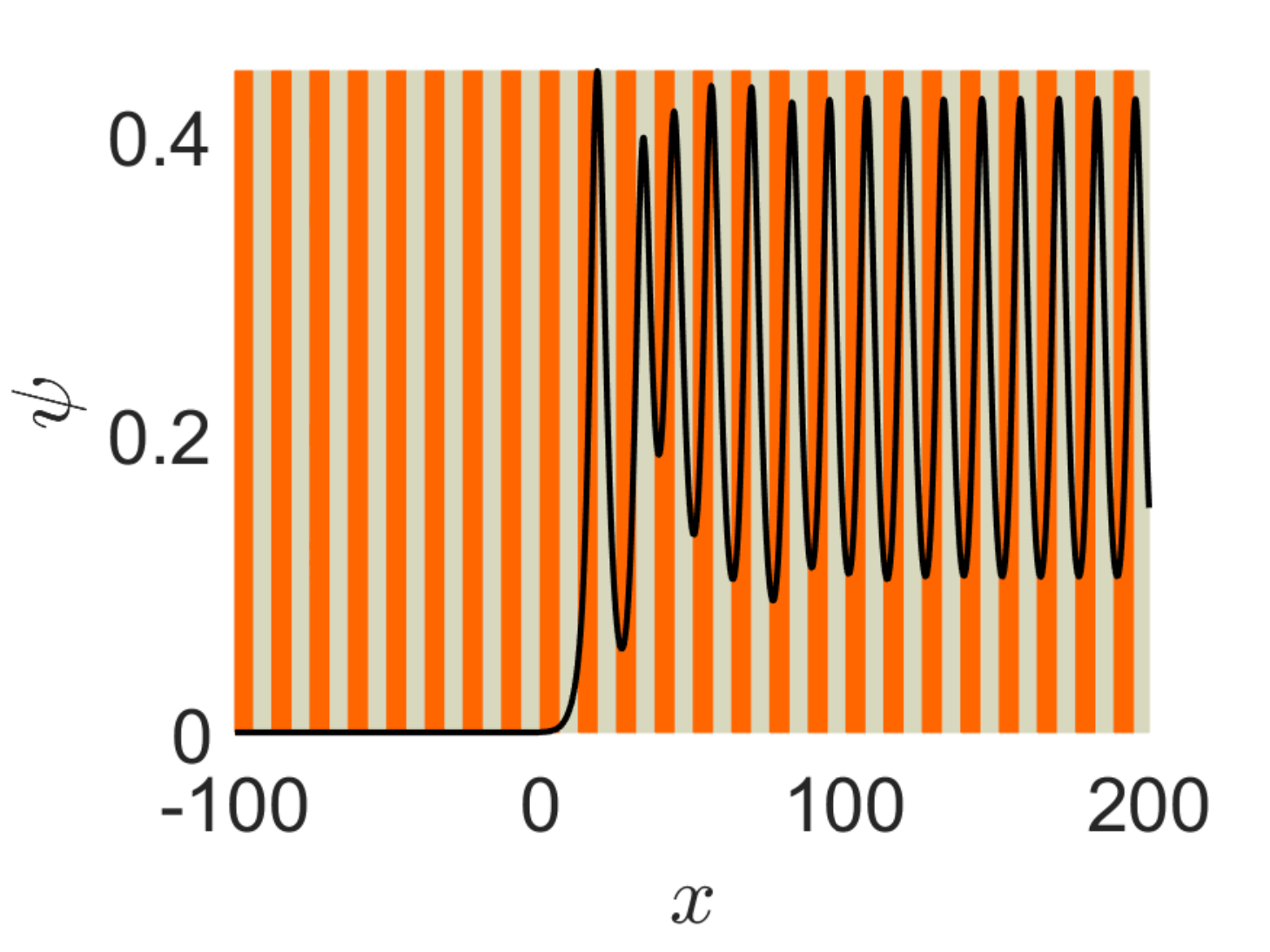}}}
 	\subfigure[]{\scalebox{\scl}{\includegraphics{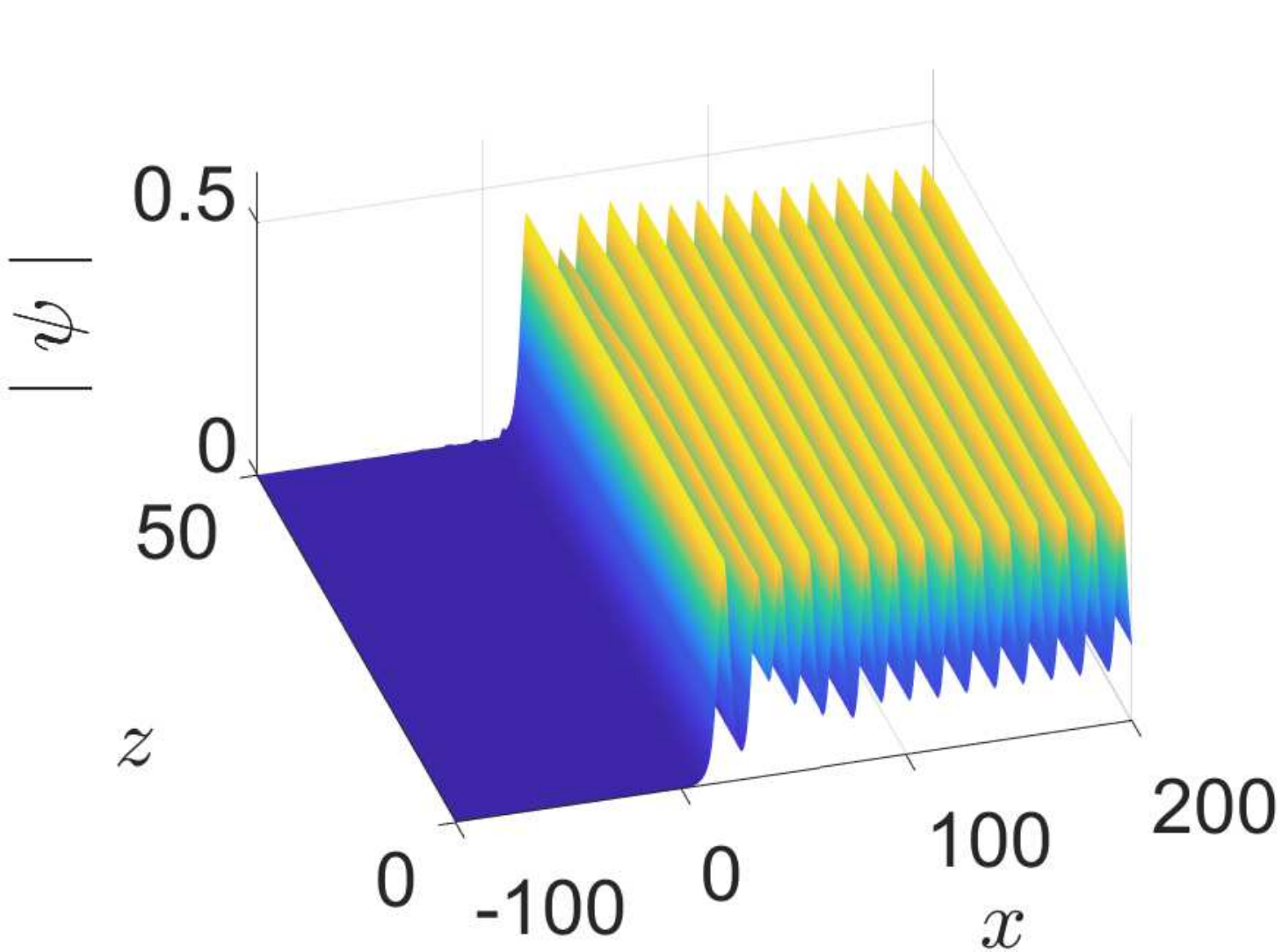}}}\\
 	\subfigure[]{\scalebox{\scl}{\includegraphics{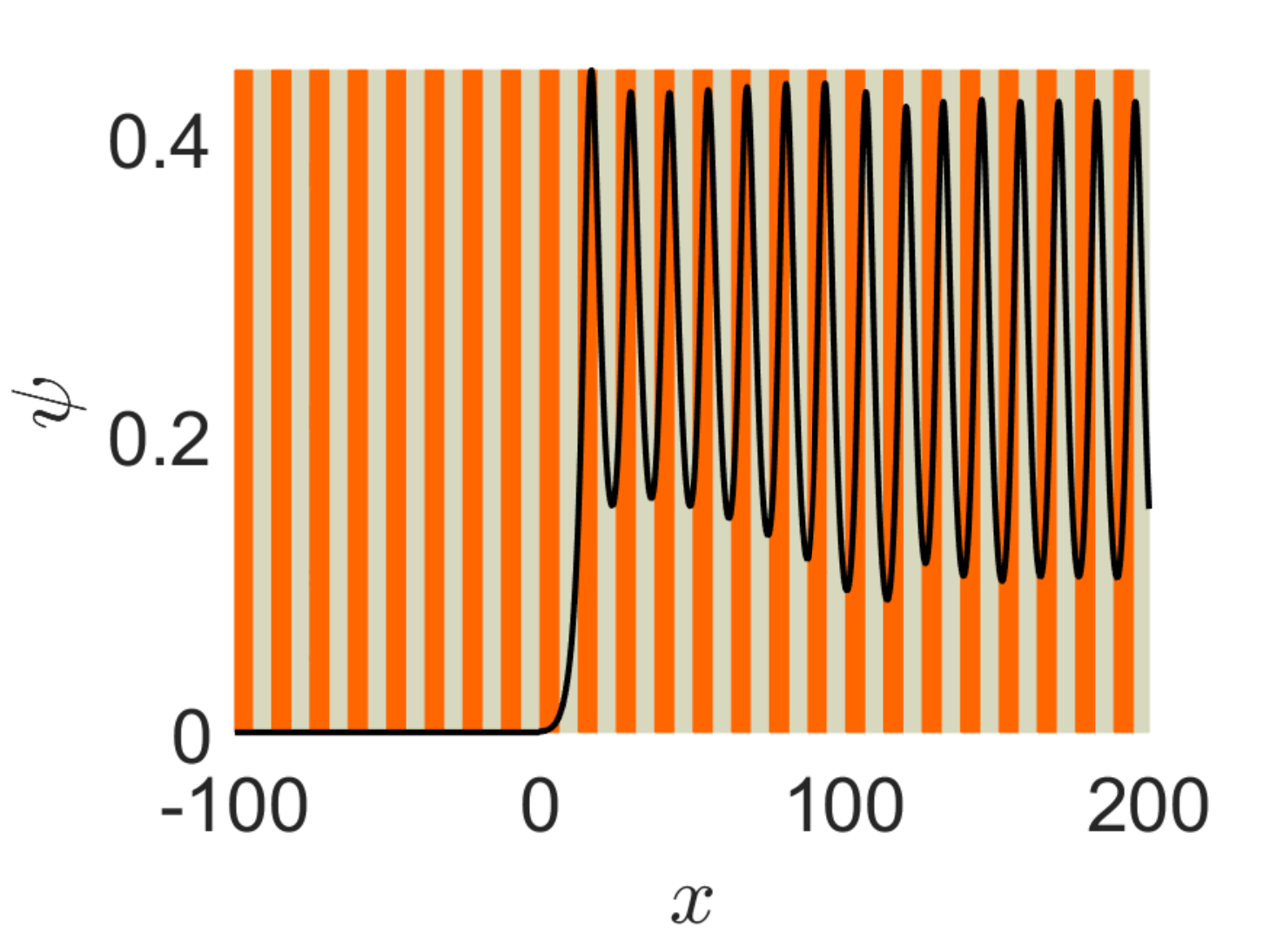}}}
 	\subfigure[]{\scalebox{\scl}{\includegraphics{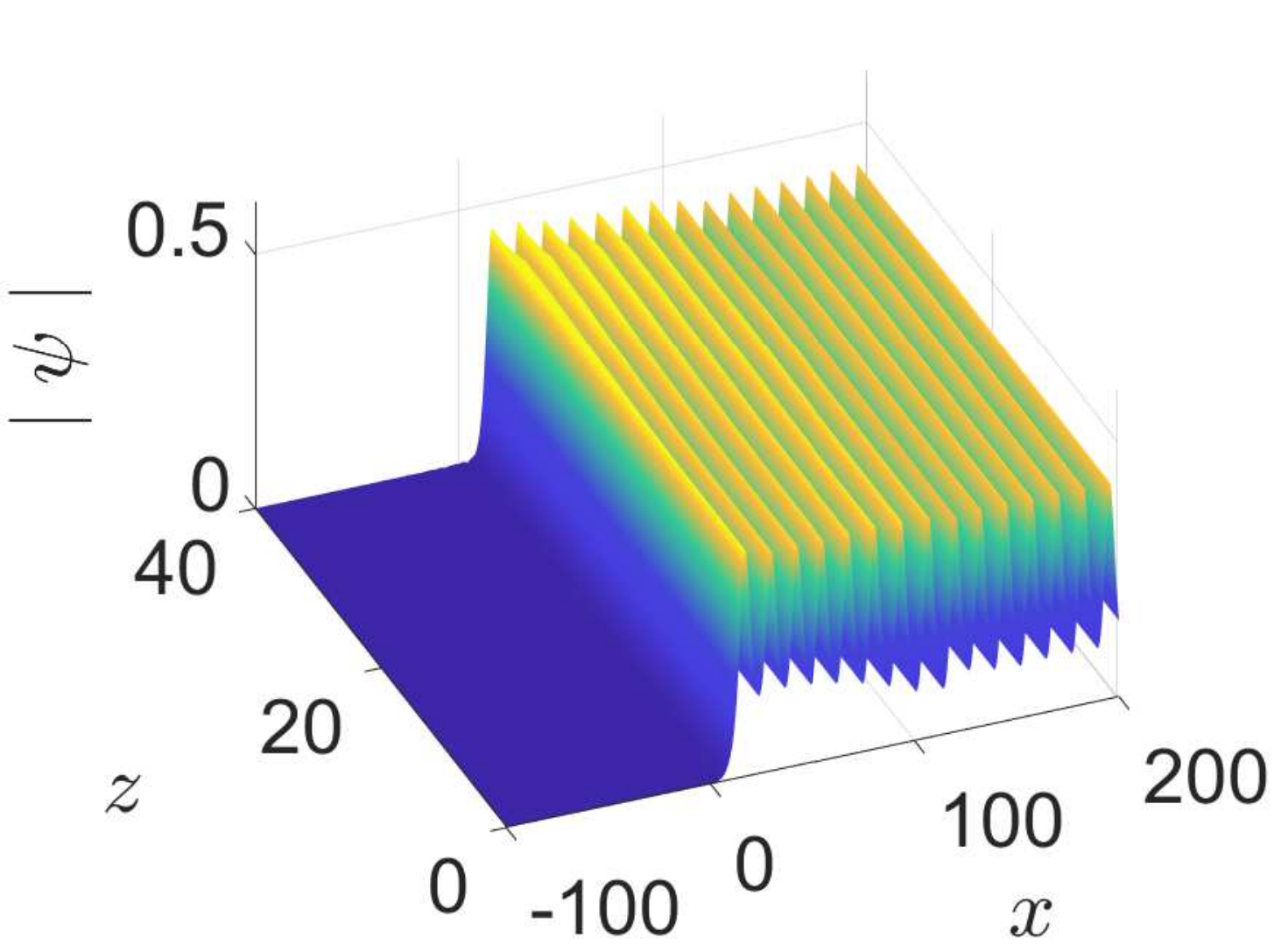}}}
  	\caption{Kink solitary wave profiles (a), (c), (e) corresponding to the parameter sets $(\beta=0.2,\gamma=0.4,A=0.05,k=0.5)$ , along with their propagation dynamics (b), (d) and (f). Dark orange and light grey regions denote the positions of minima and maxima of the refractive index variation respectively. }
 	\label{Fig:12}
\end{center}
\end{figure}
\end{document}